\documentclass[traditabstract]{aa} 

\usepackage{natbib,twoopt}
\usepackage[breaklinks=true]{hyperref} 
\bibpunct{(}{)}{;}{a}{}{,}             

\makeatletter
\newcommandtwoopt{\citeads}[3][][]{\href{http://adsabs.harvard.edu/abs/#3}%
{\def\hyper@linkstart##1##2{}%
\let\hyper@linkend\@empty\citealp[#1][#2]{#3}}}
\newcommandtwoopt{\citepads}[3][][]{\href{http://adsabs.harvard.edu/abs/#3}%
{\def\hyper@linkstart##1##2{}%
\let\hyper@linkend\@empty\citep[#1][#2]{#3}}}
\newcommandtwoopt{\citetads}[3][][]{\href{http://adsabs.harvard.edu/abs/#3}%
{\def\hyper@linkstart##1##2{}%
\let\hyper@linkend\@empty\citet[#1][#2]{#3}}}
\newcommandtwoopt{\citeyearads}[3][][]%
{\href{http://adsabs.harvard.edu/abs/#3}
{\def\hyper@linkstart##1##2{}%
\let\hyper@linkend\@empty\citeyear[#1][#2]{#3}}}
\makeatother

\usepackage[varg]{txfonts}
\usepackage{graphicx}
\usepackage{txfonts}
\usepackage{rotating}
\usepackage{lscape}
\usepackage{lipsum}
\usepackage{color}

\long\def\symbolfootnote[#1]#2{\begingroup%
\def\thefootnote{\fnsymbol{footnote}}\footnote[#1]{#2}\endgroup} 
 
\newcommand{\mic}{$\mu$m}

\newcommand{\Msun}{M$_{\odot}$}

\newcommand{\Odrie}{[O\,{\sc i}]\,63 \mic}

\begin{document} 

        \title{
              The far-infrared behaviour of Herbig Ae/Be discs:\\ \textit{Herschel}\thanks{{\it Herschel} is an ESA space 
    observatory with science instruments provided by European-led Principal Investigator consortia and with 
    important participation from NASA.} PACS photometry}

        \author{N. Pascual\inst{1,2}, B. Montesinos\inst{3}, G. Meeus\inst{1}, J.P. Marshall\inst{4,5,1} , I. Mendigut\'ia\inst{6}, \and G. Sandell\inst{7}
         }
        
        \institute{Dept. F\'isica Te\'orica, Universidad Aut\'onoma de Madrid, Campus Cantoblanco, Spain 
                        \and Department of Physical Sciences, The Open University, Walton Hall, Milton Keynes MK7 6AA, UK
                                \email{natalia.pascual@open.ac.uk}
                       \and Dept. of Astrophysics, CAB (CSIC-INTA), ESAC Campus, P.O. Box 78, 28691 Villanueva de la Ca\~nada, Spain
                       \and School of Physics, University of New South Wales, Sydney, NSW 2052, Australia
                       \and Australian Centre for Astrobiology, University of New South Wales, Sydney, NSW 2052, Australia
                       \and School of Physics \& Astronomy, University of Leeds, Woodhouse Lane, Leeds LS2 9JT, UK
                       \and SOFIA-USRA, NASA Ames Research Center, MS 232-12, Moffett Field, CA 94035-0001, USA }

\date{Received ---; accepted ---}

\abstract{Herbig Ae/Be objects are pre-main sequence stars surrounded by gas- and dust-rich circumstellar discs. These objects are in the throes of star and planet formation, and their characterisation informs us of the processes and outcomes of planet formation processes around intermediate mass stars. Here we analyse the spectral energy distributions of disc host stars observed by the Herschel Open Time Key Programme `Gas in Protoplanetary Systems'. We present \textit{Herschel/PACS} far-infrared imaging observations of 22 Herbig Ae/Bes and 5 debris discs, combined with ancillary photometry spanning ultraviolet to sub-millimetre wavelengths. From these measurements we determine the diagnostics of disc evolution, along with the total excess, in three regimes spanning near-, mid-, and far-infrared wavelengths. Using appropriate statistical tests, these diagnostics are examined for correlations. We find that the far-infrared flux, where the disc becomes optically thin, is correlated with the millimetre flux, which provides a measure of the total dust mass. The ratio of far-infrared to sub-millimetre flux is found to be greater for targets with discs that are brighter at millimetre wavelengths and that have steeper sub-millimetre slopes. Furthermore, discs with flared geometry have, on average, larger excesses than flat geometry discs. Finally, we estimate the extents of these discs (or provide upper limits) from the observations.}

         \titlerunning{Far-infrared photometry of HAeBes}
         \authorrunning{Pascual et al.}
        
         \maketitle

         \section{Introduction}

Circumstellar discs are important structures not only in the context of star formation (e.g. accretion onto a proto-star, or fragmentation), but also in the formation of planetary systems (i.e. proto-planetary or debris discs). 
Evidence of planet formation in this environment is provided by the presence of trapping of material by gravitationally induced structures, such as spirals and gaps \citep[e.g.][]{Garufi2013}. 
Understanding their evolution and structure is therefore critical to tracing the processes involved in planet formation. 
Herbig Ae/Be objects \citep[HAeBe;][]{Herbig1960} are pre-main sequence stars of intermediate mass (2 to 8 $M_{\odot}$). The nascent star is surrounded by a massive disc of gas- and dust-rich material \citep{WilliamsCieza2011}. 
This material emits strongly at infrared wavelengths because of thermal emission from the dust \citep[e.g.][]{Waters1998}. The evolution of this excess as a function of both wavelength and time provides a suite of diagnostics to trace changes in the underlying architecture of the disc and its composition, e.g. by grain growth, clearing, and settling \citep{WilliamsCieza2011}. 

HAeBe discs were identified by the \textit{InfraRed Astronomical Satellite} 
\citep[\textit{IRAS}; e.g.][]{Neugebauer1984}, which detected strong emission at mid- and far-infrared wavelengths \citep{Dong1991,Oudmaijer1992,The1994}.
With its broader wavelength coverage and spectroscopic capabilities compared
to \textit{IRAS}, the \textit{Infrared Space Observatory} \citep[\textit{ISO}; e.g.][]{Kessler1996} provided a better characterisation of these systems. Based on \textit{ISO} observations, \citet{Meeus2001} classified the spectral energy distributions (SEDs) of HAeBe discs into two 
groups based on the shape of their mid- and far-infrared excesses. Group I sources were fitted by a power-law and blackbody component, whereas the group II sources only required a power law. To distinguish between the two groups, a distinctive criterion based on the disc colour (\textit{IRAS} [12] - \textit{IRAS} [60]) and brightness ($L_{\rm NIR}/L_{\rm FIR}$) was developed by \citet{vanBoekel2003}. Initially, the two groups were explained as the result of `flat' (with a constant height vs. radius) and `flared' (with increasing height vs. radius) disc structures \citep[e.g.][]{Dominik2003}. An evolutionary scheme from flared into flat discs was proposed to explain the observations, justified as a result of grain growth and settling towards the disc mid-plane \citep{Chiang2001}.

The gas in a disc is heated by radiation re-emitted by dust grains within the disc, and the vertical height of a disc is determined by the pressure of the gas \citep[which is a function of its temperature, e.g.][]{Kamp2004}. 
Dust grains smaller than 25~$\mu$m are responsible for the disc opacity, and it is these smaller grains that dictate the geometry of the disc. By removing small grains, thereby reducing the opacity, the disc can shift from having characteristics consistent with group I to being consistent with group II.

Alternatively -- or additionally -- the flat discs
could be explained by the shadowing of a puffed-up inner rim to the disc, shadowing its outer regions from direct starlight
\citep[e.g.][hereafter DDO4]{Dullemond2004}. More recently, 
it has been shown that several of the group I sources have a dust-depleted inner region \citep[e.g.][]{Grady2009,Lyo2011, Andrews2011}, so that 
the original interpretation of evolution between flared and flat discs may no longer hold. \citet{Maaskant2013} 
even propose that all group I discs may harbour a gap. 
\cite{Meijer2008} used 2D radiative transfer models for a disc parameter study 
and concluded that an increase in the mass of small grains can make the initially
optically thin outer disc become optically thick. They also showed that, while the
mass in small ($<$ 25 \mic) dust grains determines the shape of the SED up to 60 \mic, which is the 
region used for the Meeus classification, the longer wavelength SED will change when 
larger grains (they used 2 mm-sized grains) are introduced into the midplane,
increasing the mm flux and flattening the sub-mm slope.
 
The composition of constituent dust grains in HAeBe discs can be determined through analysis of spectral features present at mid-infrared wavelengths. A sample of 53 HAeBes observed by the \textit{Spitzer Space Telescope}'s \textit{InfraRed Spectrograph} instrument \citep{Houck2004, Werner2004} was examined, revealing 45 discs with silicate dust features, and 8 with evidence of PAH emission \citep{Juhasz2010}. It was found that larger grains are more abundant in the atmospheres of flatter discs compared with those of flared discs, indicating that grain growth and sedimentation reduce disc flaring.

A recent \textit{Herschel} \citep{Pilbratt2010} study of carbon monoxide (CO) gas in HAeBe discs at far-infrared wavelengths showed 
that the degree of disc flaring influences the measured CO line strength, causing CO to be only 
detected in flaring discs \citep{Meeus2013}. It is clear that understanding the 
difference between group I and group II discs is necessary for untangling the gas 
and dust components of a disc.

In this paper we examine \textit{Herschel}/PACS far-infrared imaging observations -- at 70, 100, and 160 \mic \ -- for the 
same sample of HAeBe stars as is presented in \citet{Meeus2012}. These stars were observed 
as part of the Open Time Key Programme `GAS in Protoplanetary Systems' \citep[GASPS;][]{Dent2013}.
We also present photometry for HD 98922 as part of our sample, but have omitted this source from the subsequent analysis owing 
to its poorly defined system properties and binarity. 
The far-infrared photometric data presented here are important because they increase the overall density of coverage in the disc SEDs, 
allowing better estimates of far-infrared excesses to be made. Furthermore, they provide a way to calibrate the PACS spectra, whose absolute flux calibration is not as accurate as that of the photometric measurements.

The paper proceeds as follows. In Sect.~\ref{s_sample} we present the sample and the observations. In Sect.~\ref{s_results} 
we describe the results: photometry, infrared excesses. and radial profiles. In Sect.~\ref{s_discussion} we discuss our findings and examine the
physical interpretation of correlations identified between the disc observational properties, 
whilst finally in Sect.~\ref{s_conclusions} we draw our conclusions. We have provided the SEDs and sources of literature photometry for the 
targets in our sample in Appendix A, the \textit{Herschel} observation log in Appendix B, and the comparison of extended sources 
vs. the model PSF in Appendix C.

\section{Sample and observations}
\label{s_sample}

Here we present \textit{Herschel Space Observatory} \citep{Pilbratt2010} spectroscopic 
(63 \mic \ OI line) and photometric (70, 100, and 160 \mic)
`Photodetector Array Camera and Spectrograph' \citep[PACS;][]{Poglitsch2010} 
measurements of the target HAeBe discs. 
Our sample consists of 22 intermediate-mass HAeBe stars, and five
main-sequence A-type stars with debris discs.  The spectral types of the HAeBes
range between B9 and F3, with masses between $\sim\!4.2$ and 1.4
\Msun. The ages range from $\sim\!1$ to a few tens of Myr. In
Table \ref{table1} we list the sample of target stars along with references for their stellar parameters.

\begin{table*}[!t]\tabcolsep 2pt
\caption{General properties of the sample.The
most common name for each target is given in column 1,
(e.g. 49 Cet and CQ Tau are preferred over HD 9672 and HD 36910,
respectively), otherwise the HD identification is provided. Column 3
lists the group classification according to \citet{Meeus2001}, the
remaining columns are self-explanatory. }
\begin{center}
\begin{tabular}{llccccccr}
\hline
\hline 
\multicolumn{1}{c}{Target}   & Alternative & Group & Sp. type & $T_{\rm eff}$ & Age & $M/M_{\odot}$ & $L/L_{\odot}$ & {Refs.} \\
 & name(s) &  & & (K) & (Myr) &  & &\\
\hline               
 AB Aur*     & HD 31293      & I      & A0 Ve          & 9280       & $5\pm1$      & $2.4\pm0.2$      & $33.0\pm9.2$  & 1,2 \\
 HD 31648   & MWC 480       & II     & A3-5 Ve     & 8250       & $8.5\pm 2.0$  & $1.99$       & $13.7\pm 5.5$ & ,1,3 \\
 HD 35187  B/A * &               & II     & A2 Ve/A7       & 8990/7800       & $9.0\pm 2.0$  & $2.00$(B)   & $17.4\pm10.6$ (B)& 3,4 \\ 
 HD 36112   & MWC 758       & I      & A5 IVe     & 7750       & $3.7\pm 2.0$  & $2.17$       & $33.7\pm19.3$ & 1,3 \\
 CQ Tau     & HD 36910      & II     & F3 Ve       & 6900       & $4.0\pm 2.0$  & $1.38$       & $3.4\pm2.0$   & 1,3 \\
 \textit{HD 98922}*   &      & II     & B9 Ve       & 10600      & $<0.01$       & $>4.95$      & $>912$        & 3,5,6 \\ 
 HD 97048*   & CU Cha        & I      & A0 Ve       & 10000      & $6.5\pm 1.0$  & $2.5\pm0.2$  & $30.7\pm6.1$  & 1,3 \\
 HD 100453*  &               & I      & A9 Ve       & 7400       & $>10$         & $1.7\pm0.2$  & $8.8\pm1.4$   & 1,2 \\ 
 HD 100546  &               & I      & B9 Ve       & 10470      & $>10$         & $2.4\pm0.1$  & $22.7\pm1.9$  & 1,2 \\ 
 HD 104237*  & DX Cha        & II     & A4-5 Ve     & 8550       & $5.5\pm 0.5$  & $2.2\pm0.2$/$1.4\pm0.3$  & $28.8\pm2.4$  & 1,7 \\ 
 HD 135344B* &               & I      & F3-4 Ve     & 6810       & $10.0\pm 2.0$ & $1.6\pm0.2$  & $8.1\pm3.1$  & 1,2 \\ 
 HD 139614  &               & I      & A7 Ve       & 7400       & 8.0           & $1.7\pm0.3$  & 9.5          & 1,2 \\ 
 HD 141569*  &               & II/TO  & B9.5 Ve     & 10000      & $4.7\pm 0.3$  & $2.18$       & $29.6\pm4.2$ & 1,3 \\
 HD 142527*  &               & I      & F4 IIIe     & 6550       & $5.0\pm1.5$   & $2.0\pm0.3$  & $16.3\pm4.5$  & 8 \\ 
 HD 142666  &               & II     & A8 Ve       & 7500       & $>10$         & $1.8\pm0.3$  & 8.66     & 1,2 \\ 
 HD 144668*  & HR 5999       & II     & A7 IVe     & 7925       & $2.8\pm1.0$   & $3.2\pm0.5$  & $50.8\pm9.5$  & 1\\
 HD 150193*  & MWC 863       & II     & A2 IVe      & 9870       & $3.8\pm 2.0$  & $2.3\pm0.2$  & $48.7\pm38.0$ & 1,2 \\
 KK Oph A/B* &               & II     & A6 Ve/G5 Ve & 8000/5750  & $8\pm 2$      & $2.17$       & $13.7/2.1$    & 1,2 \\
 51 Oph     & HD 158643     & II     & B9.5 IIIe   & 10250      & $0.7\pm 0.5$  & $4.2\pm 0.3$ & $285.0\pm17$ & 1,9 \\ 
 HD 163296  & MWC 275       & II     & A1 Ve       & 9250       & $5.5\pm 0.5$  & $2.0\pm0.2$  & $33.1\pm6.2$  & 1,2 \\ 
 HD 169142  & MWC 925       & I      & A7-8 Ve     & 7500       & $7.7\pm 2.0$  & $2.0$        & $9.4\pm 5.6$ & 1,10 \\
 HD 179218*  & MWC 614       & I      & A0 IVe      & 9500       & $2.3\pm 0.3$  & $3.0\pm0.22$ & $83\pm32$ & 9,11 \\ 
\hline 
 49 Cet     & HD 9672              & Debris & A4 V     & 9500   & 40            & 2           & $21.0\pm0.7$  & 1,9,12,13 \\ 
 HD 32297   &                      & Debris & A7     & 7750   & $30$          & $1.84$      & $10.9\pm2.1$  &  1,14 \\
 HR 1998    & HD 38678, $\zeta$ Lep& Debris & A2 IV-V  & 8500   & $200\pm100$  & $1.9$        & $14\pm0.1$    &  1,15 \\
 HR 4796A*   & HD 109573A           & Debris & A0 Ve    & 9750   & $10.0\pm2.0$ & $2.4$        & $23.4\pm1.1$   & 1 \\
 HD 158352  & HR 6507              & Debris & A7 V     & 7500   & $1000\pm200$ & $1.67$       & $17.7\pm 0.6$  & 1 \\
\hline 
\hline
\end{tabular}
\tablefoot{TO = transitional object. HD 98922's properties are not well defined; the values given are orientative. *Stars marked with this symbol are known to be part of multiple systems. See text for further information.
References:
(1) \citep{Meeus2012}
(2)\citet{Boekel2005};
(3)  \citet{Manoj2006}; 
(4)  \citet{Dunkin1998}
(5)\citet{Kraus2008};
(6)\citet{vdAncker1998}; 
(7) \citet{Garcia2013};
(8) \citet{Mentigutia2014};
(9) \citet{Montesinos2009};
(10) \citet{Raman2006};
(11) This work,
(12) \citet{Zuckerman2012};
(13) \citet{Roberge2013};
(14) \citet{Donaldson2013};
(15) \citet{Moerchen2007}.
}
\end{center}
\label{table1}
\end{table*}

The available data and modes of observation for the sample are quite heterogeneous. 
For example, not all targets were observed 
in all three PACS bands, a mixture of Chop-Nod and Mini Scan Map observing modes were used, and in the case of Mini Scan Map observations, a cross-scan was not always taken.
The observations, in either Chop-Nod or Scan Map mode, were carried out using the recommended map parameters (see PACS observer's manual for further details).
A summary of the \textit{Herschel}/PACS observations
used in this work is provided in an Appendix, Table \ref{obsids}.

The observations were all reduced using the `\textit{Herschel}
Interactive Processing Environment' \citep[HIPE;][]{Ott2010} version 10.0.0 and PACS
calibration version 45. These values were current when the work was undertaken, but have since been surpassed.
We note that the refinements to data processing in more recent iterations of HIPE and PACSCal would not
fundamentally affect the analysis or conclusions of this work. The standard data reduction scripts provided
with HIPE were used to produce the scan images and mosaics for targets with both a scan and a cross-scan image. 
All available scans for each target at each wavelength were
combined to produce final images from which the source fluxes were measured (which might vary between one and six individual scans, depending on the wavelength and specific target). 
In the data processing we adopted a high-pass filter width of 15 frames in
the blue and green channels and 25 frames in the red channel,
equivalent to 62\arcsec \ at 70 and 100 \mic, and 102\arcsec \ at 160
\mic.  This allowed us to suppress 1/$f$ noise effectively without
the risk of clipping the source PSF.  To avoid biasing the background
estimation of the high-pass filter routine, a region 20\arcsec \ in
radius centred on the source peak in the input frames was masked.  To
centre the mask over the target, the location of the source in the image was
determined from SEXtractor using the level 2.5 pipeline product
supplied with each observation from the \textit{Herschel} Science Archive as a
guide.  Deglitching was carried out using the spatial deglitching
method, again using the source-centred high-pass filter mask to avoid
clipping the core of the target PSF during the image creation process.
Final image scales for maps at 70 and 100 \mic \ were 1\arcsec \ per
pixel, whilst the 160 \mic \ maps were 2\arcsec \ per pixel (compared to native
scales of 3.2 and 6.4\arcsec \ at 70/100 and 160 \mic, respectively).

We present here previously unpublished PACS photometry for HD 98922, but the
star has not been included in the subsequent analysis. This is because 
the uncertainty in its published spectral classification,
luminosity class, and absolute parameters are such
that interpreting that target is complicated as a member of the ensemble.
For example, HD 98922 has a high mass-accretion rate, $\log \dot{M_{\rm acc}}$
(M$_\odot$/yr)=$-5.76$ \citep{Garcia-Lopez2006}, an excess is
apparent in the optical part of the SED (mainly in $U$ and $B$ bands) implying
veiling in the spectrum.
A direct comparison of HD 98922's spectrum with synthetic models is therefore not straightforward. It
presents a complex Fe {\sc ii}
variable spectrum in emission and strong emissions in the first Balmer
lines.
We are currently carrying out a detailed study of the UV/optical spectrum
and SED of this object, which will be published elsewhere. Therefore,
while a reliable determination of the physical properties, extinction, 
and evolutionary status are not available, we
prefer not to give results concerning infrared excesses based on
currently published parameteres that could change substantially 
in light of a comprehensive analysis.

Some of our target stars are known to have companions at separations that fall within, or close to, the PACS beam FWHM (5.8\arcsec\ FWHM at 70~$\mu$m, 11.7\arcsec\ FWHM at 160~$\mu$m).
AB Aur's main component is known to have a companion situated at  $\sim$0.5-3\arcsec \citep{Baines2006}l;
HD 35187 is a close multiple system, with two components with similar luminosity (B, the component that hosts the disc, and A, as listed in \ref{table1}) separated by 1.38\arcsec \ and a much fainter component lying at 7.8\arcsec \ \citep{Dunkin1998};
HD 98922 has a companion at 7.8\arcsec \ \citep{Baines2006}; 
HD 100453 is a binary system, whose main star listed in Table \ref{table1} has a companion at 1.06\arcsec  \citep[M3-5, $\Delta$K = 5.1][]{Chen2006};
HD 135344B has a companion (SAO206463) 20.4\arcsec \ away \citep{Augereau2001};
HD 141569 is a triple system, with two companions at  distance of $\sim$7.6\arcsec (M2, $\Delta$K = 1.8 ) and 9\arcsec (M4, $\Delta$K = 2.4) \citep{Weinberger2000, Baines2006};
HD 142527 is a system where the main star has a faint close companion at $\sim$0.086\arcsec \ ($\Delta$K = 0.9)\citep{Close2014, Biller2012}; 
HD 144668 forms a wide ($\sim$45\arcsec)  proper motion binary system with the star HR 6000 \citep{Preibisch2006} and also presents a faint companion at  1.3\arcsec \citep{Stecklum1995};
HD 179218 has a likely companion at 2.5\arcsec \ with $\Delta$K = 6.6 \citep{Wheelwright2010, Thomas2007};
and HR 4796A has co-moving companion M star (HR 4796B) at 7.7\arcsec \ \citep{Stauffer1995}.
However, we do not see evidence of any of these companions in the \textit{Herschel} images. The total emission in the PACS wavelengths from any naked, main-sequence star at the distances to these HAeBes ($\sim$100 pc) is negligible compared to either the flux density from the disc (1 to 100 Jy) or the uncertainty on that measurement (cal. uncertainty of 5\%).
It is only in the case of HD 97048 where the flux of a companion is noticeable ($\sim$10\% brightness of the primary) but their separation (25\arcsec) is larger than our apertures (12\arcsec, 15\arcsec ,  and 20\arcsec) and 
its contribution to the measured flux of the system can be accounted for in the process.
HD~104237 has a pair of T Tauri stars within 15\arcsec \ of the primary. These stars have been identified as having infrared excesses, which is indicative of the presence of circumstellar discs \citep{Feigelson2003, Grady2004}. 
At 160  \mic, the emission from the primary is blended with the T Tauri stars (and their discs), such that this star is blended with its companions along the axis of association, and the measurement given for this star has an unquantified contribution from these companions. We therefore quote the flux measurement for HD~104237 as an upper limit.

Although our focus in this work is on presenting the far-infrared photometry of the sample and characterising the corresponding part of the SED, 
it must be noted how some of these sources have shown evidence of optical and near-IR variability. 
We summarise the variability of AB Aur, HD~31648, HD~36112, HD~35187, and CQ Tau in Section \ref{ex}, using AB Aur as an example.

\section{Results}
\label{s_results}

\subsection{Photometry}

Target fluxes were measured using an IDL-based aperture photometry
routine. The sky background and r.m.s scatter were estimated from the mean and standard deviation of 25 square sky apertures scattered randomly between 30\arcsec \ to 60\arcsec \ from the source location.
The areas of the sky apertures were chosen to match those of the flux apertures.
Aperture radii of 12\arcsec \
at 70 \mic, 15\arcsec \ at 100 \mic, and 20\arcsec \ at 160 \mic \
were used to measure the fluxes.  Measured fluxes were corrected for
the aperture size and the colour of the source (after having fitted
the dust temperature from the SED), accounting for the relative
contributions from the photosphere and dust.  A check for extended
emission from the targets was made through the shape of the aperture
corrected curve of growth of each target for flux apertures between
2\arcsec \ and 20\arcsec \ in radius, looking for a trend of increasing flux with
aperture radius.  The target photometry is given in Table
\ref{table_fluxes}. 
The dominant contribution to the uncertainty is the calibration uncertainty
of 5\%, limited by the uncertainty on the stellar photosphere models
used to calibrate the standard stars \citep{Balog2014}.

\begin{table}[!h]
\begin{center}
\caption{Photometry measured for the sample, taking the colour and
  aperture corrections into account.}
\label{table_fluxes}
\begin{tabular}{lcccc}
\hline
\hline     
  Target & $F$\ \ [70 \mic]  &  $F$\ \ [100 \mic]    &      $F$\ \ [160 \mic]             \\
        ~              & (Jy)             & (Jy)             &  (Jy)              \\
\hline         
  AB Aur   & 137.74~$\pm$~6.90 & \ldots & 65.55~$\pm$~3.29 \\
  HD 31648 & 12.30~$\pm$~0.62 & 14.30~$\pm$~0.72 & 13.18~$\pm$~0.66 \\
  HD 35187 & 5.17~$\pm$~0.26 & 4.07~$\pm$~0.20 & 2.38~$\pm$~0.12 \\ 
  HD 36112 & 20.84~$\pm$~1.04 & 18.10~$\pm$~0.91 & 12.40~$\pm$~0.62 \\
  CQ Tau & 18.14~$\pm$~0.91 & 14.09~$\pm$~0.71 & 8.69~$\pm$~0.43 \\
  HD 97048 & 67.94~$\pm$~3.41 & 69.81~$\pm$~3.53 & 59.31~$\pm$~2.98 \\ 
  HD 98922 & 3.73~$\pm$~0.19 & \ldots & 0.81~$\pm$~0.04 \\ 
  HD 100453 & 36.26~$\pm$~1.82 & 27.54~$\pm$~1.38 & 15.86~$\pm$~0.79 \\
  HD 100546 & 160.6$\pm8.7$ & 115.97~$\pm$~5.81 & 53.88~$\pm$~2.69 \\
  HD 104237 & 10.17~$\pm$~0.51 & \ldots & 4.64~$\pm$~0.24 \\
  HD 135344B & 30.45~$\pm$~1.53 & 29.05~$\pm$~1.46 & 20.82~$\pm$~1.04 \\
  HD 139614 & 18.68~$\pm$~0.94 & 16.94~$\pm$~0.85 & 12.79~$\pm$~0.64 \\
  HD 141569 & \ldots & 3.29~$\pm$~0.17 & 1.28~$\pm$~0.07 \\
  HD 142527 & 113.25~$\pm$~5.67 & 98.74~$\pm$~4.95 & 63.71~$\pm$~3.19 \\
  HD 142666 & 6.56~$\pm$~0.33 & 5.91~$\pm$~0.30 & 4.33~$\pm$~0.22 \\
  HD 144668 & 5.47~$\pm$~0.27 & 3.38~$\pm$~0.18 & 1.50~$\pm$~0.15 \\
  HD 150193 & \ldots & 4.30~$\pm$~0.22 & 2.34~$\pm$~0.12 \\
  KK Oph    & 4.65~$\pm$~0.23 & 3.29~$\pm$~0.16 & 1.72~$\pm$~0.09 \\
  51 Oph    & 0.92~$\pm$~0.05 & 0.35~$\pm$~0.03 & 0.15~$\pm$~0.02 \\
  HD 163296 & 18.91~$\pm$~0.95 & \ldots & 21.00~$\pm$~1.06 \\
  HD 169142 & 27.84~$\pm$~1.53 & \ldots & 11.91~$\pm$~0.64 \\
  HD 179218 & 22.74~$\pm$~1.14 & \ldots & 6.84~$\pm$~0.34 \\
  49 Cet    & 2.21~$\pm$~0.11 &    1.95~$\pm$~0.10    &  1.02~$\pm$~0.05  \\ 
  HD 32297  & 1.10~$\pm$~0.06 & 0.86~$\pm$~0.04 & 0.41~$\pm$~0.02 \\
  HR 1998   & 0.24~$\pm$~0.03 & 0.13~$\pm$~0.01 & 0.03~$\pm$~0.02 \\
  HR 4796A  & 6.34~$\pm$~0.32 & 3.93~$\pm$~0.20 & 1.71~$\pm$~0.09 \\
  HD 158352 & 0.23~$\pm$~0.02 & \ldots & 0.15~$\pm$~0.01 \\
\hline
\end{tabular}
\end{center}
\end{table}

\subsection{SED and excesses}\label{ex}

We compiled published observations to create the SEDs of our targets (see Table \ref{literature}),  
combining our new PACS observations with data spanning ultraviolet to millimetre wavelengths, including spectra from the
{\em International Ultraviolet Explorer}\footnote{http://sdc.cab.inta-csic.es/ines/}
and {\em Spitzer}/IRS\footnote{http://cassis.astro.cornell.edu/atlas/}, where available. 
To determine the stellar contribution to the total emission a specific model photosphere for each star was extracted 
or computed by interpolation from the grid of PHOENIX/GAIA models \citep{Brott2005}.
Beyond 45 \mic, the stellar photospheric contribution was extrapolated to mm wavelengths from the
Rayleigh-Jeans regime. For a given star the model photosphere was reddened with several values of $E(B\!-\!V)$
(values given in table \ref{Av}, assuming
$R_V$ = 3.1) 
and normalized to the measured flux in $V$ band, until a best (least-squares) fit to
the optical photometry was obtained. 
In Appendix A and Figs. \ref{seds-HAeBe} 
and \ref{seds-debris}, we show the SEDs for all 27 targets.

We calculate the total fractional excess, $F_{\rm IR}/F_{\star}$ ,
and three partial excesses, $F_{\rm NIR}/F_{\star}$, $F_{\rm MIR}/F_{\star},$ and
$F_{\rm FIR}/F_{\star}$, by subtraction of the stellar photosphere model 
from the SED. In this work, `N', `M', and `F' denote near (2--5 \mic),
mid (5--20 \mic), and far (20--200 \mic) infrared regimes, respectively. To calculate the excess, 
the dereddened SED was integrated between the corresponding limits, and the 
photospheric contribution according to the stellar model was subtracted from this total. Table 
\ref{table_excesses} lists the excesses; in column 2 we give the
value of $\lambda_0$ for each target, the wavelength from which the
SED departs from a pure photospheric behaviour. Stars hosting a debris 
disc show only weak excesses, starting at longer wavelengths than the 
HAeBe stars.

\begin{table*}[ht!]
\begin{center}
\caption{Wavelength $\lambda_0$ at which the excess starts in the target SED and the fractional excesses in the near-, mid-, and far-infrared.}
\label{table_excesses}
\begin{tabular}{lccccc}
\hline\hline\noalign{\smallskip}
     Star               &  $\lambda_0$  &  $F_{\rm IR}$/$F_{\star}$  &  $F_{\rm NIR}/F_{\star}$ &  $F_{\rm MIR}/F_{\star}$  &  $F_{\rm FIR}/F_{\star}$ \\ 
                         &   (\mic)    &   ($\lambda_0$-mm)    &  (2--5 \mic)   &   (5--20 \mic) & (20--200 \mic) \\
\hline
AB aur              &  0.55  &  6.77$\times10 ^{-1}$     &  2.10$\times10 ^{-1}$     &  1.18$\times10 ^{-1}$   &  1.82$\times10 ^{-1}$  \\  
HD 31648            &  0.77  &    4.77$\times10 ^{-1}$   &  1.31$\times10 ^{-1}$     &  1.65$\times10 ^{-1}$   &  6.23$\times10 ^{-2}$ \\  
HD 35187            &  1.12  &    1.09 $\times10 ^{-1}$   &  3.29$\times10 ^{-2}$   &  3.50$\times10 ^{-2}$    &  2.39$\times10 ^{-2}$ \\ 
HD 36112            &  0.77  &    6.24$\times10 ^{-1}$   &  1.86$\times10 ^{-1}$      &  1.27$\times10 ^{-1}$    &  1.52E-01  \\  
CQ Tau                       &               &                   &                    &                    &            \\
 \hspace*{0.2cm}\textit{bright state}  &  0.63   &    7.25$\times10 ^{-1}$  &  1.82$\times10 ^{-1}$    &  2.34$\times10 ^{-1}$  &  2.57$\times10 ^{-1}$\\      
 \hspace*{0.2cm}\textit{faint state}   &  0.70   &    1.06                  &  2.68$\times10 ^{-1}$ &  3.14$\times10 ^{-1}$   &  3.40$\times10 ^{-1}$\\  
HD 97048                     &  0.70         &    3.23$\times10 ^{-1}$  &  5.29$\times10 ^{-2}$ &  6.41$\times10 ^{-2}$   &  1.53$\times10 ^{-1}$  \\     
HD 100453                    &  0.80         &    6.14$\times10 ^{-1}$    &  1.56$\times10 ^{-1}$   &  1.57$\times10 ^{-1}$      &  2.39$\times10 ^{-1}$ \\ 
HD 100546                    &  1.00         &    6.26$\times10 ^{-1}$   &  4.83$\times10 ^{-2}$    &  2.42$\times10 ^{-1}$      &  3.06$\times10 ^{-1}$ \\ 
HD 104237                    &  0.70         &    2.64$\times10 ^{-1}$     &  8.93$\times10 ^{-2}$  &  6.37$\times10 ^{-2}$      &  2.24$\times10 ^{-2}$ \\      
HD 135344B                   &  0.60         &    5.56$\times10 ^{-1}$  &  1.68$\times10 ^{-1}$    &  7.14$\times10 ^{-1}$       &  1.61$\times10 ^{-1}$ \\
HD 139614                    &  1.00         &    4.09$\times10 ^{-1}$    &  6.92$\times10 ^{-2}$  &  1.12$\times10 ^{-1}$       &  1.96$\times10 ^{-1}$ \\
HD 141569                    &  1.15         &    1.03$\times10 ^{-2}$  &  8.14$\times10 ^{-4}$    &  2.15$\times10 ^{-3}$       &  5.70$\times10 ^{-3}$ \\
HD 142527                    &  0.70         &    1.09                   &  2.89$\times10 ^{-1}$    &  1.86$\times10 ^{-1}$      &  3.94$\times10 ^{-1}$ \\
HD 142666                    &  1.00         &    2.67$\times10 ^{-1}$   &  9.63$\times10 ^{-2}$   &  7.61$\times10 ^{-2}$   &  3.93$\times10 ^{-2}$ \\
HD 144668                    &  0.77         &    4.82$\times10 ^{-1}$   &  1.86$\times10 ^{-1}$   &  1.08$\times10 ^{-1}$   &  1.88$\times10 ^{-2}$  \\ 
HD 150193                    &  0.60         &    4.76$\times10 ^{-1}$   &  1.27$\times10 ^{-1}$   &  1.61$\times10 ^{-1}$  &  4.53$\times10 ^{-2}$ \\j 
KK Oph                       &  0.70         &    2.14                   &  8.78$\times10 ^{-1}$   &  8.71$\times10 ^{-1}$    &  1.21$\times10 ^{-1}$  \\ 
51 Oph                       &  1.10         &    2.16$\times10 ^{-2}$ &  8.66$\times10 ^{-3}$   &  9.10$\times10 ^{-3}$    &  1.00$\times10 ^{-3}$ \\    
HD 163296                    &  1.00         &    2.61$\times10 ^{-1}$    &  9.99$\times10 ^{-2}$   &  7.57$\times10 ^{-2}$   &  2.96$\times10 ^{-2}$\\  
HD 169142                    &  0.75         &    3.99$\times10 ^{-1}$   &  6.22$\times10 ^{-2}$  &  7.45$\times10 ^{-2}$  &  1.99$\times10 ^{-1}$ \\   
HD 179218                            &  1.00         &    3.32$\times10 ^{-1}$     &  5.47$\times10 ^{-2}$    &  1.51 $\times10 ^{-1}$  &  9.93$\times10 ^{-2}$ \\ 
\hline
49 Cet                       &  7.00         &    7.17$\times10 ^{-4}$   & \ldots   &  5.75$\times10 ^{-5}$     &  6.45$\times10 ^{-4}$\\   
HD 32297                     &  5.40         &    6.09$\times10 ^{-3}$   & \ldots   &  4.10$\times10 ^{-4}$  &  5.60$\times10 ^{-3}$\\  
HR 1998                      &  6.45         &    1.31$\times10 ^{-4}$   & \ldots  &  8.93$\times10 ^{-5}$   &  4.10$\times10 ^{-5}$ \\ 
HR 4796A                     &  4.00         &    4.16$\times10 ^{-3}$  &  1.69$\times10 ^{-5}$     &  2.55$\times10 ^{-4}$   &  3.84$\times10 ^{-3}$\\ 
HD 158352                    &  0.75         &    2.14$\times10 ^{-2}$  &  1.51$\times10 ^{-3}$    &  5.23$\times10 ^{-4}$    &  8.93$\times10 ^{-5}$\\
\hline
\end{tabular}
\end{center} 
\end{table*} 

There is some evidence of near-infrared variability for some of our targets, namely AB Aur \citep{Shenavrin2012}, HD31648 \citep{Sitko2008}, HD 36112 \citep{Beskrovnaya1999}, HD 35187 \citep{Dunkin1998}, and CQ Tau \citep{Shenavrin2012},  but there is no further evidence of infrared variability at longer wavelengths caused by circumstellar matter. Still, for the purpose of further  studies, the observation date for our targets is given in the last column of Table \ref{obsids}.
In this respect, a test was performed using AB Aur and the variable photometric values given by \citet{Shenavrin2012} in J, H, K, L, and M bands, where the  $F_{\rm NIR}/F_{\star}$ is 0.13 for their faintest case and 0.34 for the brightest, and a ratio of 0.21 for the mean photometric values, the same as in this paper. We do not have simultaneous photometry for any of our targets, and we do not show any data from different epochs.  Therefore, there might be subtle inaccuracies in the near-infrared excesses for the enumerated objects because no information about brightness is available for our near-infrared bibliographic data. We advise caution in relation to the near-infrared excess while reading the rest of the paper, since the flux at  4.6 \mic \ might vary up to 20\%.

\subsection{Correlations}

We carried out a statistical analysis on the sample to look for correlations amongst the measured properties of our target stars. 
The relations among parameters were analysed with their corresponding `$p$-values' (see Table. \ref{table_stat}).
These coefficients give the probability that the two variables compared are not correlated. We obtained $p$-values 
from three different tests (Spearman, Kendall, and Cox-Hazard). Two parameters are classified as `correlated' 
when their $p$-values $<1\%$, `tentatively correlated' when 0.01$<,p,<$0.05 \citep[e.g][]{Bross1971}, and `not 
correlated' when $p\geqslant $ 0.05. When a correlation is present, we show the linear fit in Table \ref{table_stat}, and the corresponding figures 
in the next section.

\begin{table*}
\newcommand{\mc}[3]{\multicolumn{#1}{#2}{#3}}
\begin{center}
\caption{Correlation coefficients from the statistical analysis.}
\begin{tabular}{llcclcc}
\multicolumn{1}{c}{Parameter 1} & Parameter 2 & \mc{3}{c}{$p$-value} & Correlation & $r$ \\
\hline
  &    & Spearman & Kendall & Cox-Hazard & \\
\hline\hline
F[70 \mic]  & F[0.85 mm] & 0.329 & 0.0282 & 0.3106 & T & \ldots \\
F[70 \mic]  & F[1.3\phantom{0} mm]  & 0.0082 & 0.0045 & 0.0475 & Yes & 0.81 \\
F[100 \mic] & F[0.85 mm]& 0.0034 & 0.0005  & 0.0011 & Yes$^{a}$& 0.89 \\
F[100 \mic] & F[1.3\phantom{0} mm] & 0.0005 & 0.0001 & 0.0312 & Yes &  0.91 \\
F[160 \mic] & F[0.85 mm] & 0.0017 & 0.0004 & 0.0605 & Yes & 0.88 \\
F[160 \mic] & F[1.3\phantom{0} mm] & 0.0003  & $<$ 0.001 & 0.0255 & Yes & 0.89 \\
FIR excess & mm slope & 0.0417 & 0.0251 & 0.0462 & T  & \ldots \\
IR excess & mm slope & 0.0295 & 0.0124 & 0.0147 & T & \ldots \\
FIR/NIR & mm flux & 0.1812 & 0.1194 & 0.9476 & No & \ldots \\
FIR/NIR & mm slope & 0.0564 & 0.0555 & 0.1781 & No &\ldots  \\
FIR/NIR& OI (63 \mic) & 0.1627 & 0.1730 & 0.0611 & No &\ldots  \\
\hline
\end{tabular}
\tablefoot{(a) This metric included only 11 targets since not all had (sub-)mm photometry. T stands for tentatively correlated (see text).}
\label{table_stat}
\end{center}
\end{table*}

\subsection {Radial profiles and extended far-infrared emission}
\label{rprof}

\begin{figure}[h!]
\begin{center}
\includegraphics[scale=0.40]{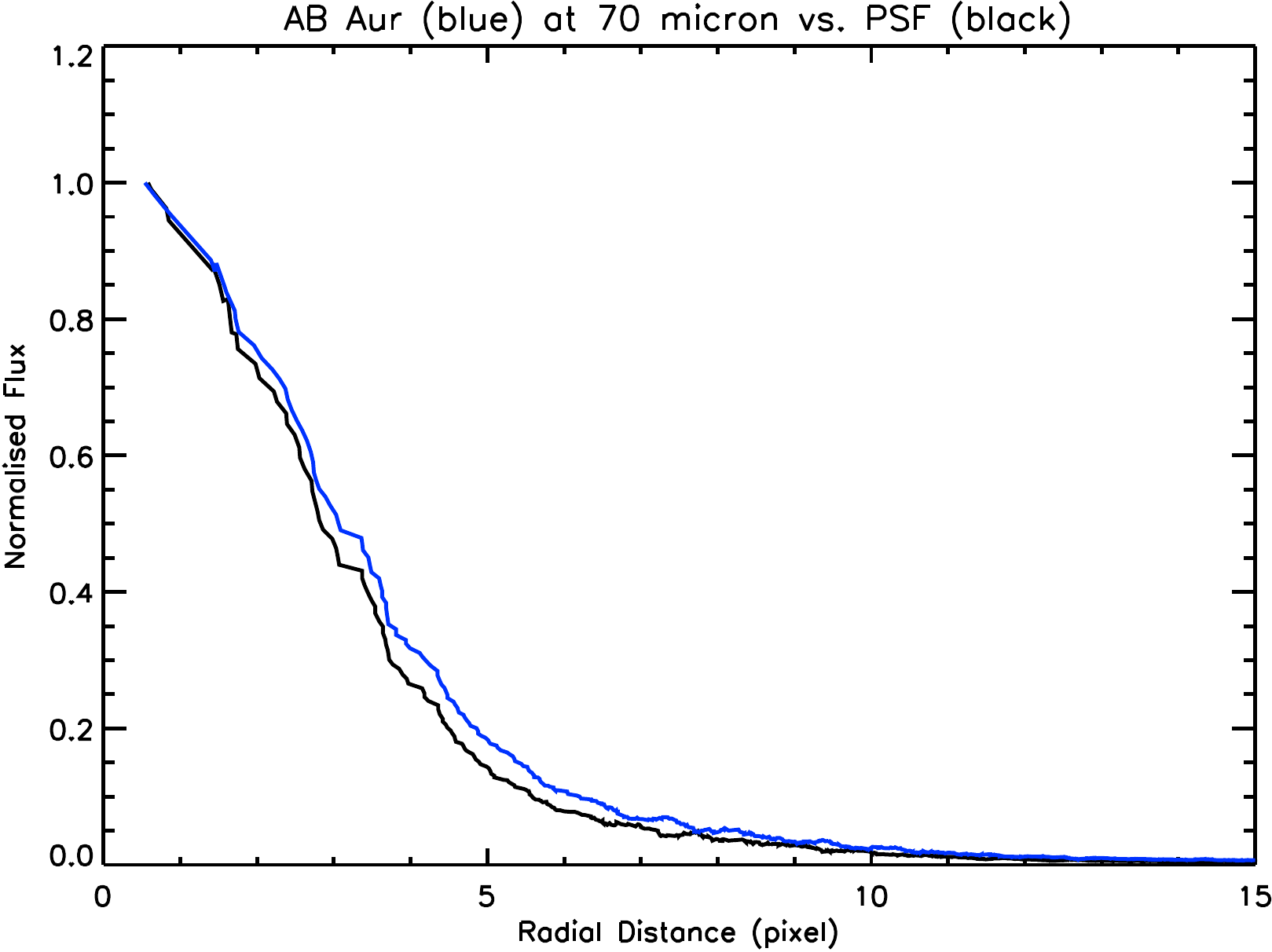}
\caption{Radial profile for AB aur in blue at 70 \mic, PSF ($\alpha$ Boo) in black. 
Object emission is more extended than the emission from a point source obtained 
using the same observation mode and reduction procedure. 
This a showcase of our marginally resolved sources.}
\label{ABAur}
\end{center}
\end{figure}

Some Herbig AeBe stars have large discs ($R_{\rm disc}~\sim$500~au). Even though the 70 \mic \ flux density 
is dominated by emission from warm dust ($\sim$100 K) in the inner part and upper surface layers 
of the disc, it is still likely that some of these discs can be resolved by PACS at 70 or even at 
100 \mic. To check for extended emission, we first did single Gaussian fits of the whole 
sample and compared them to $\alpha$ Boo, which is a point source at all PACS wavelengths. We 
measured the full-width half-maximum (FWHM) of $\alpha$ Boo to be 6.1\arcsec\ at 70 \mic \ and 6.3\arcsec\ 
at 100 \mic. The extent of most of the stars in our sample were consistent with the FWHM of $\alpha$ 
Boo, confirming that their discs are unresolved, except for AB~Aur, HD~100546, HD~104237, HD~141569A, 
and HD~142527, which are `marginally resolved' (i.e. only extended along its major axis compared to 
$\alpha$ Boo) and the debris discs 49~Cet, which is resolved (i.e. extended along both axes compared to 
$\alpha$ Boo). Since this is only a crude way to check for extended emission and can give spurious 
results, we also created azimuthally averaged radial profiles for those sources and compared them to 
the radial profile for $\alpha$ Boo. These radial profiles are shown in Fig. \ref{ABAur} for AB Aur 
and in Fig. \ref{extended_profiles} for the remainder. 

\begin{table}[h!]
\caption{Gaussian fit to our sample. R/U column indicates whether the object is resolved (R) or unresolved (U) compared to point source $\alpha$ Boo. 
Last column states wavelength at which the FWHM was measured.} 
\begin{center}
\begin{tabular}{lcccr}
\hline 
\hline 
Star & $d$   &   FWHM   &  R/U & \multicolumn{1}{c}{$\lambda$}  \\
     & (pc)  & (\arcsec)&      & \multicolumn{1}{c}{(\mic)}     \\
\hline          
 AB Aur    & $ 139.3\pm 19.0$ & $6.7\pm0.2$ & R   & 70 \\ 
 HD 31648 & $137.0\pm 26.2$  & $6.1\pm0.2$   & U & 70 \\                        
 HD 35187  & $114.2\pm 32.4$  &$6.1\pm0.2$  & U & 70  \\                        
 HD 36112  & $279.3\pm 75.0$  &$6.1\pm0.2$  & U & 70  \\                        
 CQ Tau    & $113.0\pm 24.0$  &$6.1\pm0.2$  &  U &70  \\                        
 HD 97048  & $159.5\pm 15.7$  &$6.1\pm0.2$  & U & 70 \\ 
 HD 98922  & {\em200}          &$6.1\pm0.2$  &    U & 70\\
 HD 100453 & $121.5\pm 9.7$   &$6.1\pm0.2$  & U & 70  \\                
 HD 100546 & $96.9\pm 4.0$    & $6.5 \pm0.2$ &  R & 70  \\      
 HD 104237 & $114.7\pm 4.7$   &$6.4\pm0.3$   & R & 70  \\               
 HD 135344B& $142.0\pm 27.0$  &$6.1\pm0.2$   & U & 70  \\                       
 HD 139614 & $140.0\pm 42$    &$6.1\pm0.2$  &  U & 70  \\                       
 HD 141569 & $116.1\pm 8.1$   &$7.2\pm0.2$   & R & 100 \\       
 HD 142527 & $233.1\pm 56.2$  &$6.5\pm0.2$  & R & 70  \\                        
 HD 142666 & $145.0\pm 43$    &$6.1\pm0.2$   & U & 70  \\ 
 HD 144668 & $162.9\pm 15.3$  &$6.1\pm0.2$   & U & 70  \\
 HD 150193 & $216.5\pm 76.0$  &$6.3\pm0.1$  & U & 100 \\        
 KK Oph    & {\em 260}        &$6.1\pm0.2$  & U & 70  \\                        
 51 Oph    & $124.4\pm 3.7$   &$6.1\pm0.2$ & U &    70  \\                      
 HD 163296 & $118.6\pm 11.1$  &$6.1\pm0.2$   & U & 70  \\                       
 HD 169142 & $145.0\pm 43$    &$6.1\pm0.2$ & U &    70  \\
 HD 179218 &    $253.8\pm44.7 $ &$6.1\pm0.2$&  U &   70  \\
\hline 
 49 Cet $^{(a)}$   & $59.4\pm 1.0$    &$7.8\pm0.6$  & R & 70  \\ 
 HD 32297 $^{(b)}$ & $112.4\pm 10.8$  &$6.1\pm0.2$& U & 70   \\ 
 HR 1998   & $21.6\pm 0.1$    &$6.1\pm0.1$ & U & 70   \\
 HR 4796A  & $72.8\pm 1.8$    &$7.0\pm0.4$ & U & 70  \\ 
 HD 158352 & $59.6\pm 0.9$    &$6.1\pm0.2$ & U & 70    \\
\hline
\end{tabular}
\tablefoot{Distances are from the revised parallaxes by \citet{vanLeeuwen2007}, except for 
HD 135344B \citep{Muller2011}, HD 98922 \citep{Manoj2006},  HD 139614, HD 142666, and 
HD 169142 \citep{vanBoekel2005}, HD 89822 and  \citep{Blondel2006}.\\
(a) has been resolved in \cite{Roberge2013}, with a disc size of 200 au and (b) in \cite{Donaldson2013}, with a disc size of 110 au.}
\end{center}
\label{sizes}
\end{table}

Three of the five HAeBe stars with spatially resolved discs have large inner gaps 
and prominent dust rings at distances greater than 100 au. At 1.3 mm, AB Aur has a dust 
ring with the peak of the dust emission at 145 au \citep{Tang2012}, HD~141569A has three 
major rings seen in scattered light at 15, 185, and 300 au \citep{Clampin2003, Thi2014}, 
while at mm wavelengths, HD~142527 has a large gap between 10 and 140 au, which contains 
gas but very little dust \citep{Casassus2012}. The only debris disc which we 
resolved, 49 Cet, is also know to have an outer disc with large grains with an inner radius of 
60 au, which dominates the emission at far-infrared wavelengths \citep{Wahhaj2007}. Its
disc was also resolved with PACS by \citet{Roberge2013}, who did a deconvolution of 
their 70 \mic \ image and found a half-width at half maximum along the disc major axis 
of $\sim$ 200 au, consistent with measurements of the CO disc \citep{Hughes2008}, 
but no sign of a central clearing, likely due to the angular resolution of \textit{Herschel}/PACS.

A necessary condition for seeing spatially extended far-infrared emission therefore seems to be 
that the disc has to have large inner gaps and dust rings at large radii, which contribute 
to or dominate the observed emission. This is almost certainly the case for HD~104237, 
which has not been studied in as much detail as the other three stars. An attempt 
to image the disc with the \textit{Hubble Space Telescope} `Space Telescope Imaging Spectrograph' instrument was unsuccessful \citep{Grady2004}. 
However, high contrast imaging has advanced greatly in the past ten years, in particular
the advent of high contrast, high angular resolution imaging provides exciting avenues for further exploration
and characterisation of such systems. It is likely that HD~104237 will show features similar to those seen for the other three HAeBe stars resolved with PACS. Another option is to 
observed the star with ALMA, which has unprecedented angular resolution and sensitivity 
at millimetre and sub-millimetre wavelengths, tracing the largest and coldest dust grains 
in the circumstellar disc.

\section{Discussion}\label{s_discussion}
In the following section we only discuss the HAeBes. We neglect the debris discs in the discussion
because their discs are fundamentally different, because it is much less massive and optically thin and has rare examples of gas emission 
\citep[e.g.][]{Donaldson2013,Moor2011,Roberge2013}.
As previously noted in \cite{Meijer2008}, there is no natural dichotomy in
the appearance of the source SEDs. Rather, a smooth transition exists between group I
and II sources. This is also seen in our results. However, making a distinction
is a useful tool for studying the disc geometry. For each group, we have calculated the 
mean value of the total fractional excesses, the far-infrared excesses, the ratios 
$F_{\rm NIR}/F_{\rm FIR}$, the mm slope, and 1.3 mm flux. (see Table \ref{t_means}). 

\begin{table}[ht!]
\begin{center}
\caption{Mean, total, and far-infrared excess, FIR/NIR flux ratio, mm slope, mm flux, and  \Odrie \ line flux
 for each group of sources. 
KK Oph is not included because it shows an exceptionaly high far-infrared excess compared to the rest of the sample,
probably due to the presence of a companion.
The $\sigma$ columns show the standard deviation for the mean values for each group.}

\begin{tabular}{lrrrr}
\hline
\hline
      & \multicolumn{1}{c}{Group I}  & \multicolumn{1}{c}{$\sigma_{\rm Gp I}$} & \multicolumn{1}{c}{Group II}  &  \multicolumn{1}{c}{$\sigma_{\rm Gp II}$}\\
\hline
$\langle F_{\rm IR}/F_{\star}\rangle$                    & 0.565    & 0.227   & 0.312  & 0.236 \\
$\langle F_{\rm FIR}/F_{\star}\rangle$                   & 0.218    & 0.09    & 0.051  & 0.075 \\
$\langle F_{\rm FIR}/F_{\rm NIR}\rangle$           & 2.266    & 1.690   & 1.170  & 2.258 \\
$\langle$ mm slope $\rangle$                       & -4.41    & 0.47    & -4.05  & 0.54  \\
$\langle F_{\rm 1.3mm}\rangle$ (mJy)               & 545.4    & 975.8   & 136.4  & 178.9 \\
$\langle F_{\rm{[OI]}}\rangle$ ($10^{-17}$W/m$^2$) & 69       & 99      & 8.3    & 6.4   \\
 \hline
\end{tabular}
\label{t_means}
\end{center}
\end{table}

\begin{figure*}[h!]
\begin{center}
\includegraphics[scale=0.35]{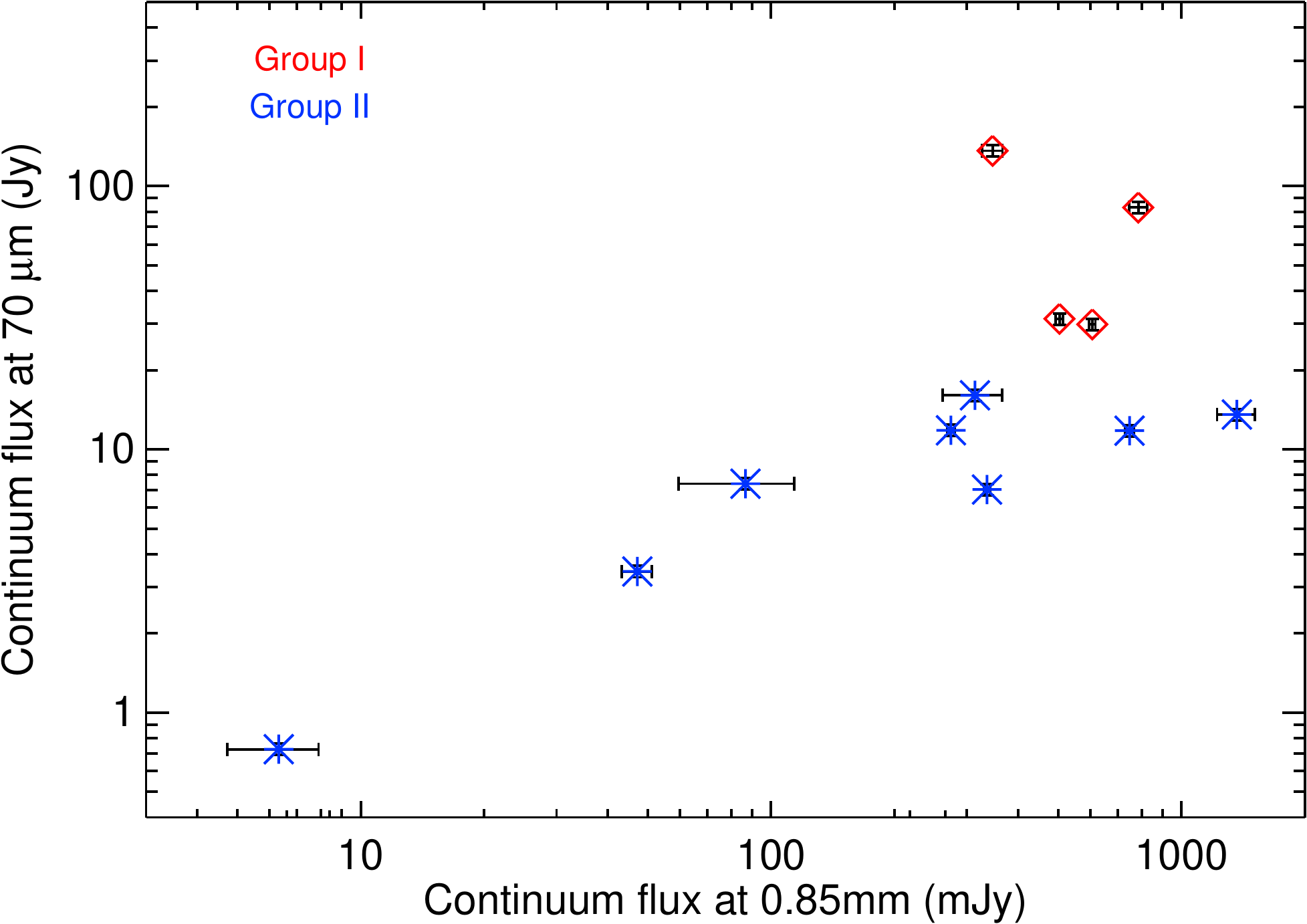}
\includegraphics[scale=0.35]{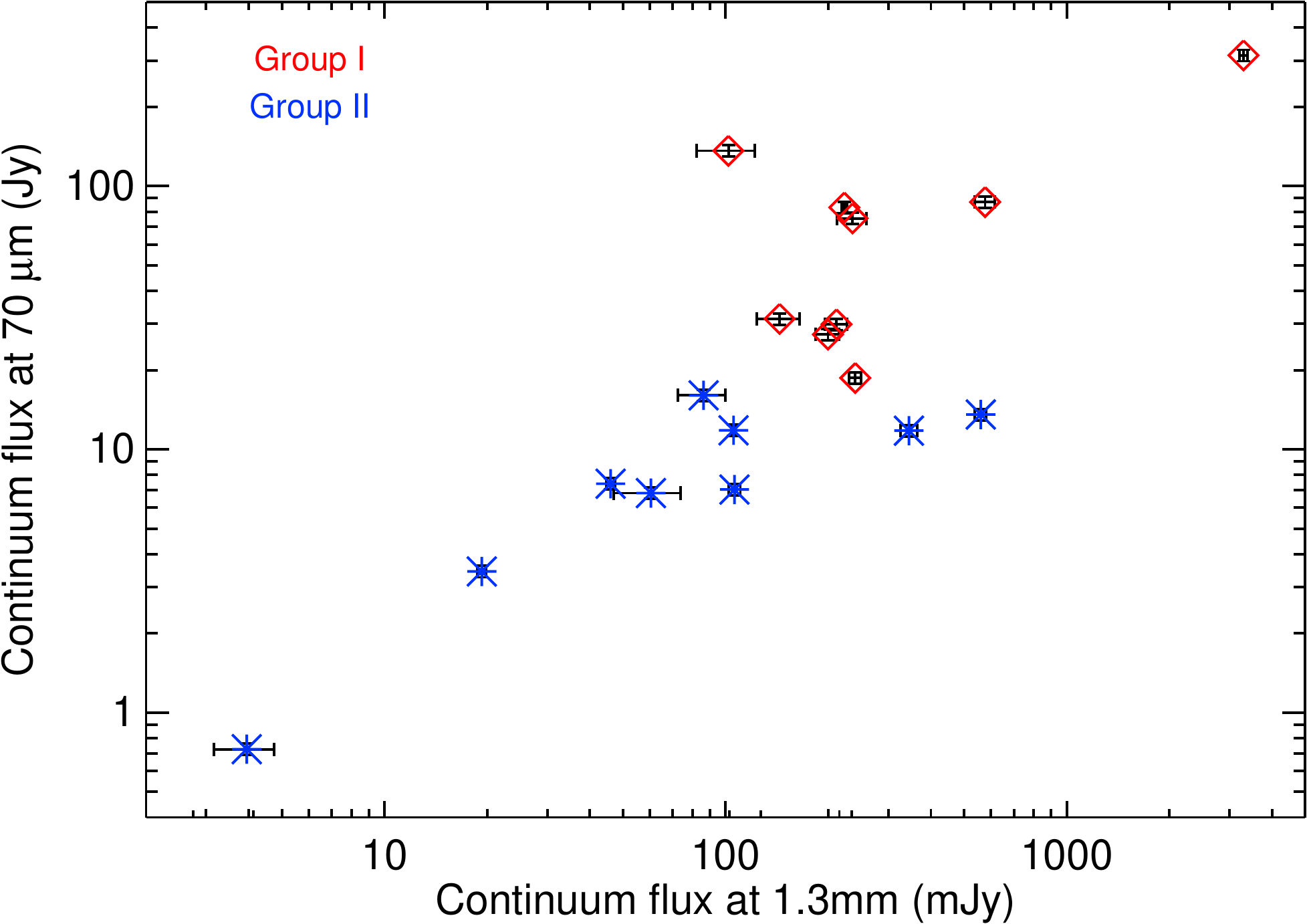}
\includegraphics[scale=0.35]{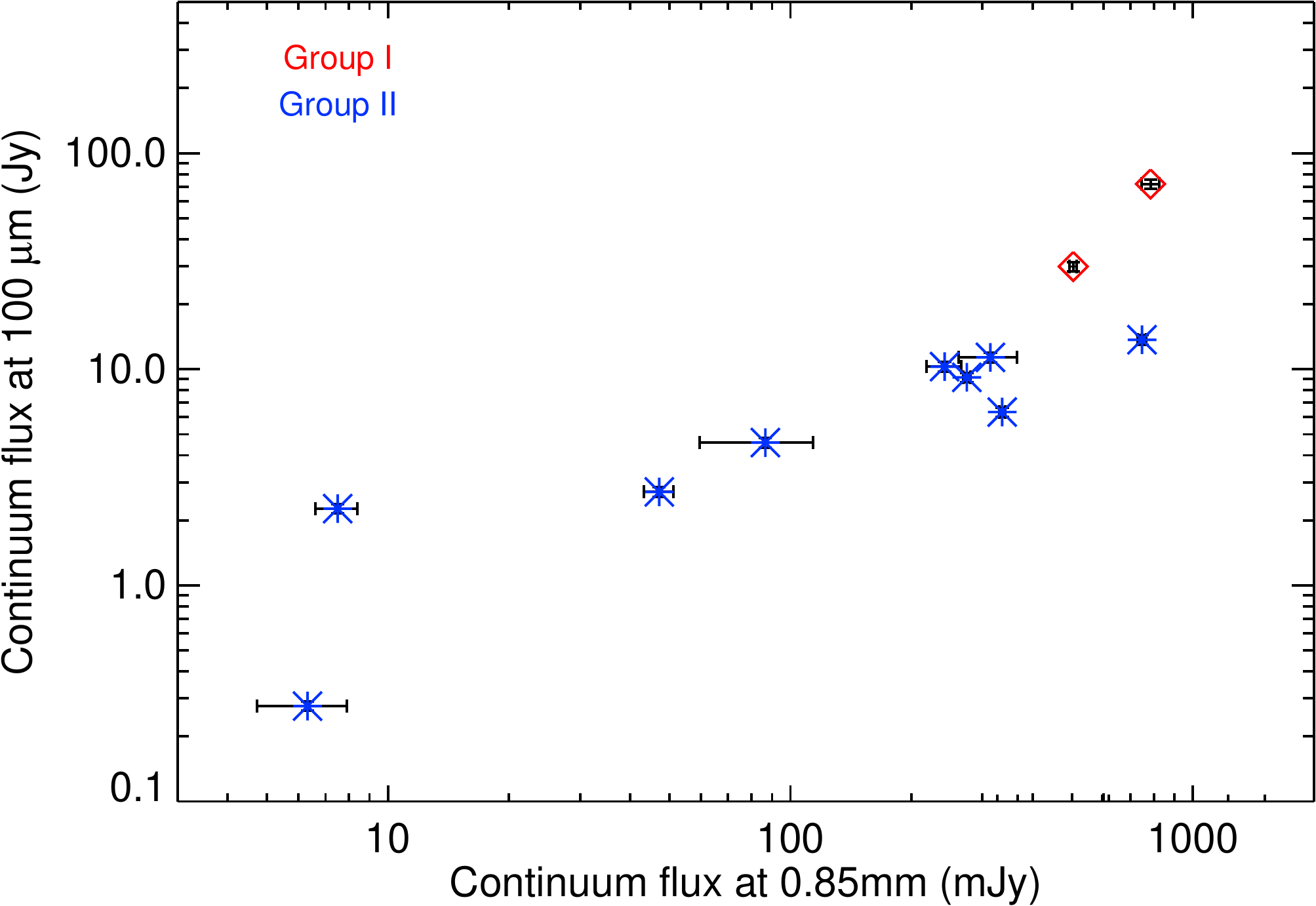}
\includegraphics[scale=0.35]{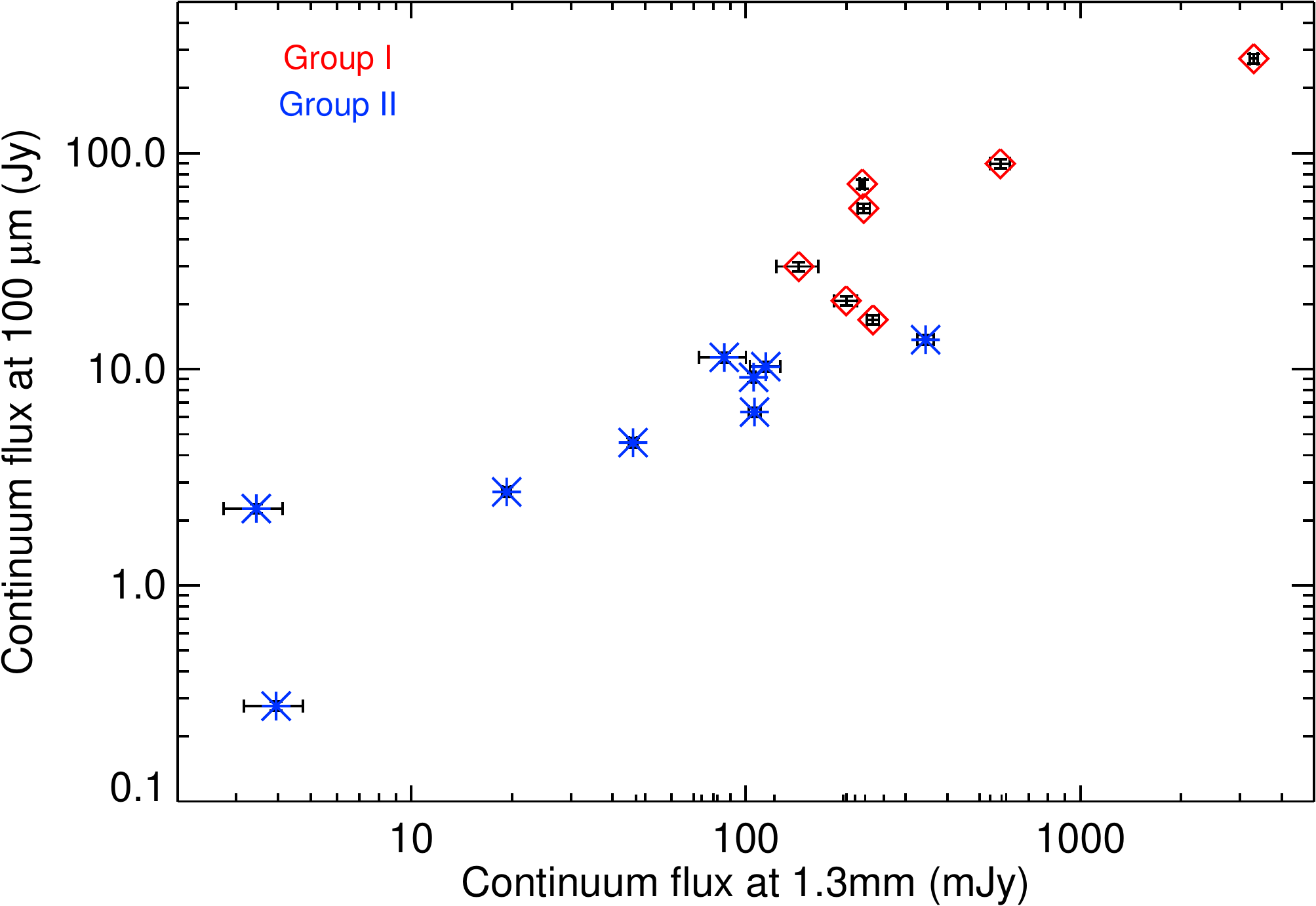}
\includegraphics[scale=0.35]{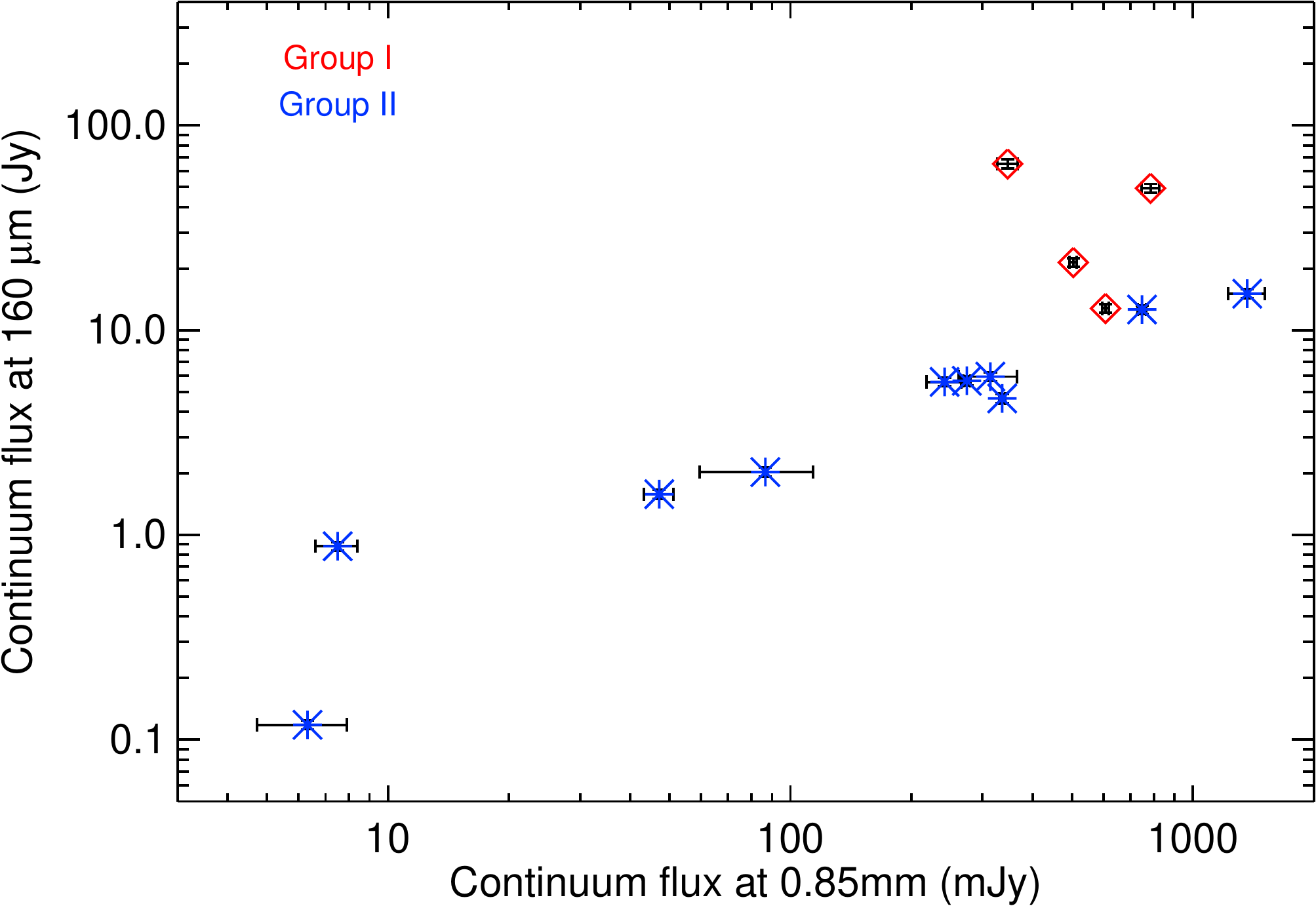}
\includegraphics[scale=0.35]{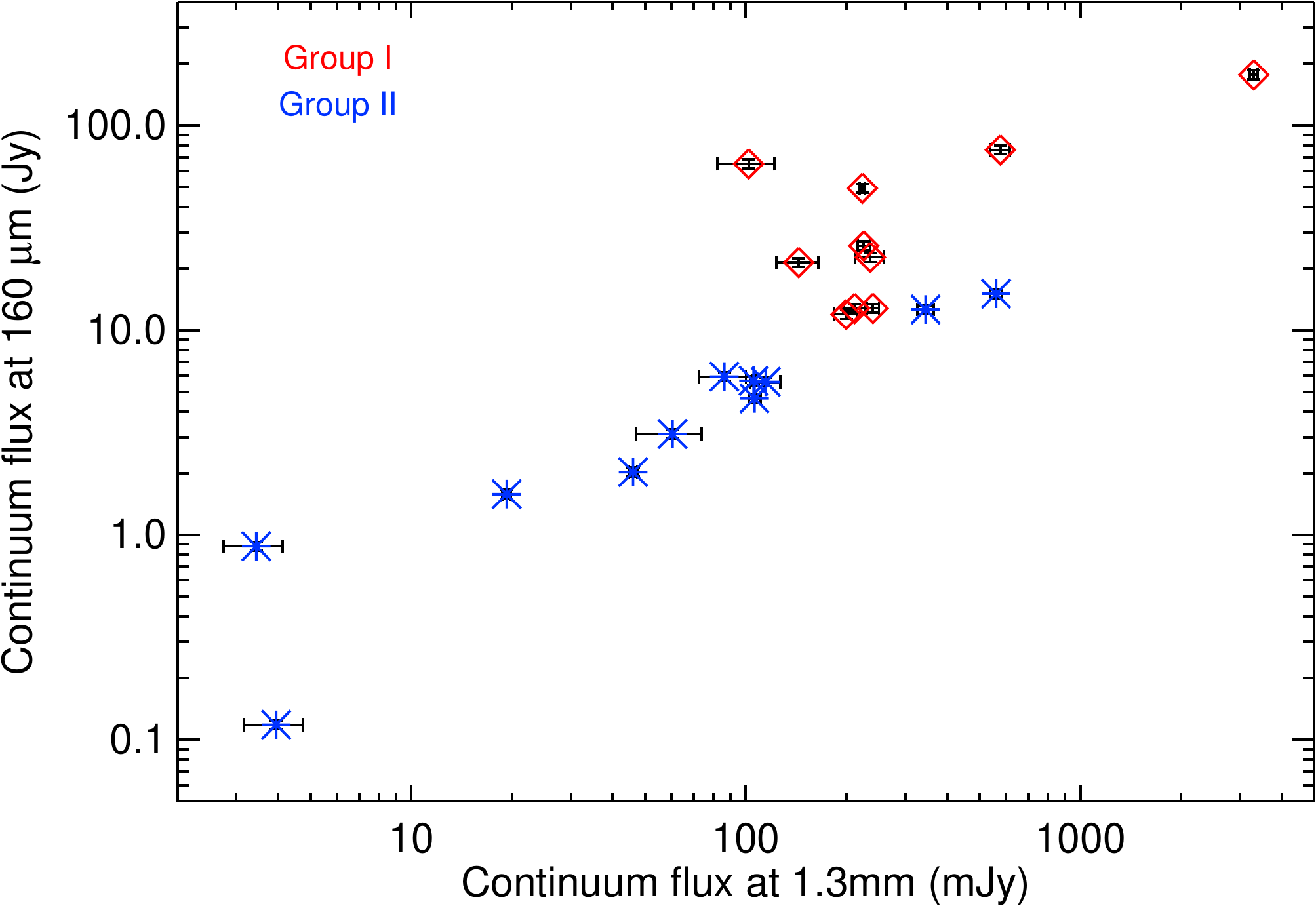}
\caption{PACS 70, 100, and 160~$\mu$m fluxes (top to bottom) plotted against sub-mm data (left hand column) and mm data (right hand column). 
The far-infrared correlates with the mm region with a tighter correlation for longer wavelengths. 
Statistical analysis is presented in Table \ref{table_stat}.}
\label{f_flux}
\end{center}
\end{figure*}

From the parameter study by \cite{Meijer2008} it became clear that, while the SED 
shortward of 60 \mic \ is determined by the mass in small ($<$ 25 \mic) dust grains,
the SED longward of 100 \mic \ is mainly determined by the larger dust grains
that are located in the disc midplane. Adding larger dust grains to the disc will 
increase the mm flux, as well as change the slope of the mm SED. We now use 
our PACS photometry to study these effects in our sample of HAeBes. Therefore,
we need to keep the following in mind:

\begin{enumerate}
\item{The SED classification (group I/II) is based on wavelengths up to 60 \mic \
and is determined by the mass in small grains;}
\item{The SED at $\lambda > 100$ \mic \ is determined by the larger dust grains;}
\item{At mm wavelengths, the emission is optically thin, hence the flux relates to the 
total dust mass (excluding larger bodies such as pebbles);}
\item{The slope at mm wavelengths gives an indication of grain size.}
\end{enumerate}

In Fig.\ref{f_flux} we plot the source fluxes in each PACS waveband as a function of the flux
at 850 and 1300 \mic. The fluxes are scaled to a distance of 140 pc
to correct for distance effects. For group II sources there is a clear correlation between the far-infrared 
and mm fluxes, and the correlation is stronger at longer wavelengths. 
This means that an increase in disc mass is accompanied by an increase in far-infrared 
flux, as already predicted by \cite{Meijer2008}, the effect being stronger 
at longer wavelengths, where the disc becomes more and more optically
thin. It is interesting to note that the correlation is only seen for the group II sources, 
the flared discs do not show a correlation. This is likely due to a greater importance
of their UV luminosities and PAH emission, contributing to the heating of the gas,
visible at IR wavelengths \citep[see discussion in][]{Meeus2012}.

\begin{figure*}[h!]
\begin{center}
\includegraphics[scale=0.4]{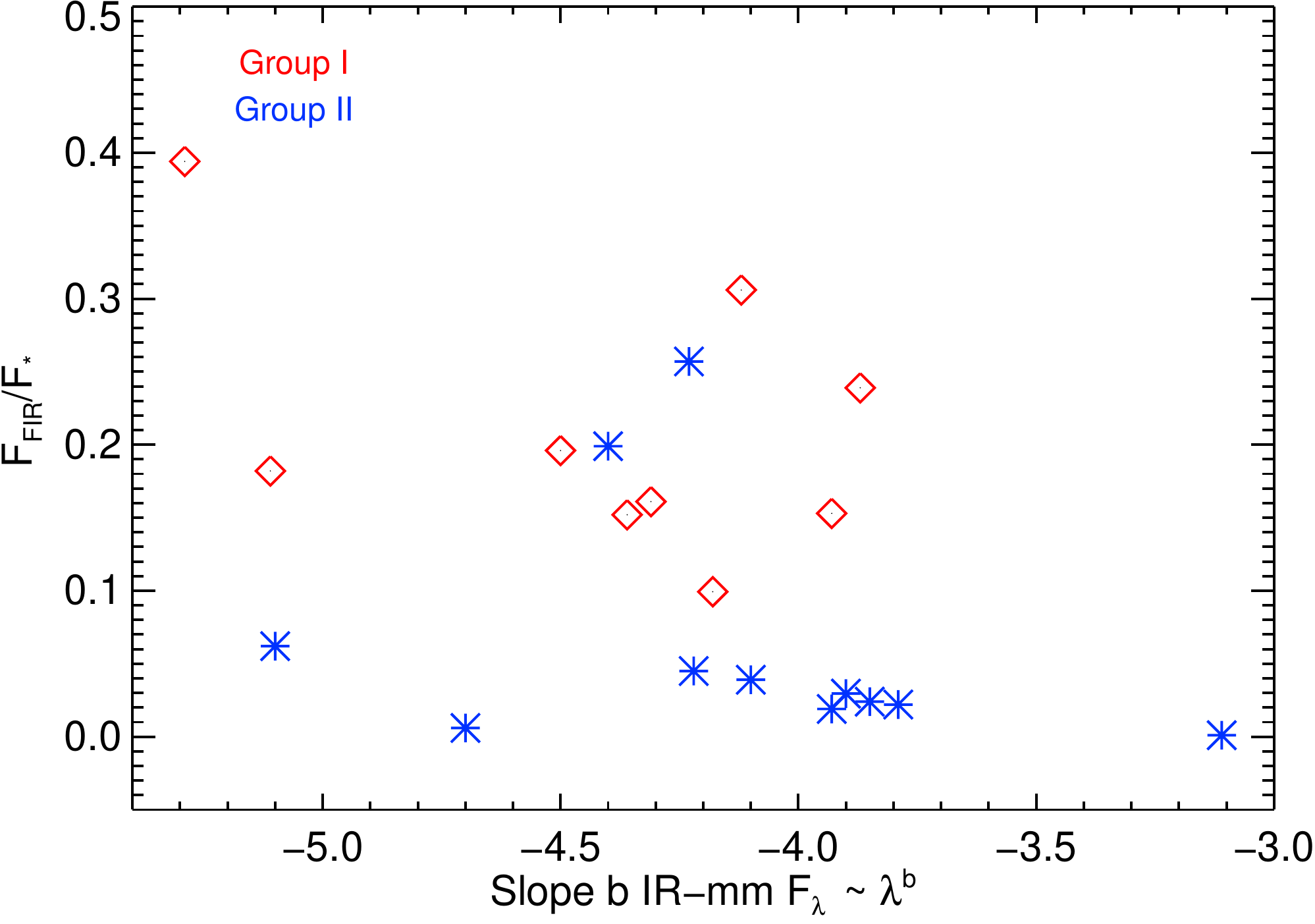}
\includegraphics[scale=0.4]{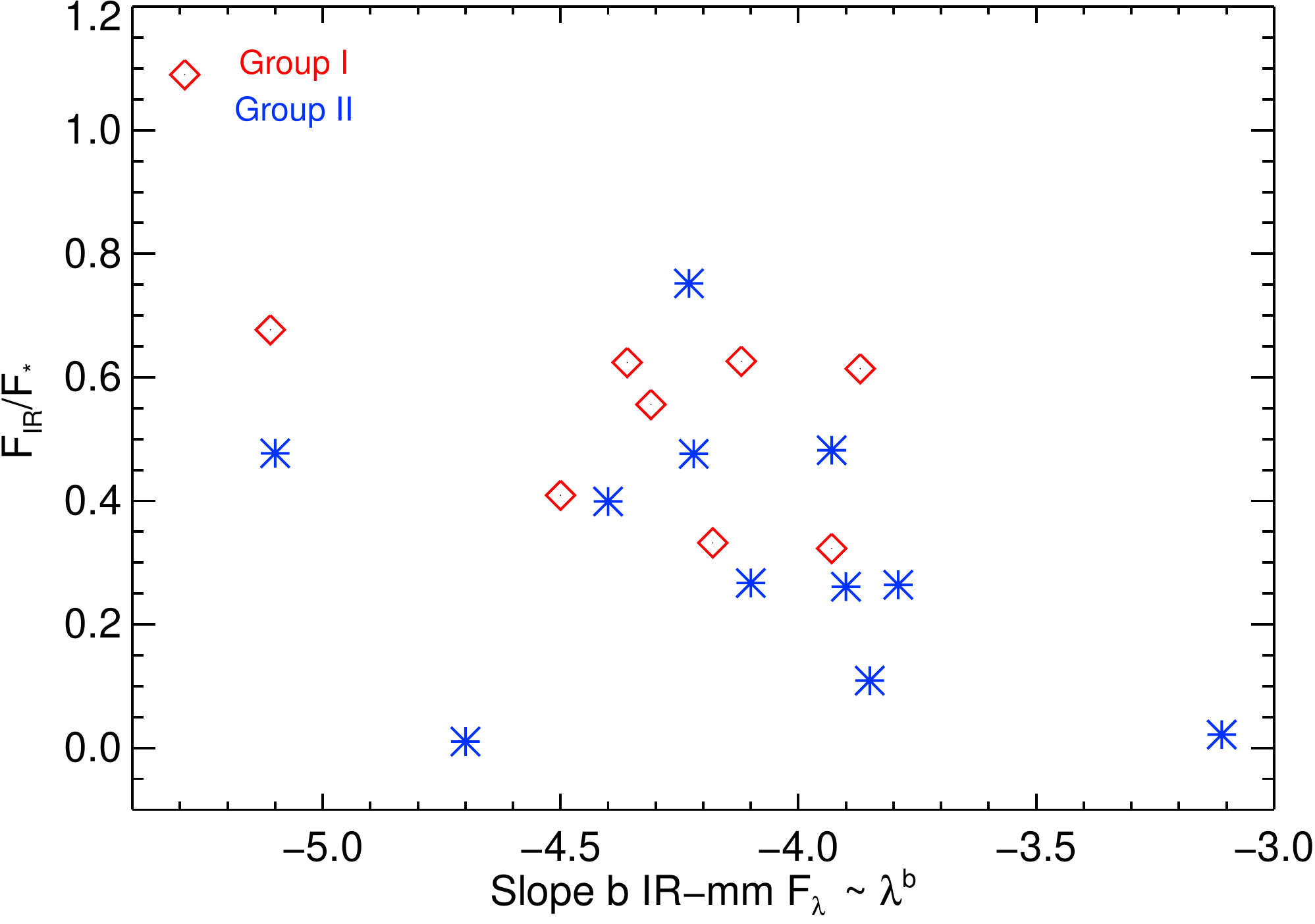}
\caption{Far-infrared (left) and infrared (right) excess versus mm continuum and slope, as shown in Table \ref{table_stat}. 
Only far-infrared excess is tentatively correlated to the mm slope (plot on the left).}
\label{f_excess}
\end{center}
\end{figure*}

In Fig. \ref{f_excess}, we plot the excesses as a function of the mm slope. We 
do not observe a correlation, but on average, the excesses are larger for 
sources with steeper mm slopes (see Table \ref{t_means}). This indicates that 
a larger number of small grains (steeper slope) increases the (far-)infrared excess, 
confirming the prediction by \cite{Dullemond2004}: since small grains dominate the opacity, an 
increase in their mass will lead to an increased amount of flaring.  

\begin{figure*}[h!]
\begin{center}
\includegraphics[scale=0.4]{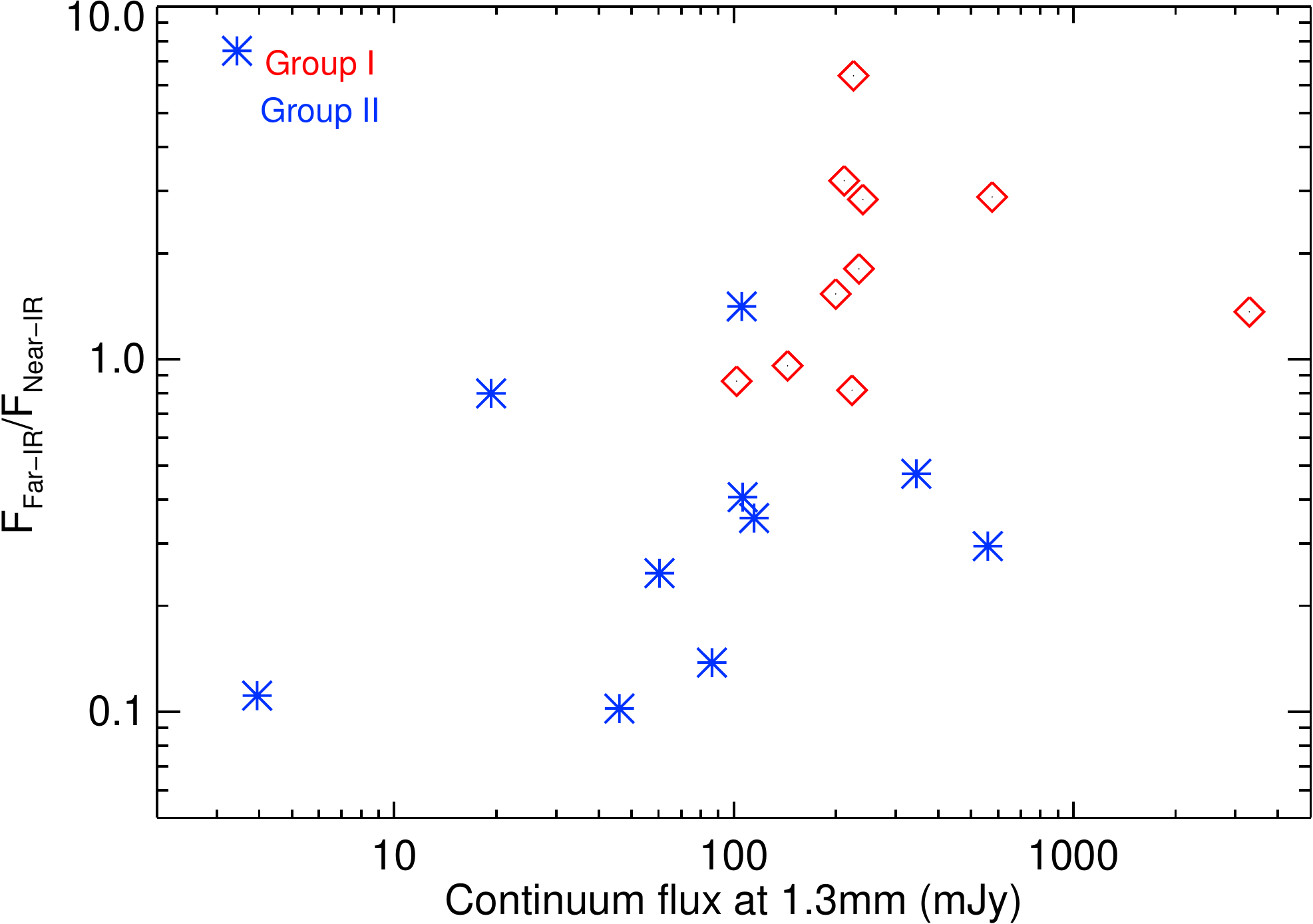}
\includegraphics[scale=0.4]{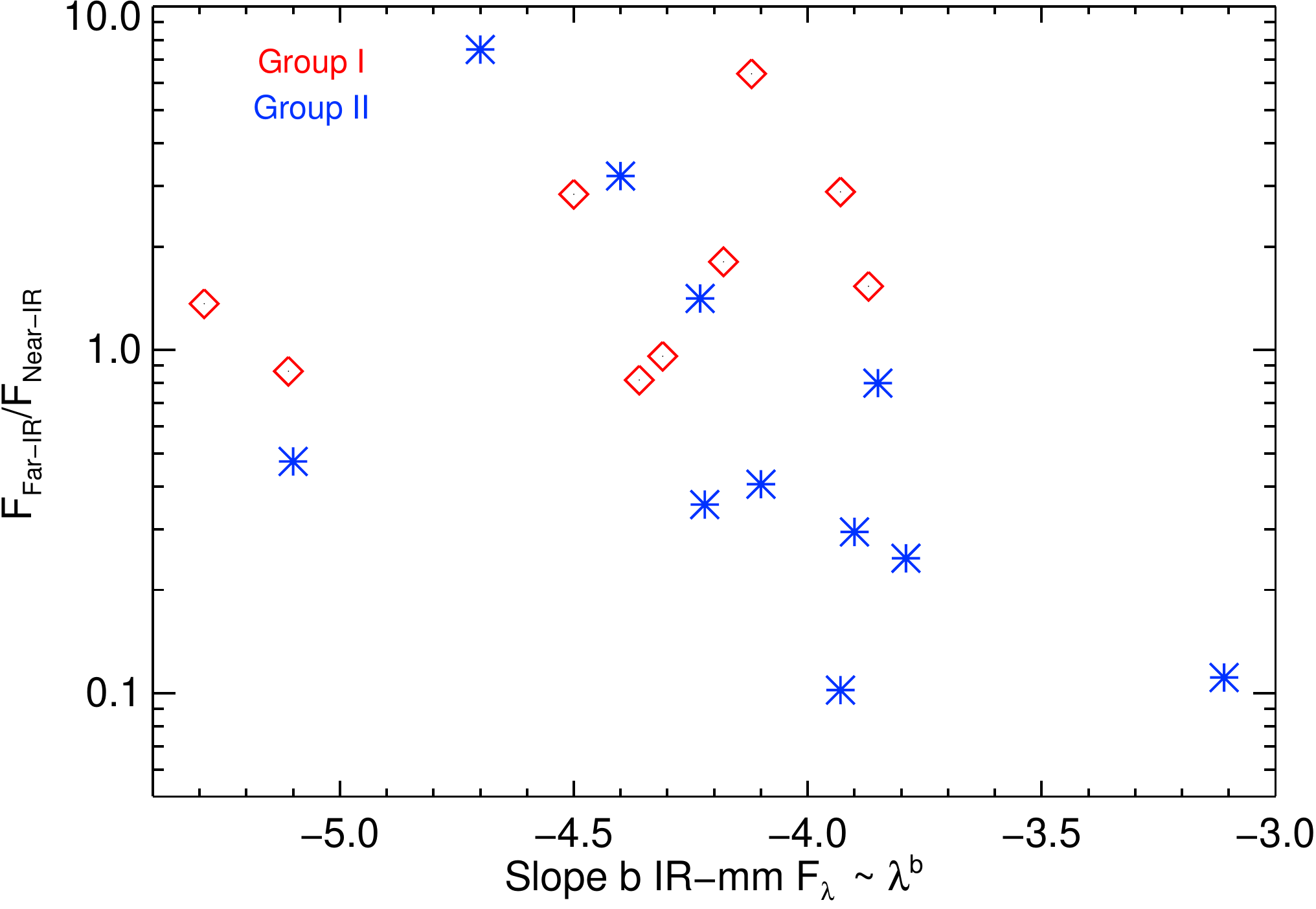}
\caption{Excess ratio versus 1.3 mm continuum flux density (left) and infrared-millimetre slope (right). 
There are no correlations between these parameters. }
\label{f_ratio}
\end{center}
\end{figure*}

\begin{figure*}[h!]
\begin{center}
\includegraphics[scale=0.4]{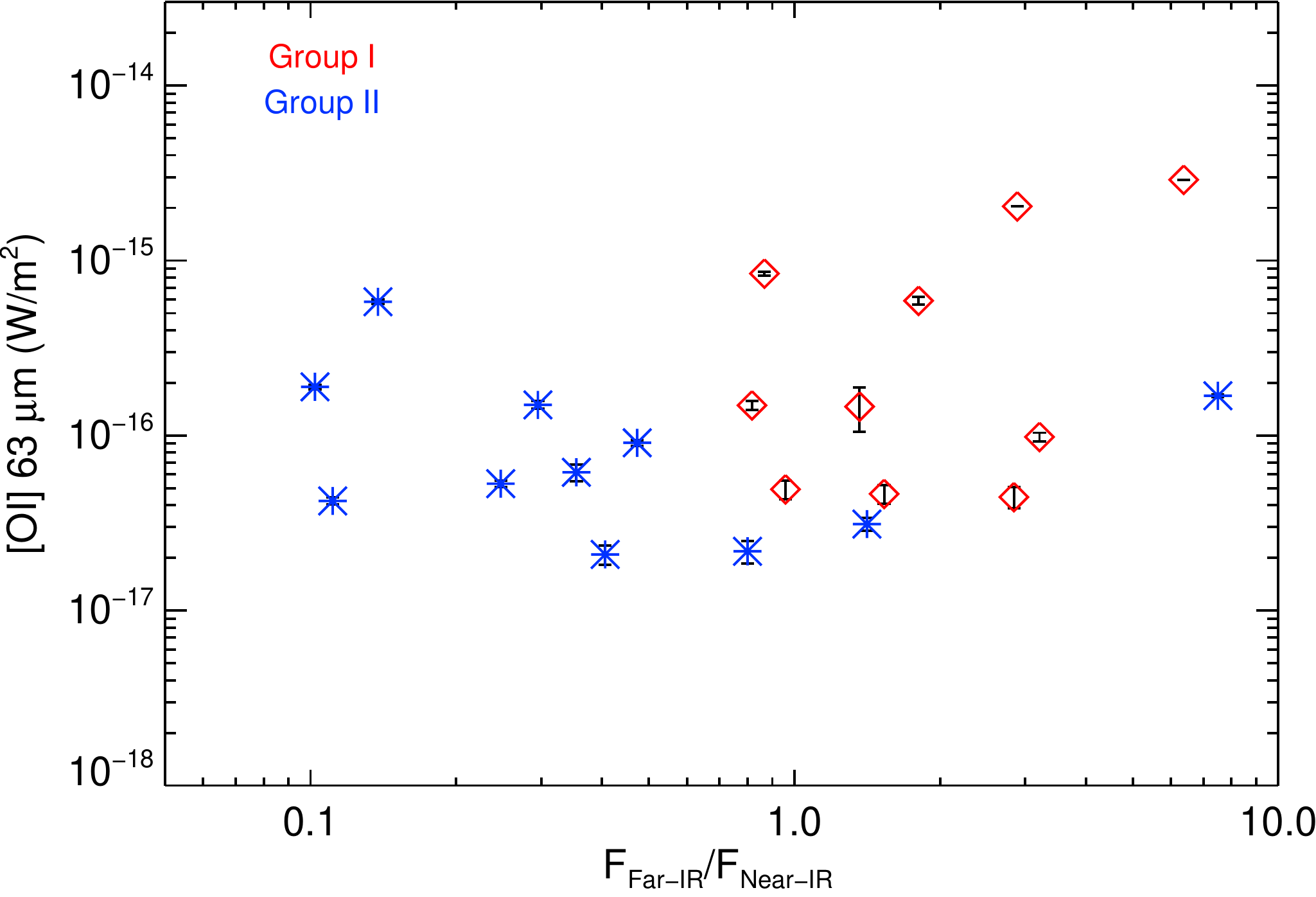}
\caption{\Odrie \ line flux as a function of the far-infrared /near-IR ratio, where a correlation is not observed.}
\label{f_OI}
\end{center}
\end{figure*}

In Fig. \ref{f_ratio}, we plot the excess ratio as a function of the 1.3 mm flux density and the 
infrared to mm slope. Here we observe the following: the ratio of far-infrared to near-infrared flux (FIR/NIR)
correlates with the mm flux:  flaring discs have a higher dust mass than flat discs 
(excluding the mass in larger grains that might be hiding in the disc midplane. 
The FIR/NIR ratio also tends to be larger for sources with a steeper mm slope, again 
indicating that an increase in lower grain mass leads to an increase in flaring.

In Fig. \ref{f_OI} we plot the [OI] 63 \mic \ flux density as as function of the far-
to near-IR ratio. The three sources with the highest [OI] flux densities are the FUV bright stars,
AB Aur, HD 97048, and HD 100546. Excluding those three sources, there is no
correlation between the \Odrie \ line flux density and the far-infrared to near-infrared flux ratio. 
On average, the group I sources do have higher \Odrie \ line flux density than 
group II sources (see Table~\ref{t_means}).

\section{Conclusions}
\label{s_conclusions}

In this paper, we present new \textit{Herschel/PACS} far-infrared photometry and imaging observations obtained at 70  (and/or) 100, and 160 \mic \
for a sample of 22 HAeBes and five debris discs. We combine these new measurements with literature data across a broad range of wavelengths 
to complement the far-infrared photometry and construct SEDs.  We 
calculated the discs' excesses in three regimes that span the near-, mid-, and far-infrared, as well as the total excess
due to disc emission. 
This is the first time the group I/II discs have been studied at far-infrared wavelengths; previous
studies concentrated on the SED up to 60 \mic, with a comparison to the mm region. 
The region between 100 and 160 \mic \ is important because it is where the 
disc evolves from an optically thick to an optically thin regime \citep[see discussion in][]{Meijer2008}.

As has been previously noted \citep[e.g.][]{Dominik2003}, group I sources have, on average, 
larger infrared excesses than group II sources, as well as steeper mm slopes \citep{Acke2004}.
This suggests that a higher mass of small dust grains is present in these group I sources. 
In our observations, we observe a similar trend, including the excesses at far-infrared wavelengths. 
These results could be consistent with an evolution from group I to group II. However, we note that this
evolutionary scenario is currently under debate \citep[e.g.][]{Mendigutia2012, Maaskant2013}.

Relating the far-infrared emission of HAeBe discs with other observational properties, we found the following:

\begin{enumerate}
\item{For group II sources, the far-infrared flux density correlates with the mm flux density,
the correlation being stronger for longer wavelengths. This suggests 
that the emission in the far-infrared correlates with the dust mass. The far-infrared flux densities of 
group I sources do not correlate with the corresponding mm flux densities and are likely more influenced
by the stellar UV luminosity and heating by PAHs.}
\item{On average, the far-infrared excess is greater for sources with steeper mm slopes:
a larger number of small grains increases the far-infrared excess.}
\item{The far-infrared to near-infrared excess ratio is greater for sources with 
a higher 1.3 mm flux, implying a greater degree of flaring, while the sub-mm slope is steeper for larger excess ratios.}
\item{We do not find a correlation between the \Odrie \ line flux and the
far-infrared to near-infrared excess ratio, but on average group I sources have higher
\Odrie \ line fluxes.}
\end{enumerate}

We also studied the spatial extent of our sources in the far-infrared. Several of our sources 
show a larger spatial extent than expected of a point source, namely: AB Aur, HD 100546, HD 104237, HD 141569A, 
HD 142527, and the debris disc 49 Cet. 

Finally, the photometric data set presented here is important for cross-calibrating \textit{Herschel}/PACS spectra, 
whose absolute flux calibration is not as certain. This aspect of our work has already 
been presented in several papers from the GASPS consortium \citep[e.g.][]{Fedele2013, Meeus2013}.

\begin{acknowledgement}

We would like to thank the PACS instrument team for their dedicated support and M. van den Ancker for the bibliographic photometry data.
G. Meeus, J.P. Marshall, and B. Montesinos are partly supported by AYA-2011-26202. 
G. Meeus is supported by RYC-2011-07920. 
J.P. Marshall is supported by a UNSW Vice Chancellor's Fellowship.
This research made use of the SIMBAD database, operated at the CDS, Strasbourg, France.

\end{acknowledgement}

\clearpage

\bibliography{references_final}
\bibliographystyle{aa}
\nocite{*}

\begin{appendix}

\section{SEDs}

\begin{figure*}
\caption{SEDs of GASPS Herbig Ae/Be stars. PACS fluxes reported in this work can be identified with red triangles. Red arrows are $3\sigma$ upper limits.
Blue circles correspond to literature data from Table \ref{literature}. The solid dark red line shows IUE spectrum and purple line for the {\em Spitzer}/IRS spectrum.
The solid black line is the PHOENIX/GAIA model fitted to the stellar photospheric emission.}
\vspace*{1.0cm}
\hspace*{0.5cm}\includegraphics[scale=0.4]{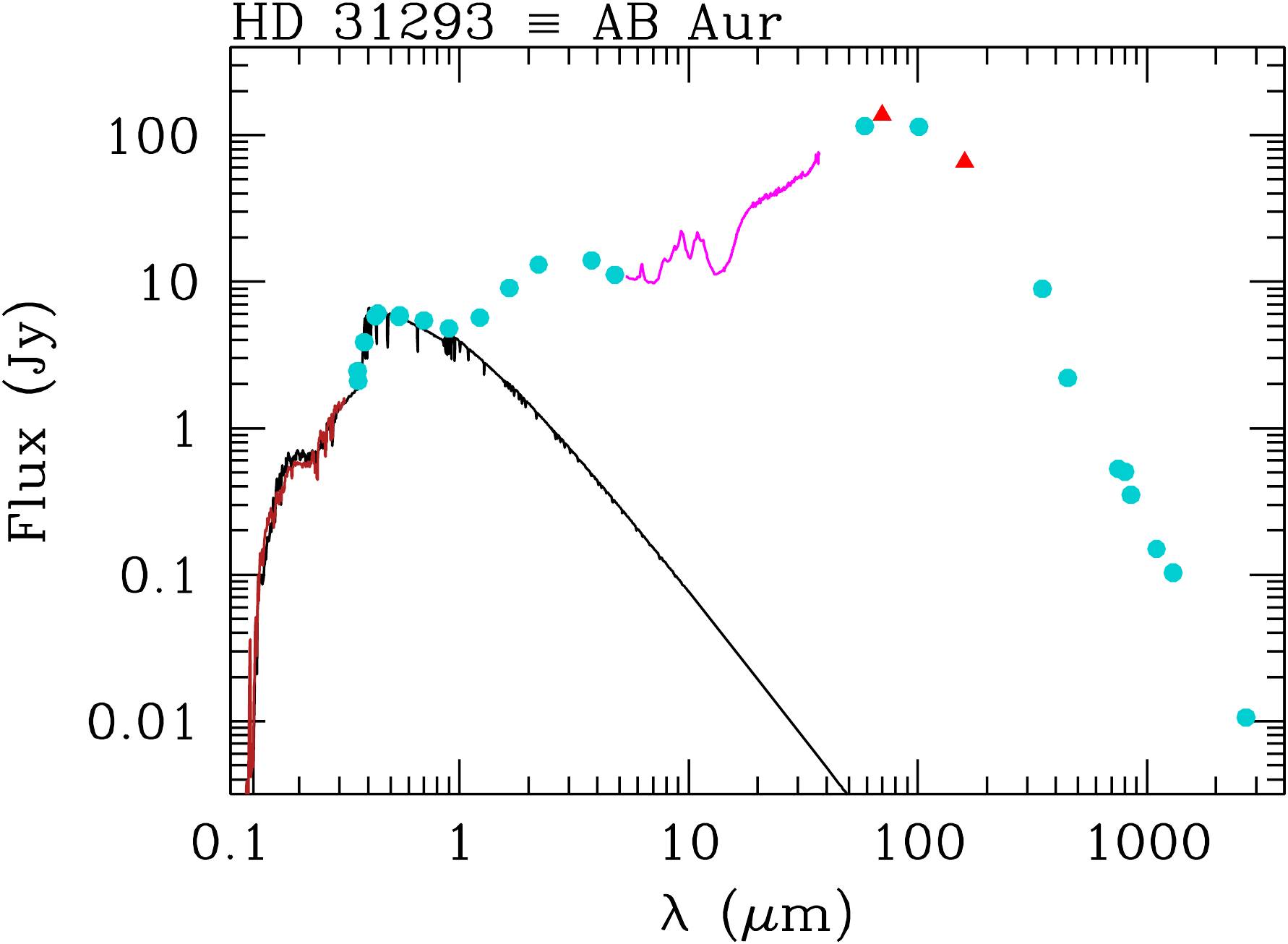}\hspace*{0.5cm}\includegraphics[scale=0.4]{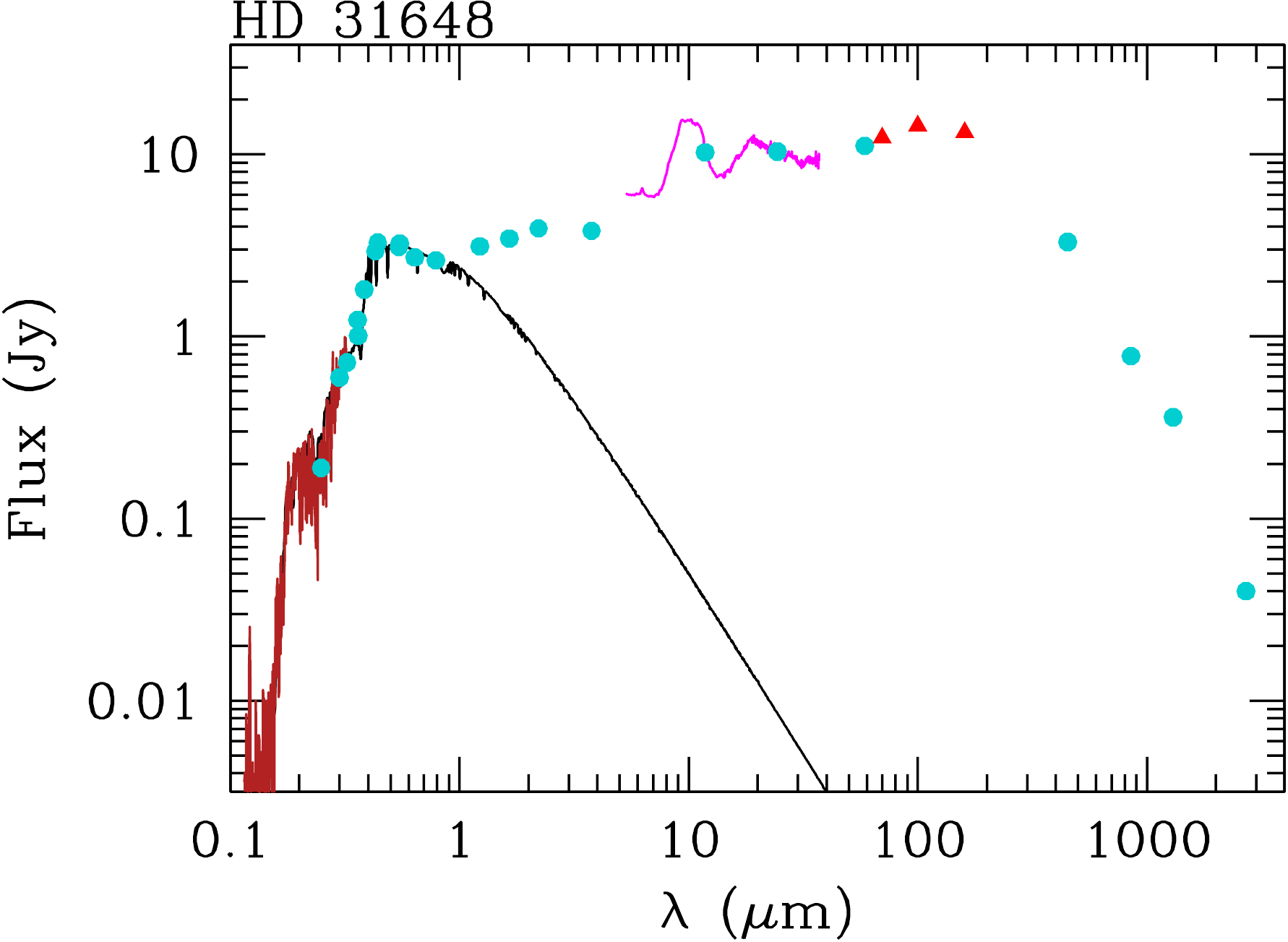}
\hspace*{0.3cm}\includegraphics[scale=0.4]{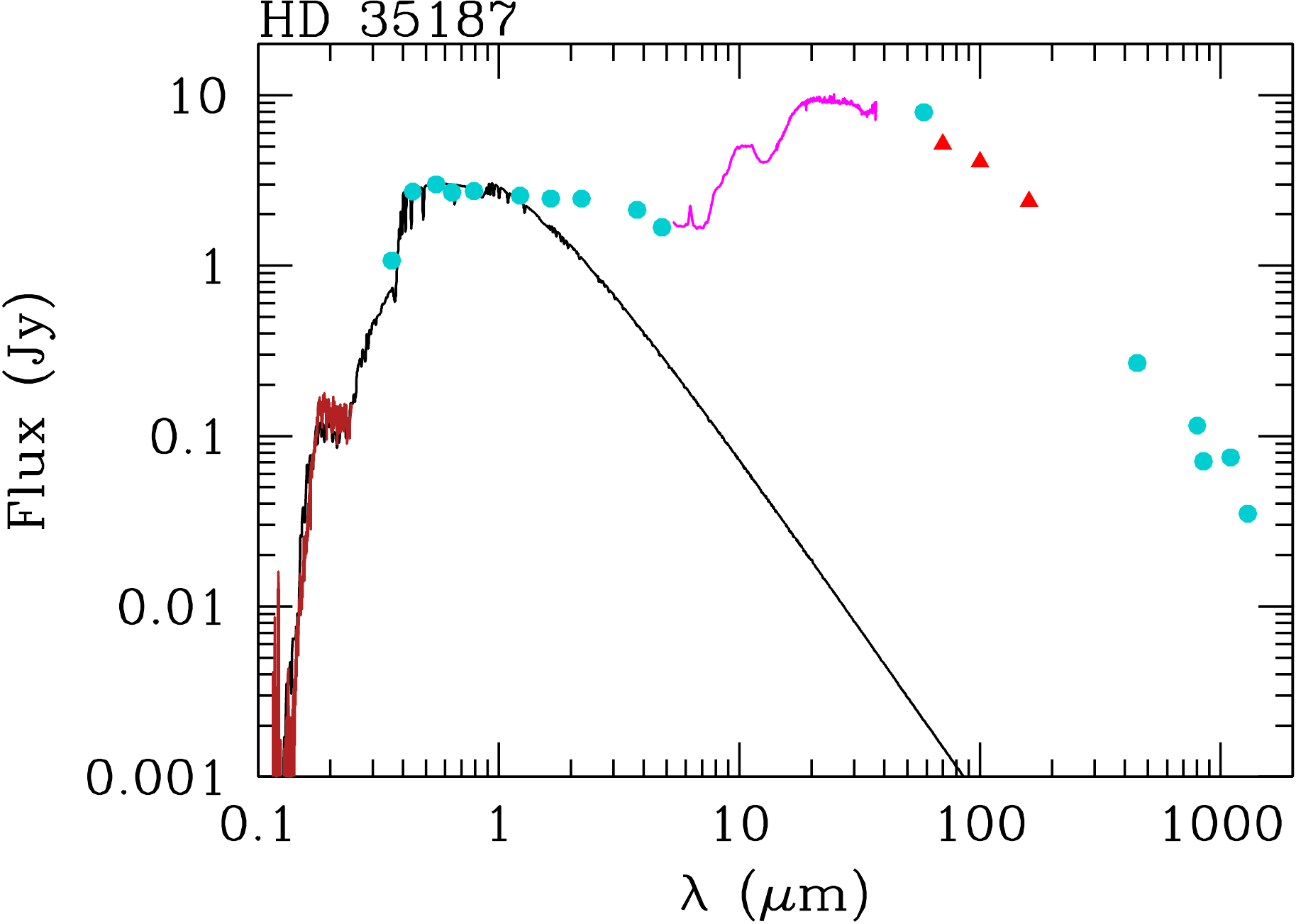}\hspace*{0.3cm}\includegraphics[scale=0.4]{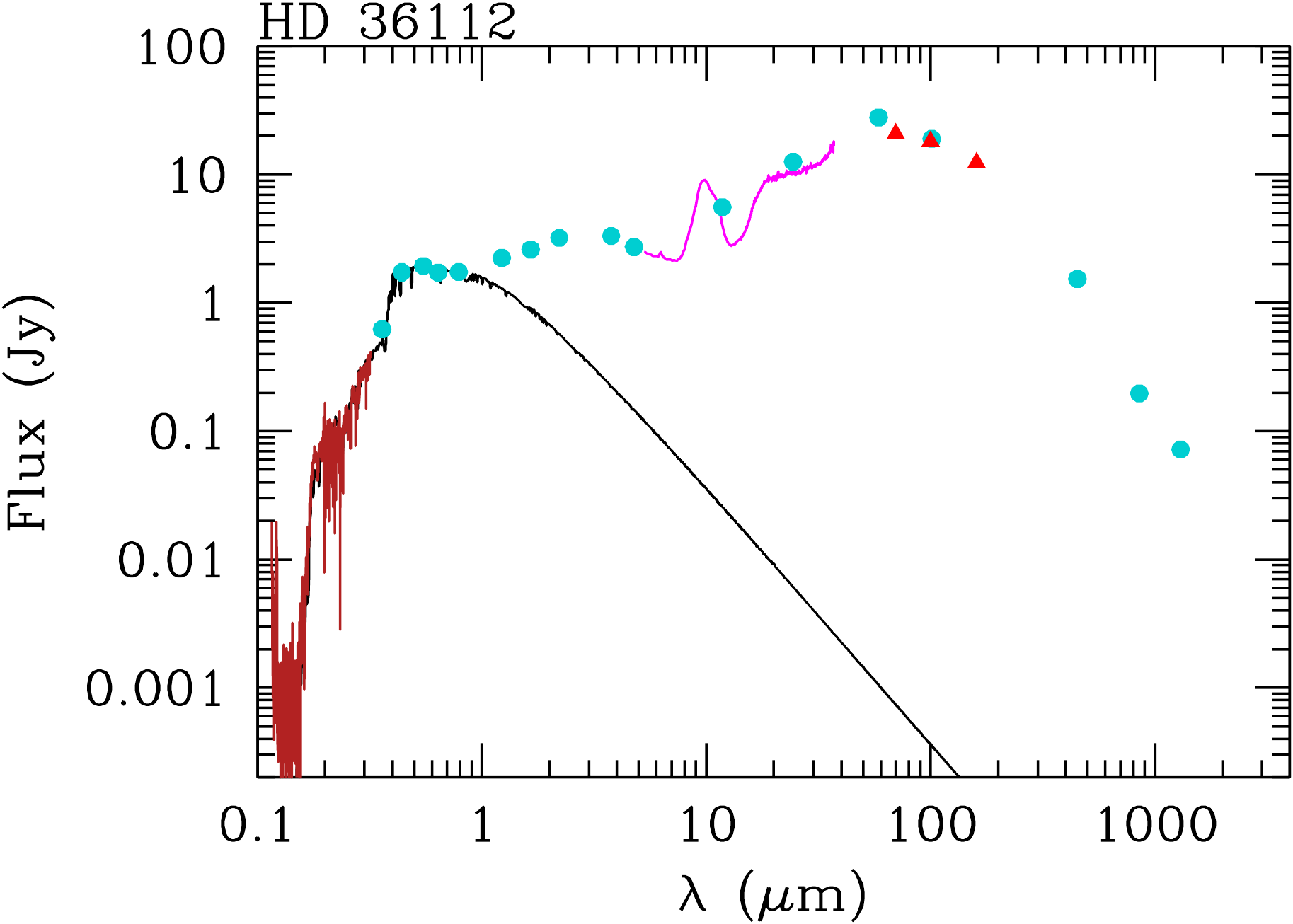}
\hspace*{0.3cm}\includegraphics[scale=0.4]{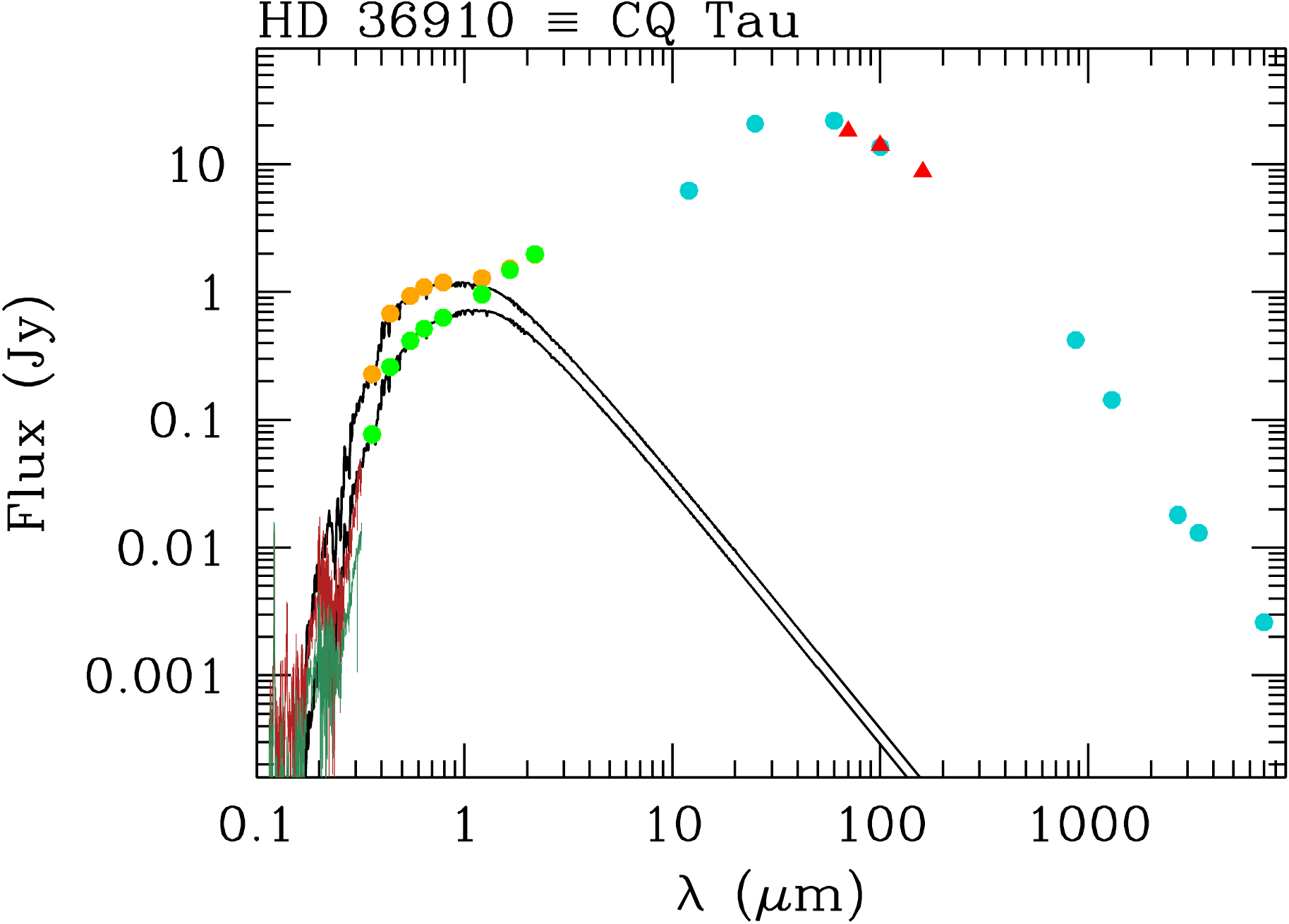}\hspace*{0.5cm}\includegraphics[scale=0.4]{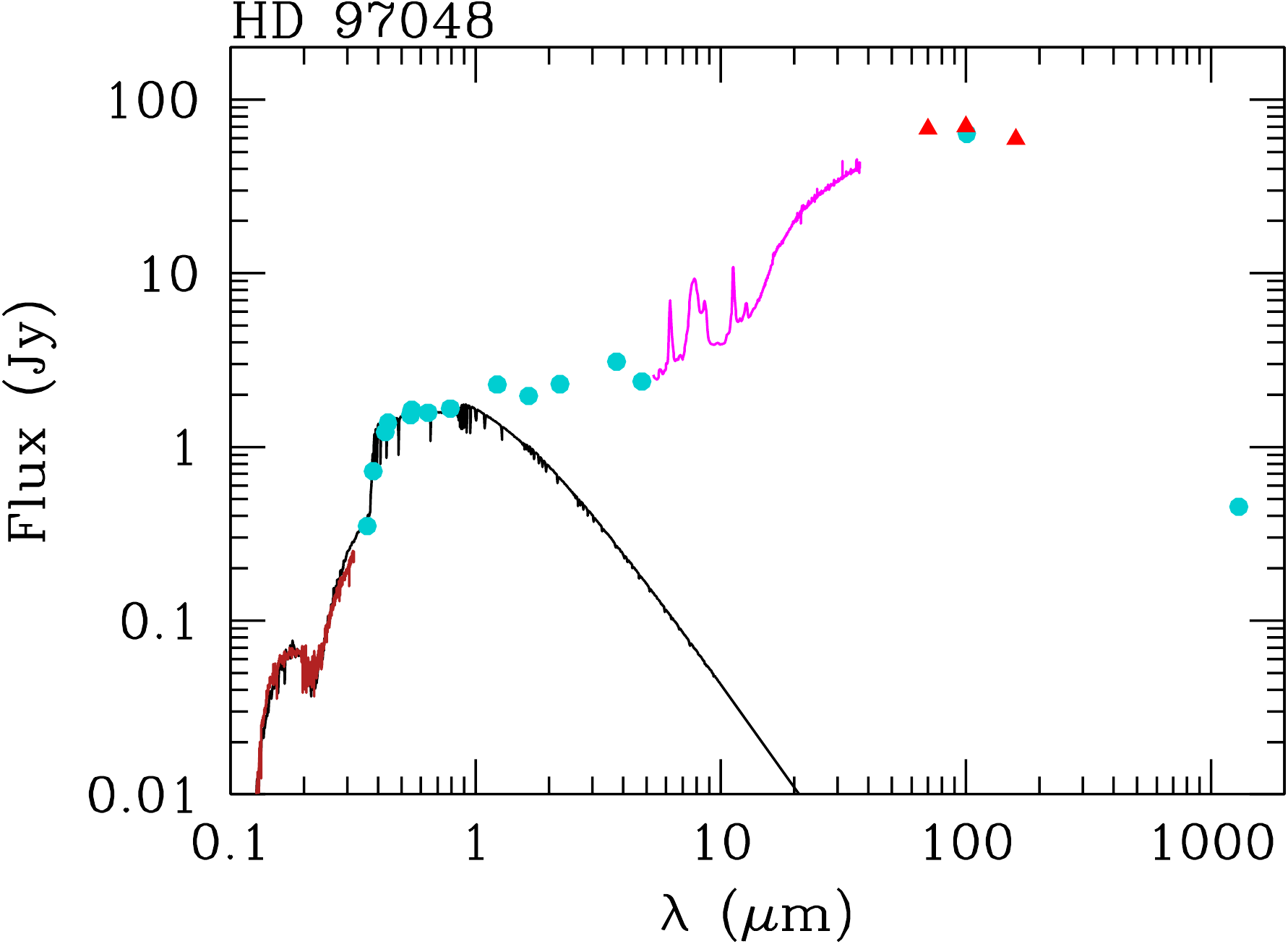}
\hspace*{0.5cm}\includegraphics[scale=0.4]{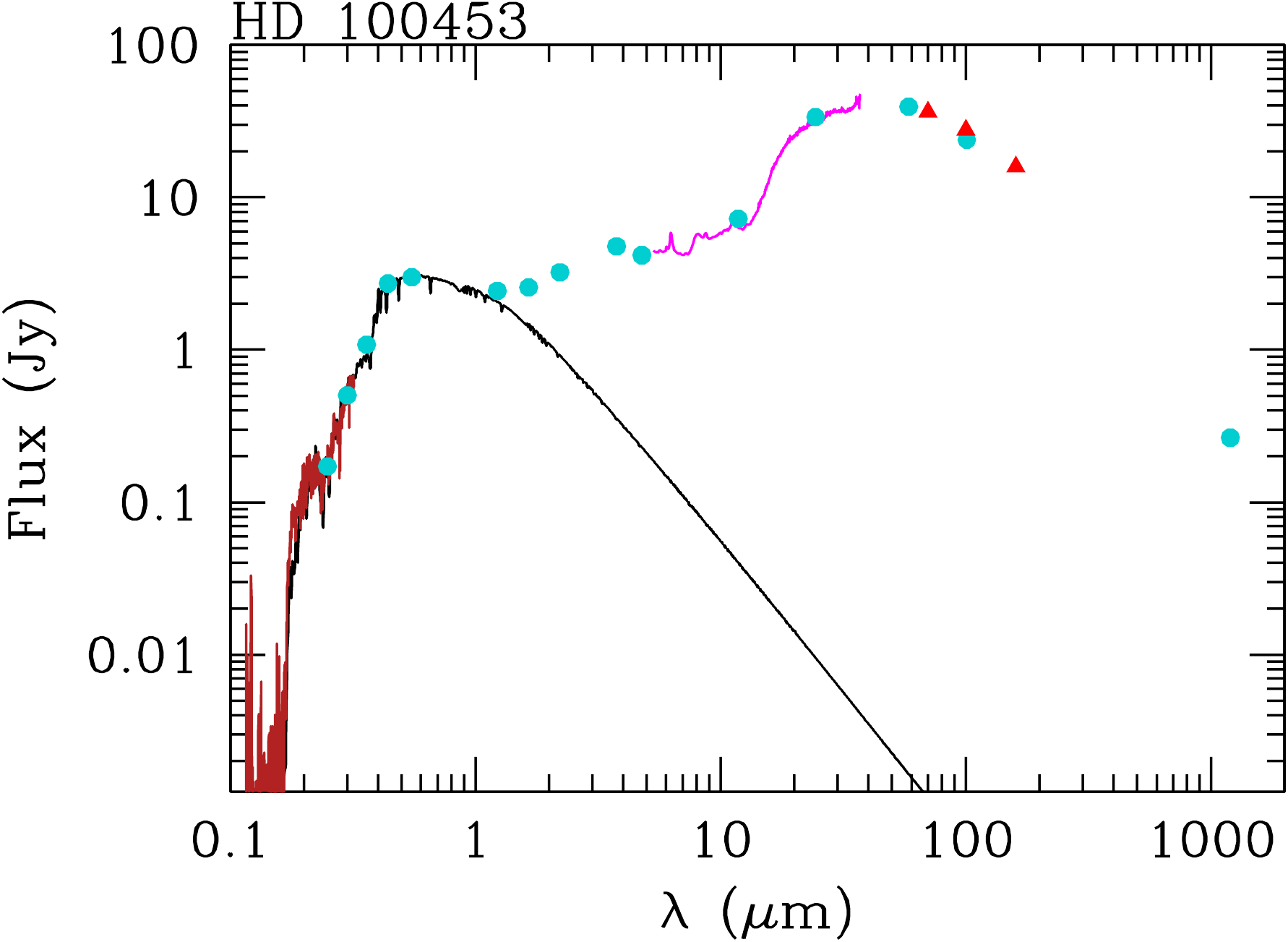}\hspace*{0.6cm}\includegraphics[scale=0.4]{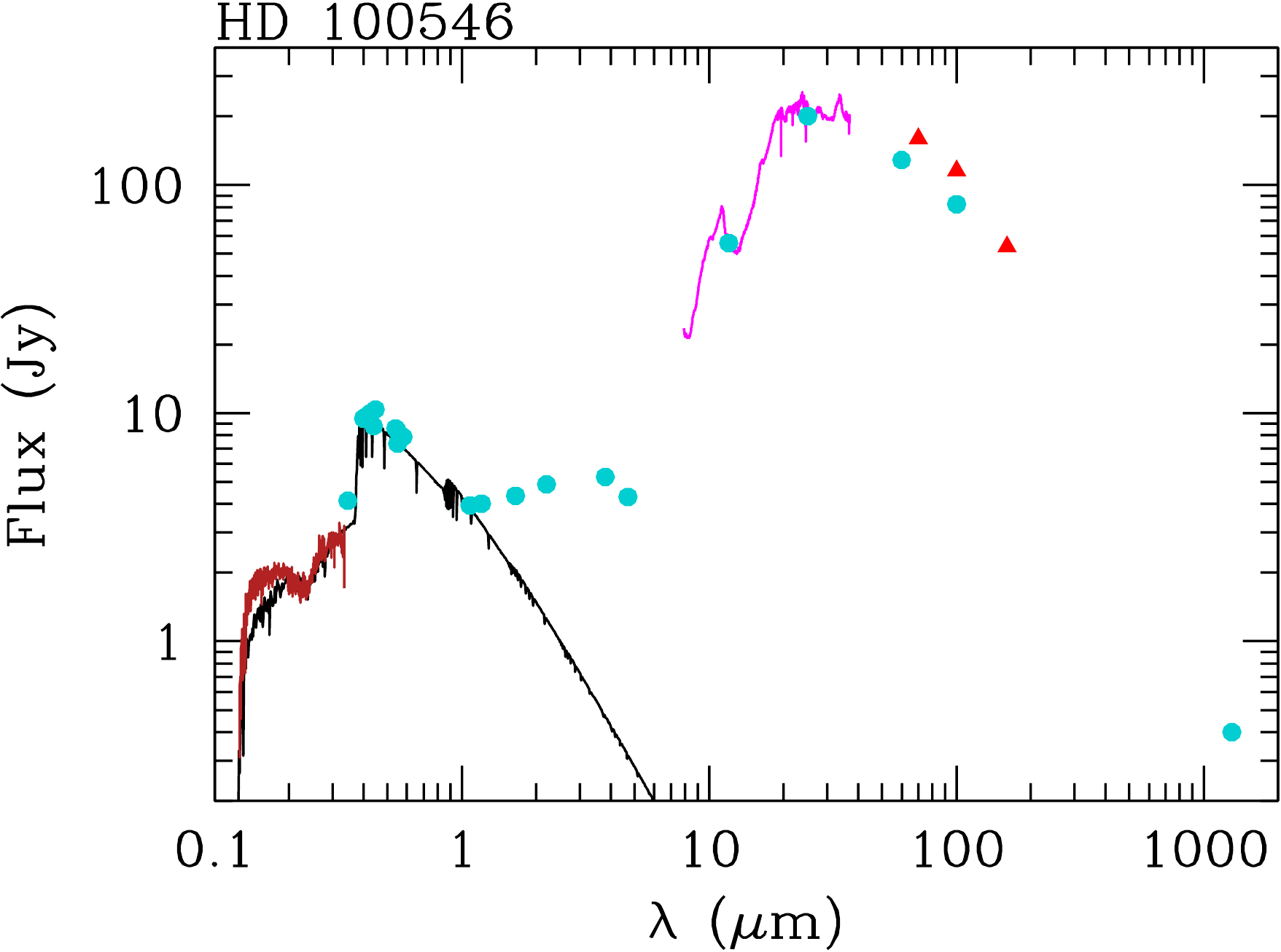}
\label{seds-HAeBe}
\end{figure*}

In Figs. \ref{seds-HAeBe} and \ref{seds-debris} we show the spectral
energy distributions of the objects studied in this paper. The red
triangles show the PACS fluxes, red arrows are $3\sigma$ upper limits.
The literature data used to build the SED are plotted as blue
circles. When available, the IUE spectrum is plotted as a solid dark
red line and the {\em Spitzer}/IRS spectrum as a purple line. The
solid black line is the PHOENIX/GAIA model fitted to the stellar
photospheric emission. Table \ref{literature} gives the references for
the photometry collected to build the SEDs.

\clearpage
\clearpage

\addtocounter{figure}{-1}
\begin{figure}
\caption{SEDs of GASPS Herbig Ae/Be stars (continued).}
\vspace*{1.0cm}
\hspace*{0.5cm}\includegraphics[scale=0.4]{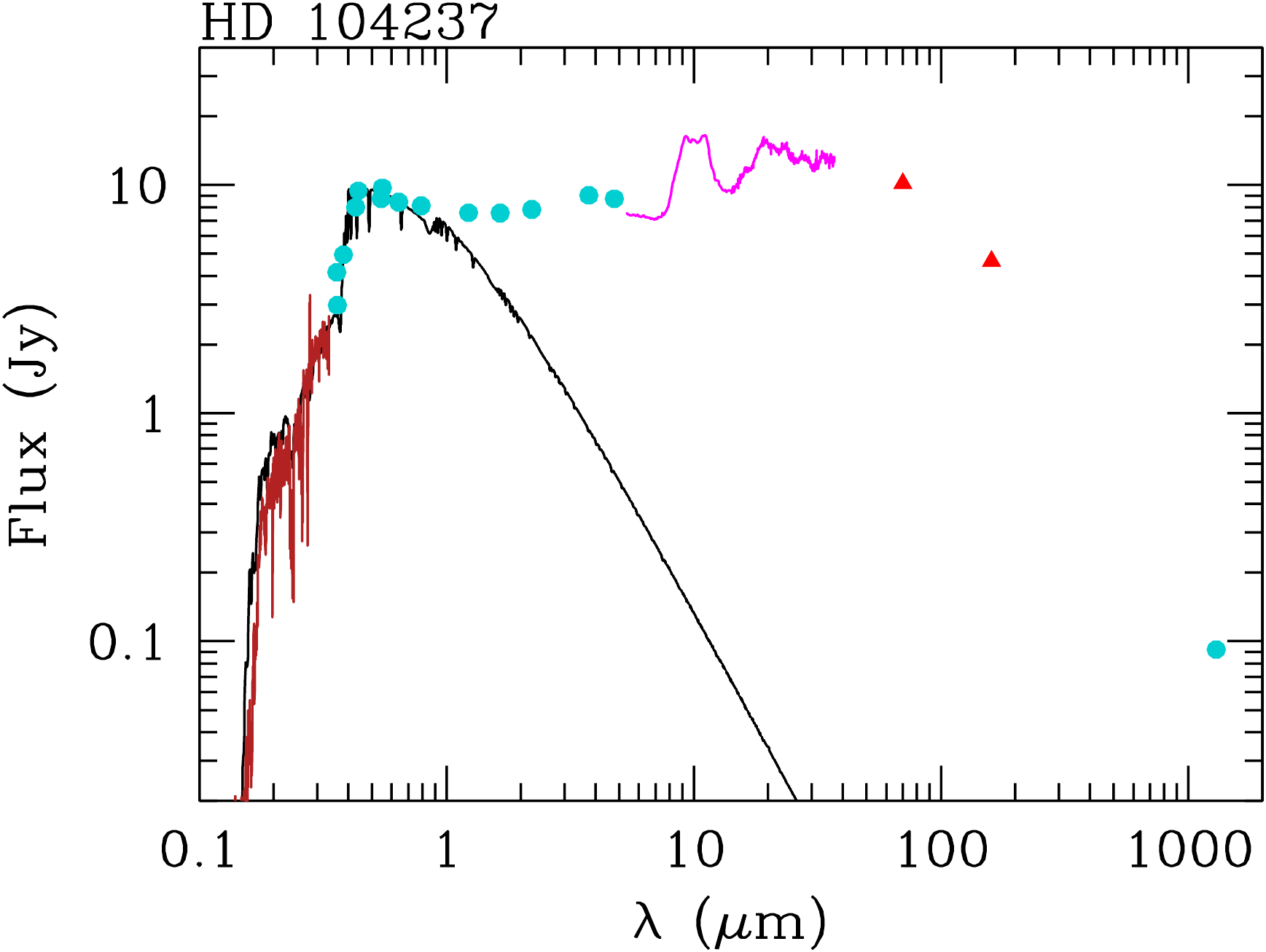}\hspace*{0.3cm}\includegraphics[scale=0.4]{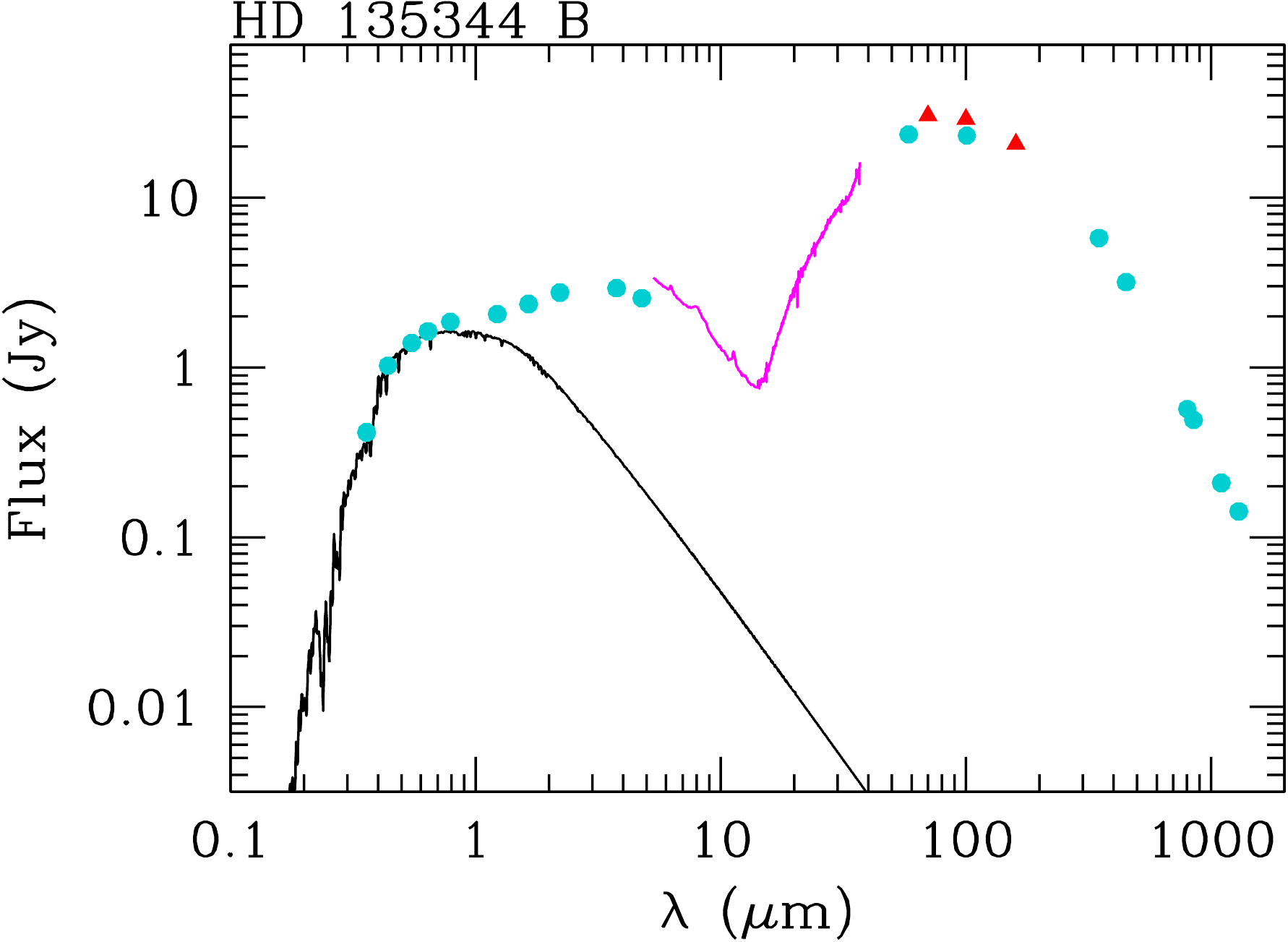}
\hspace*{0.3cm}\includegraphics[scale=0.4]{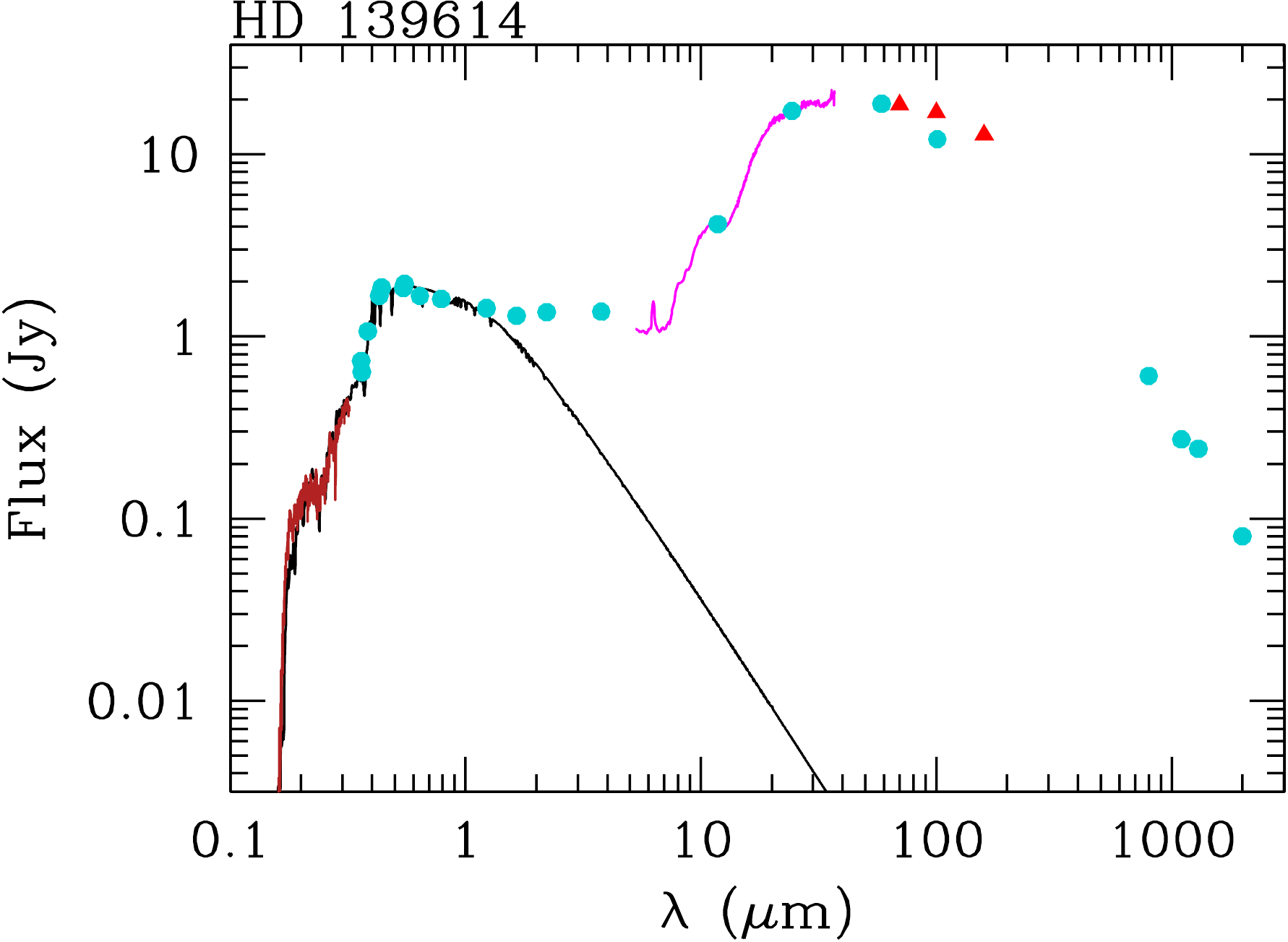}\hspace*{0.3cm}\includegraphics[scale=0.4]{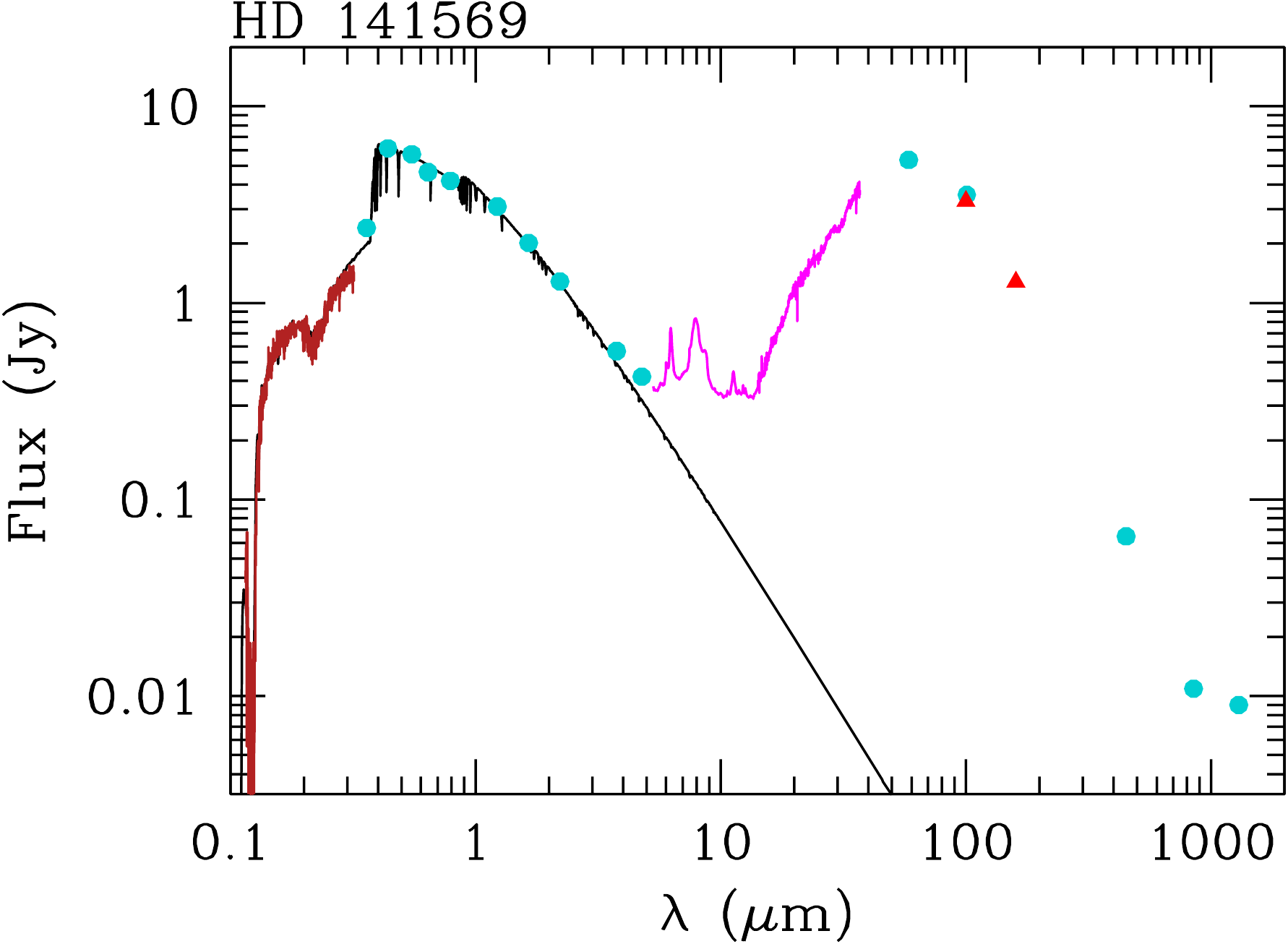}
\hspace*{0.3cm}\includegraphics[scale=0.4]{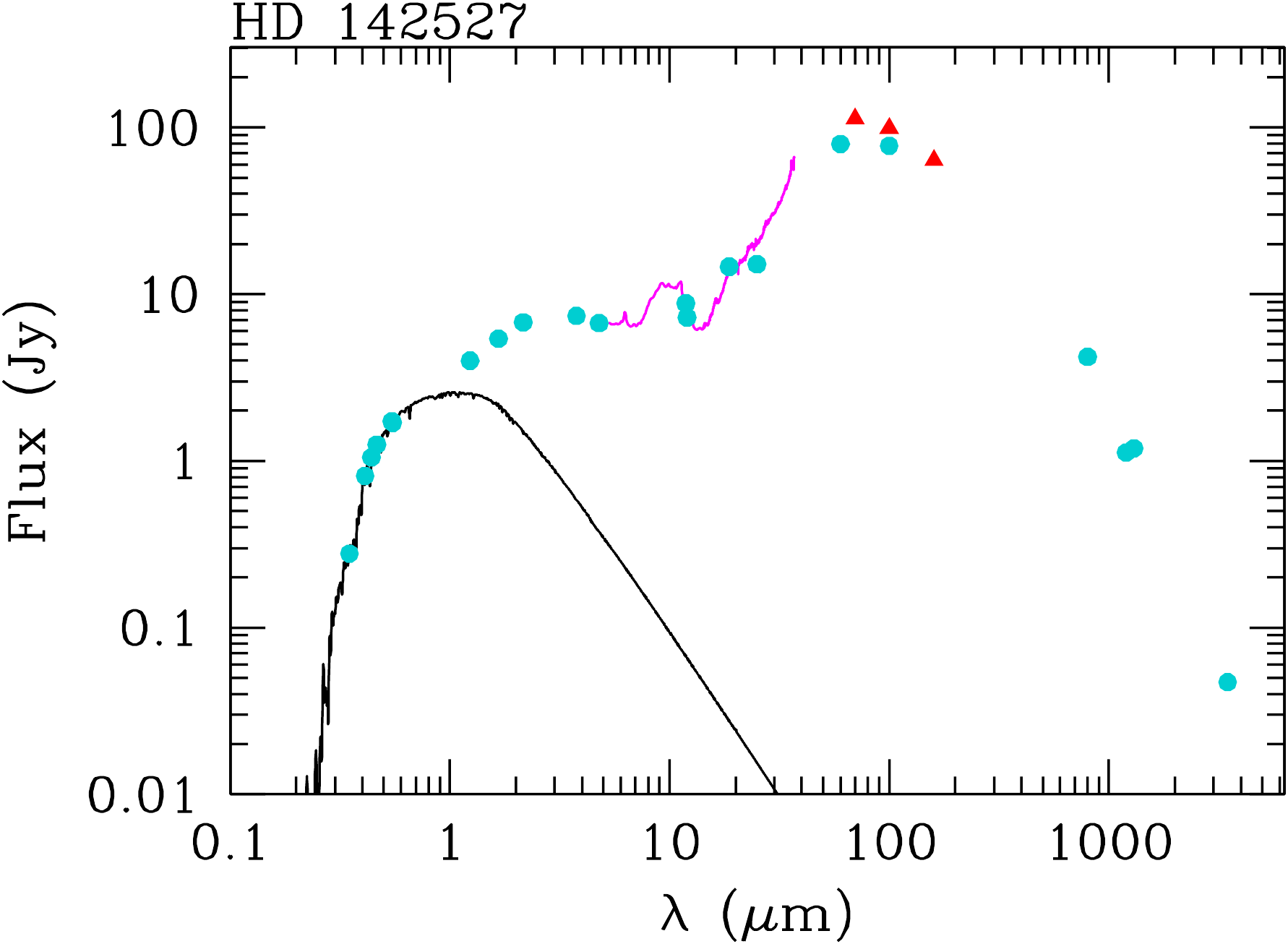}\hspace*{0.3cm}\includegraphics[scale=0.4]{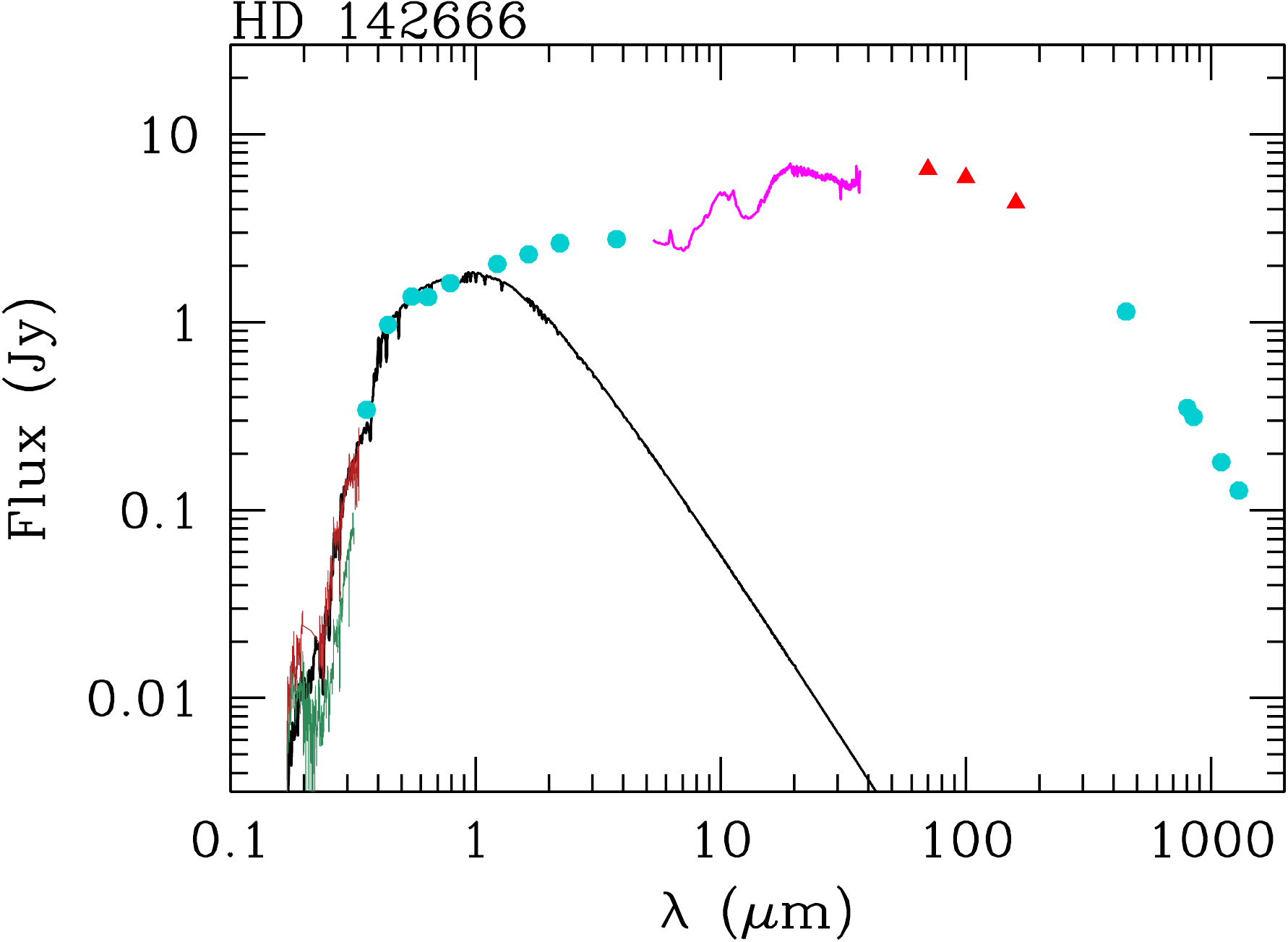}
\hspace*{0.3cm}\includegraphics[scale=0.4]{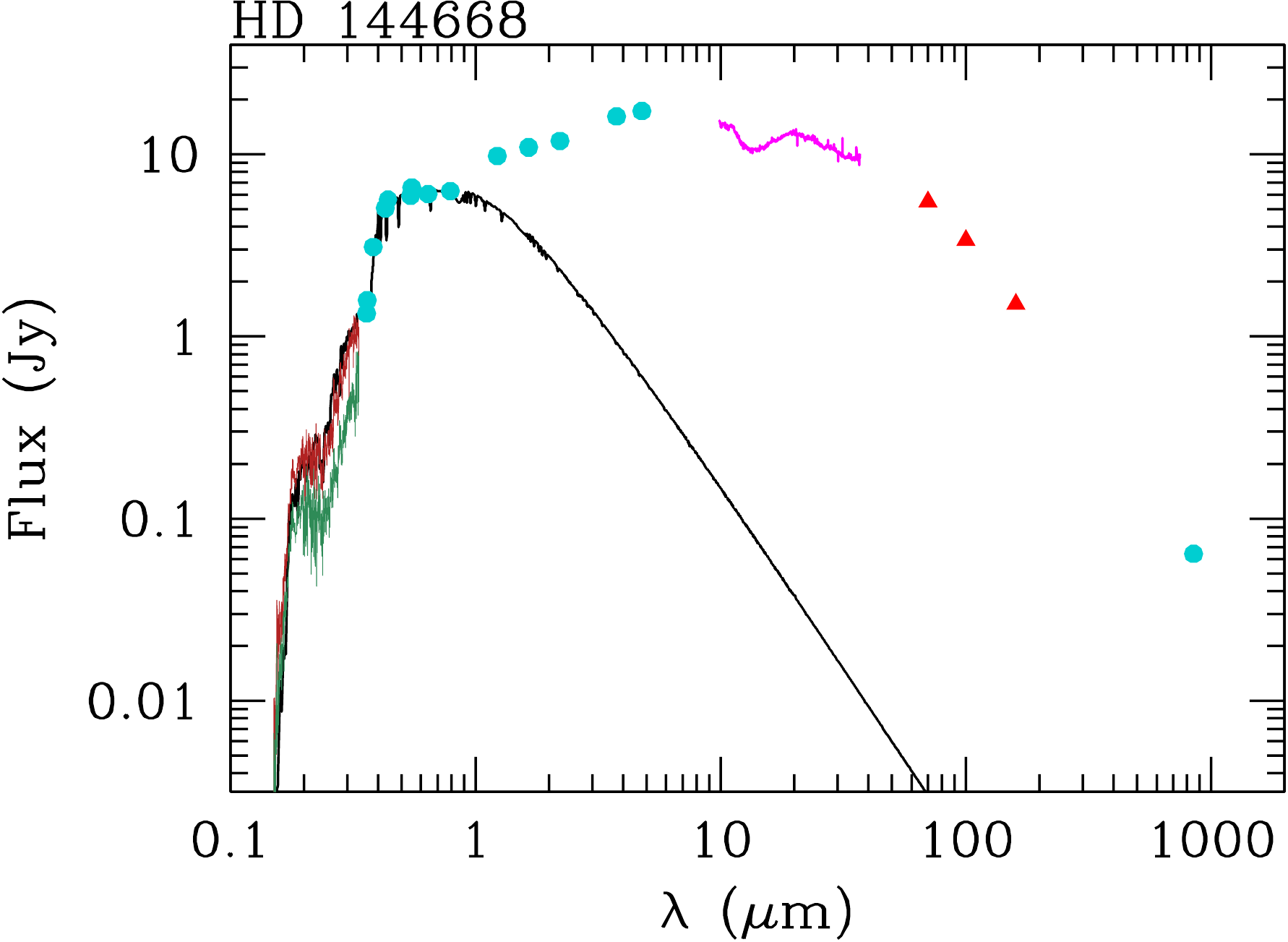}\hspace*{0.3cm}\includegraphics[scale=0.4]{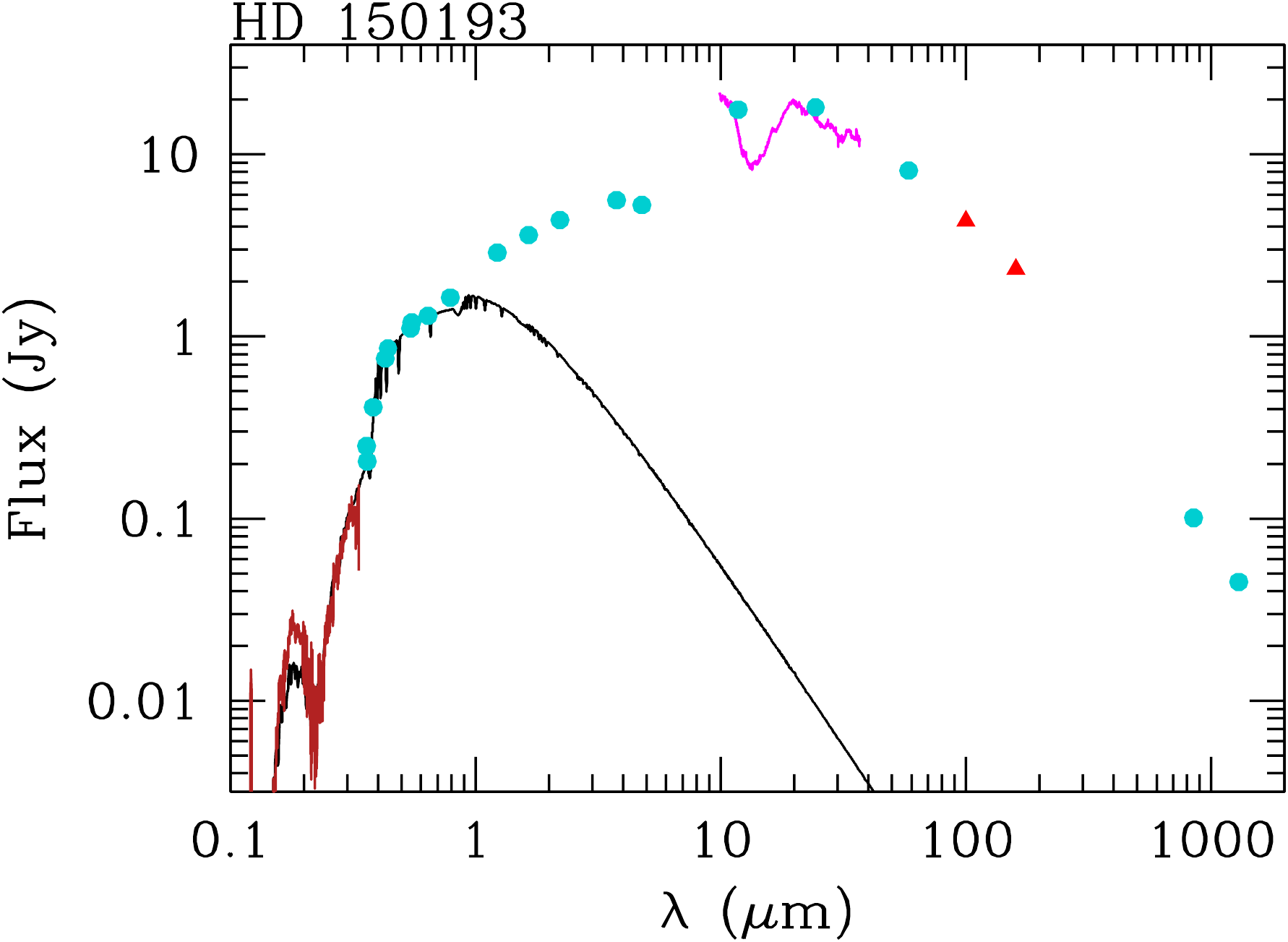}

\end{figure}
\clearpage

\addtocounter{figure}{-1}
\begin{figure*}
\caption{SEDs of GASPS Herbig Ae/Be stars (continued). The KK Oph fit was made manually and is shown in two wavelength intervals: 0.1-2.3 \mic \ and 0.1-4000 \mic. The models for the cool and hot components are in red and green, respectively, shown scaled according to their spectral types and luminosity classes. With only five photometry points in the optical, it was assumed that the contribution to the UV U and B flux comes mostly from the hot component, and the extinction was adjusted for this component. Then, the model for the cool component was reddened with different amounts of A$_{V}$ until the total flux at V and R was matched by sum of the models of the hot and cool components (both of them reddened); the final composite model is shown in black.}
\vspace*{1.0cm}
\hspace*{0.5cm}\includegraphics[scale=0.4]{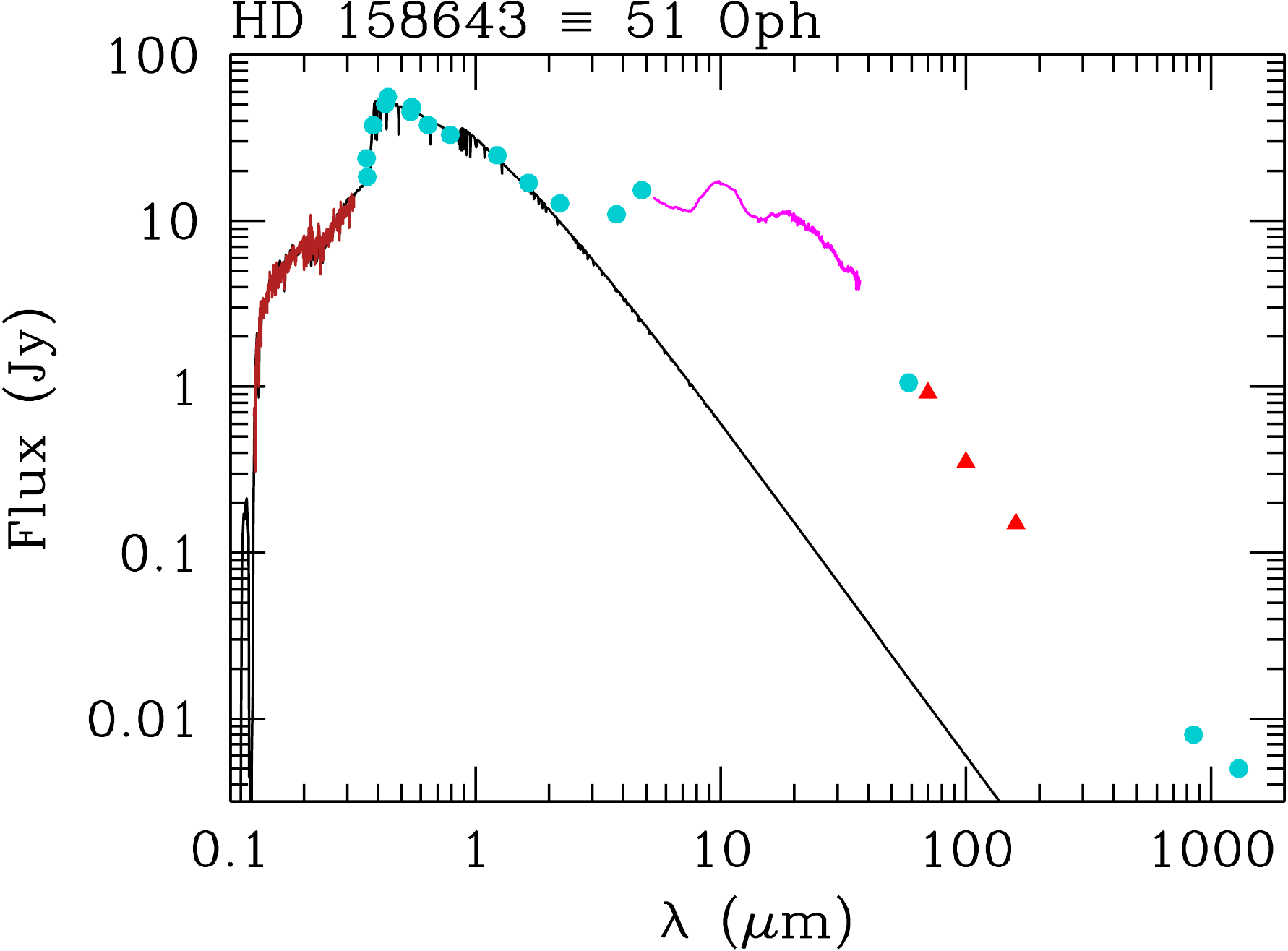}\hspace*{0.3cm}\includegraphics[scale=0.4]{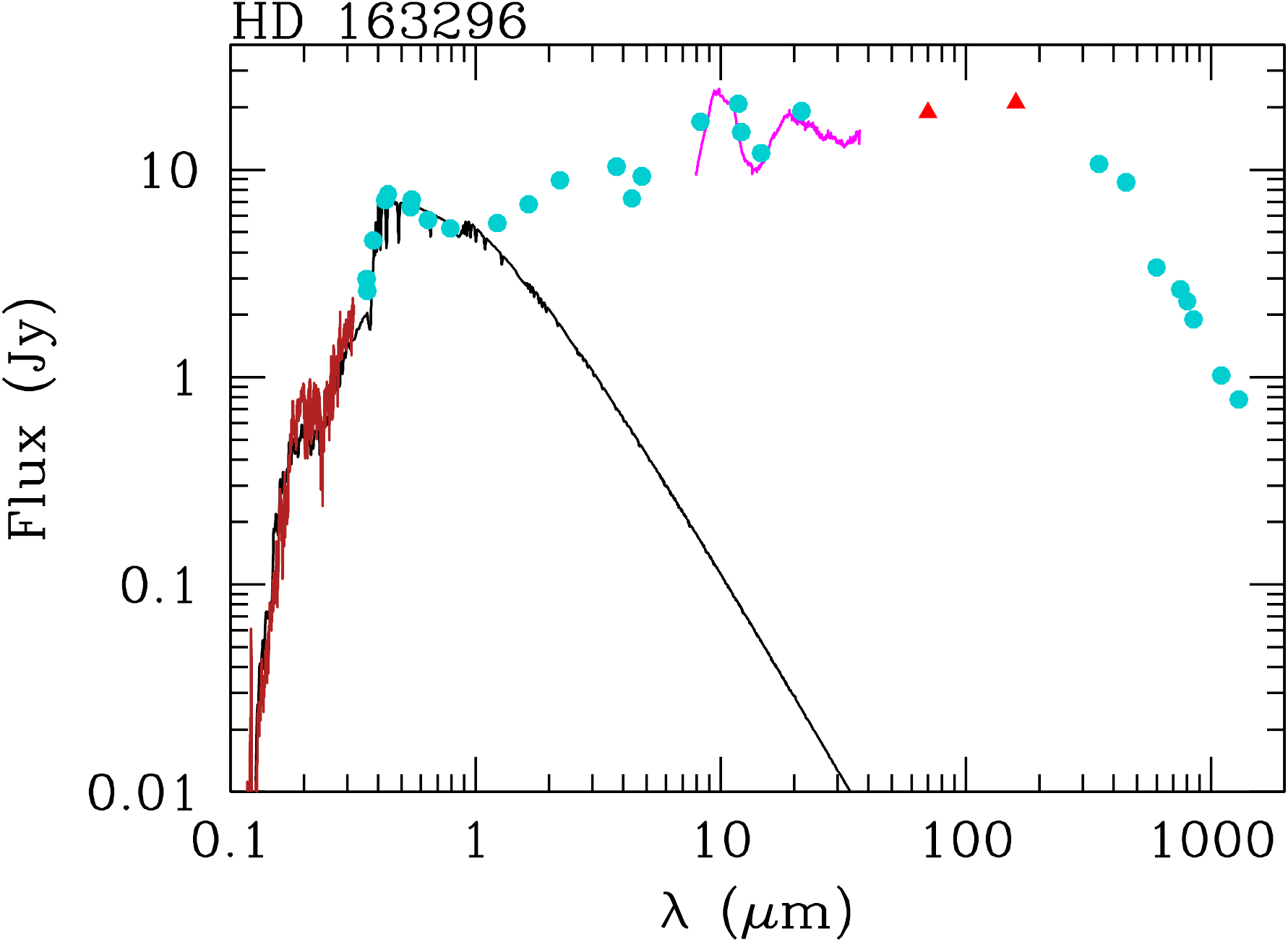}
\hspace*{0.5cm}\includegraphics[scale=0.4]{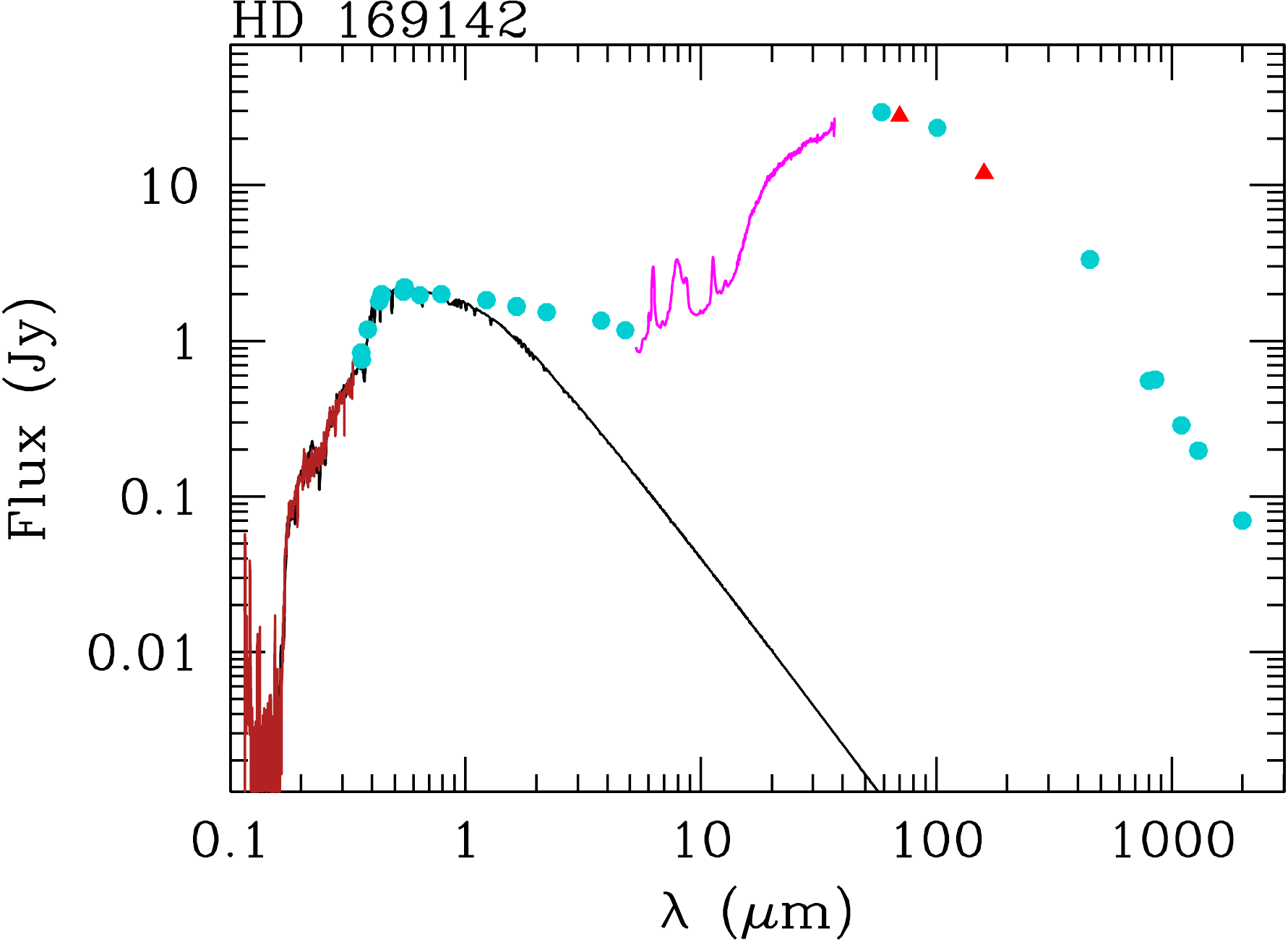}\hspace*{0.5cm}\includegraphics[scale=0.4]{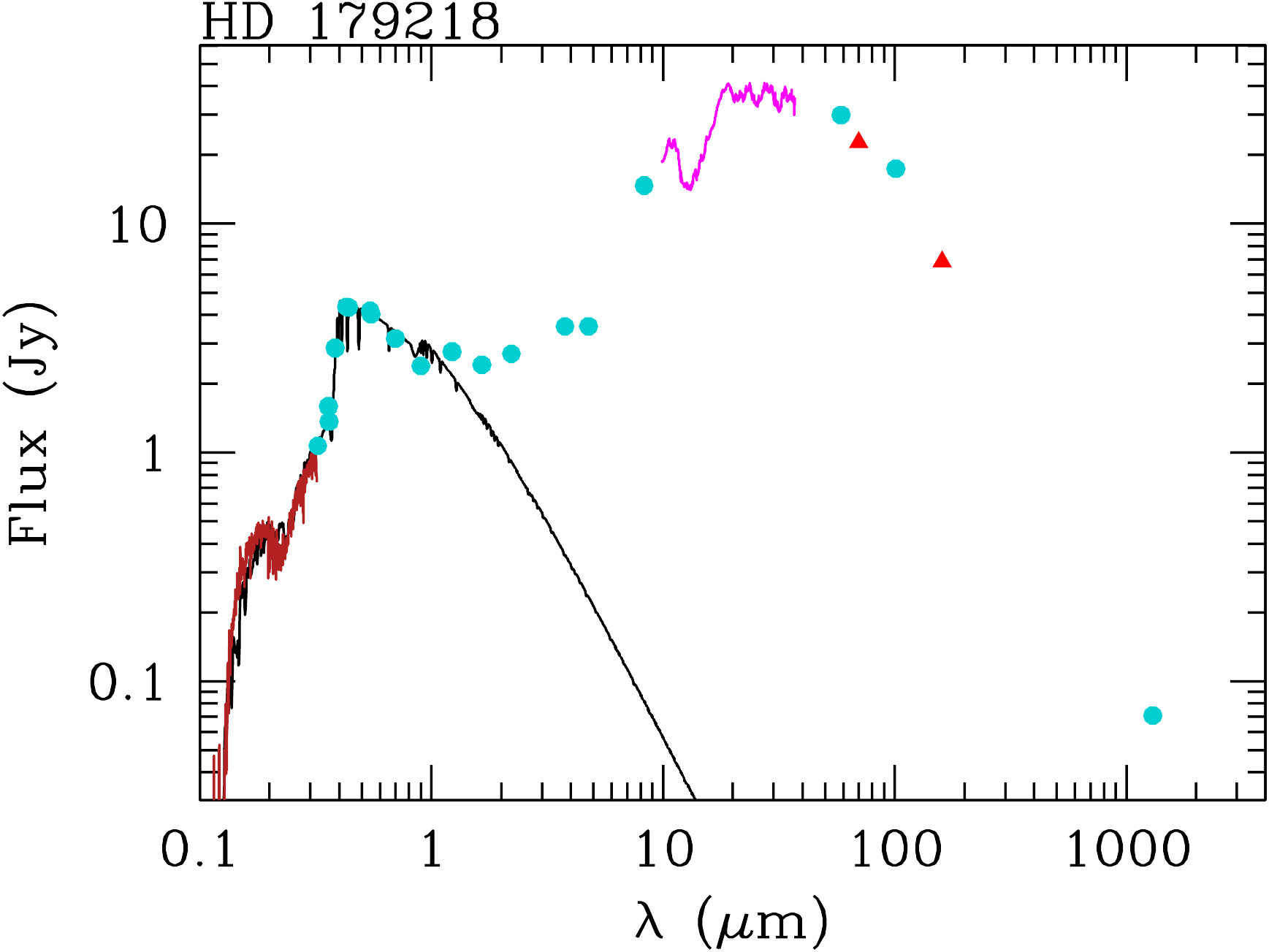}
\hspace*{0.3cm}\includegraphics[scale=0.6]{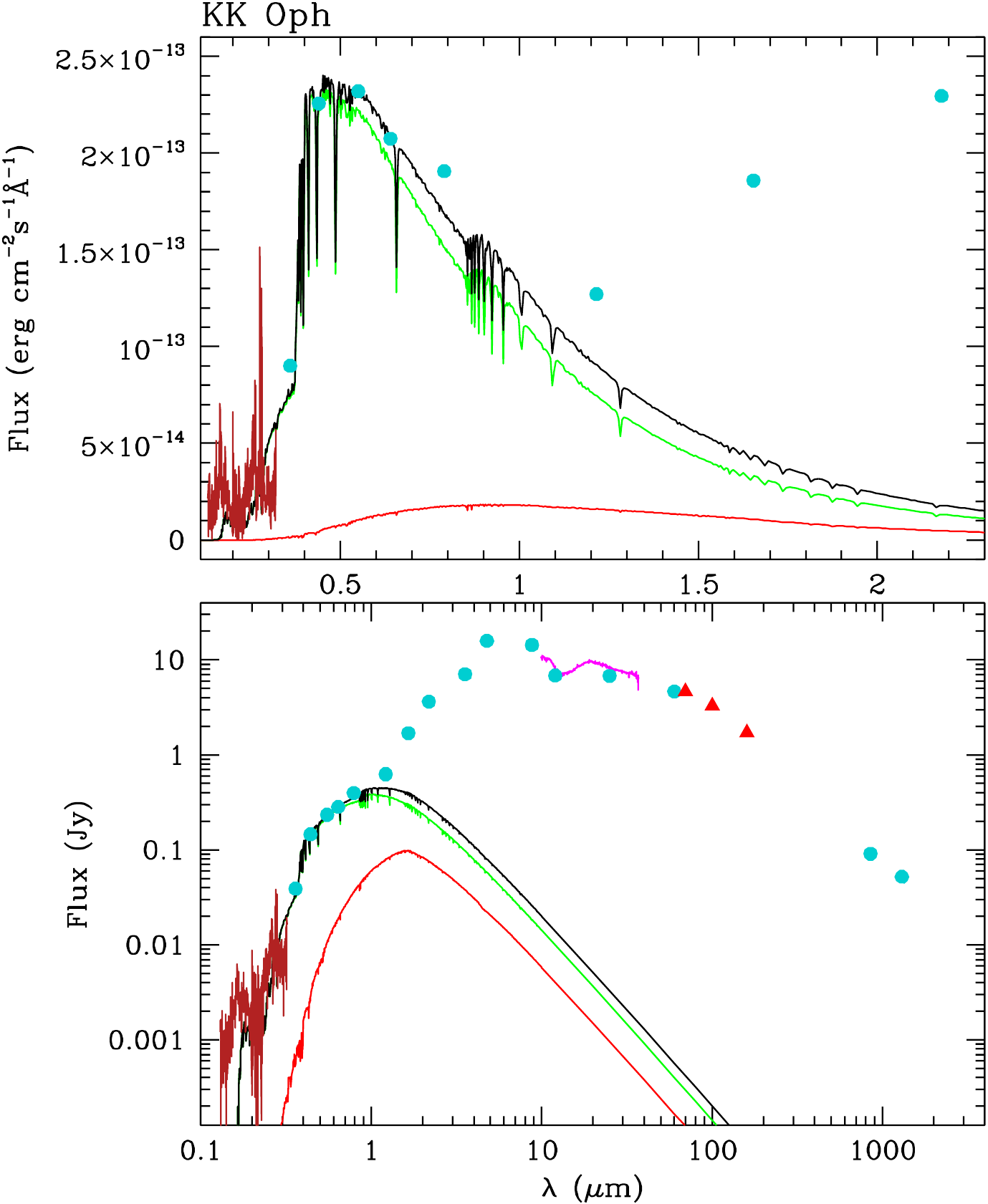}
\end{figure*}
\clearpage

\begin{figure}
\caption{SEDs of GASPS debris-disc host stars.}
\vspace*{1.0cm}

\hspace*{0.3cm}\includegraphics[scale=0.4]{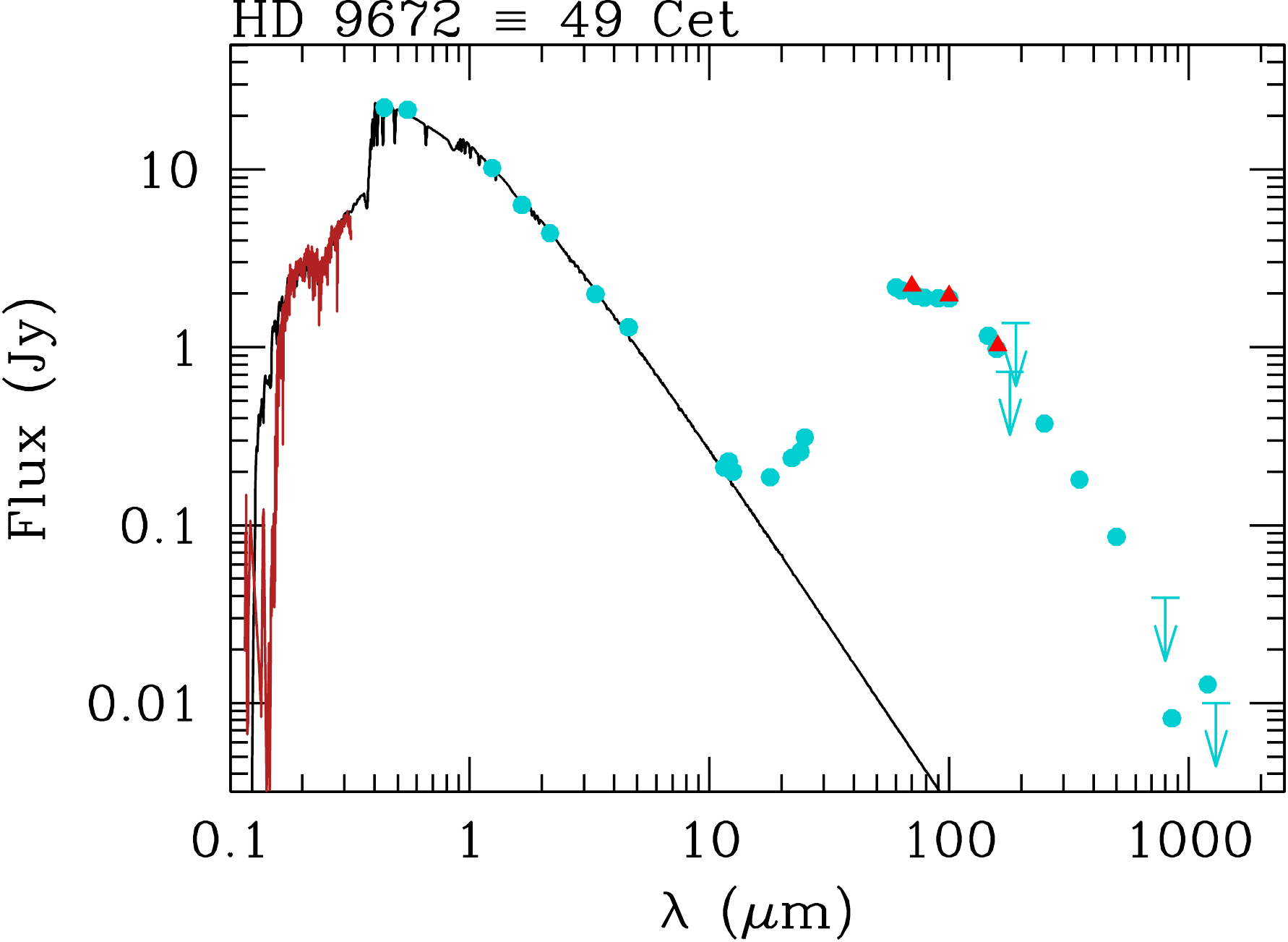}\hspace*{0.3cm}\includegraphics[scale=0.4]{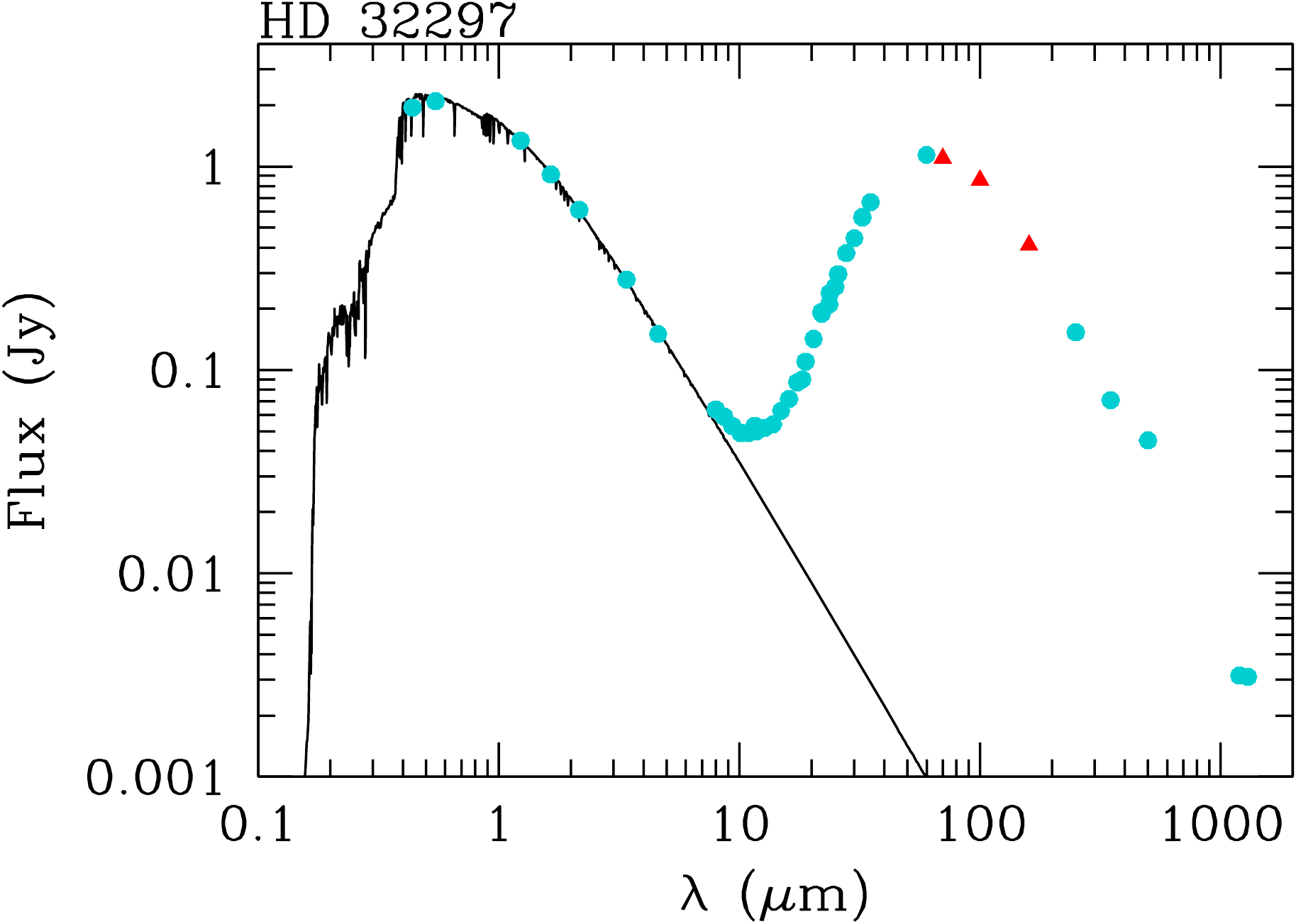}
\hspace*{0.3cm}\includegraphics[scale=0.4]{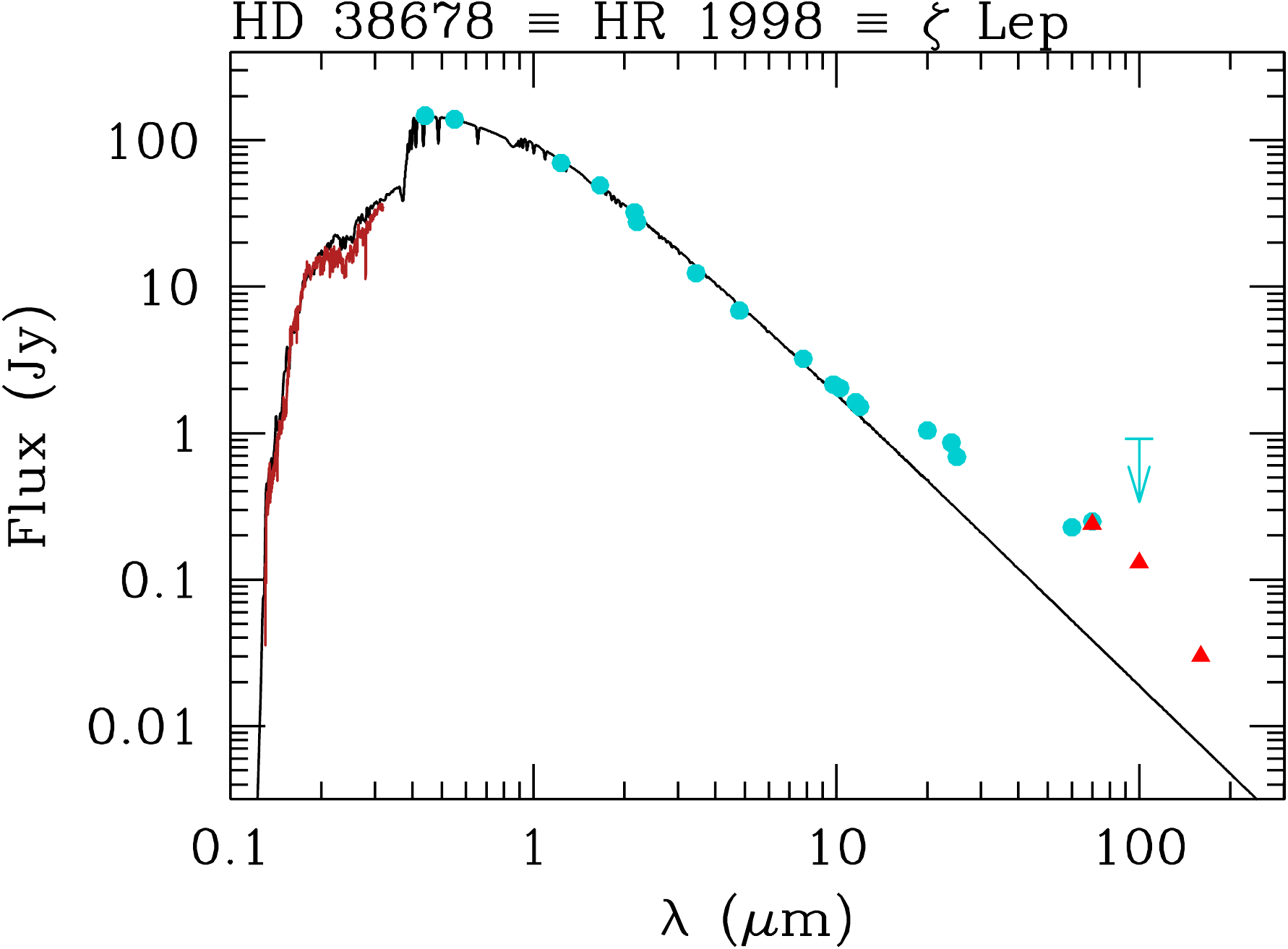}\hspace*{0.5cm}\includegraphics[scale=0.4]{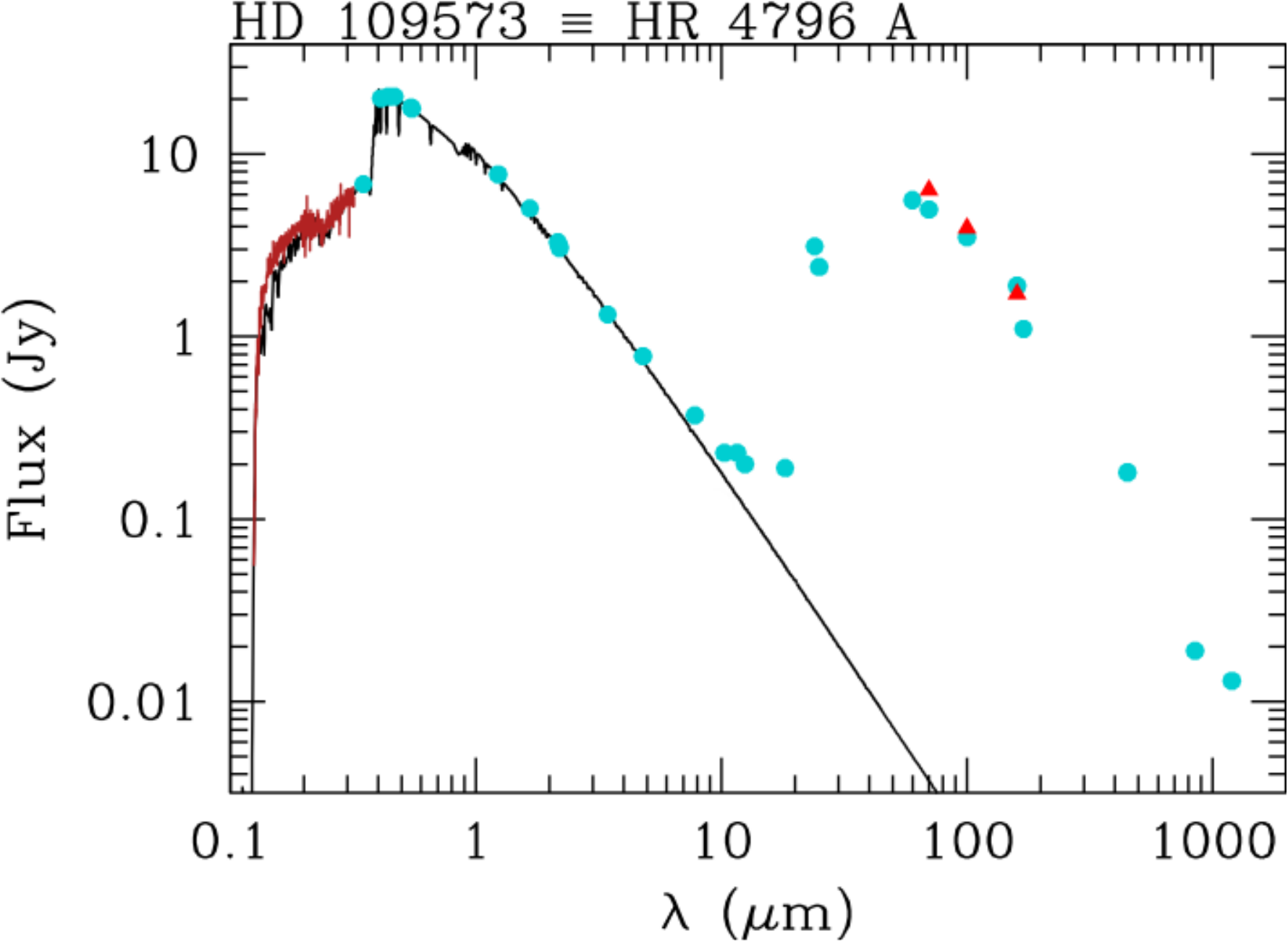}
\hspace*{0.3cm}\includegraphics[scale=0.4]{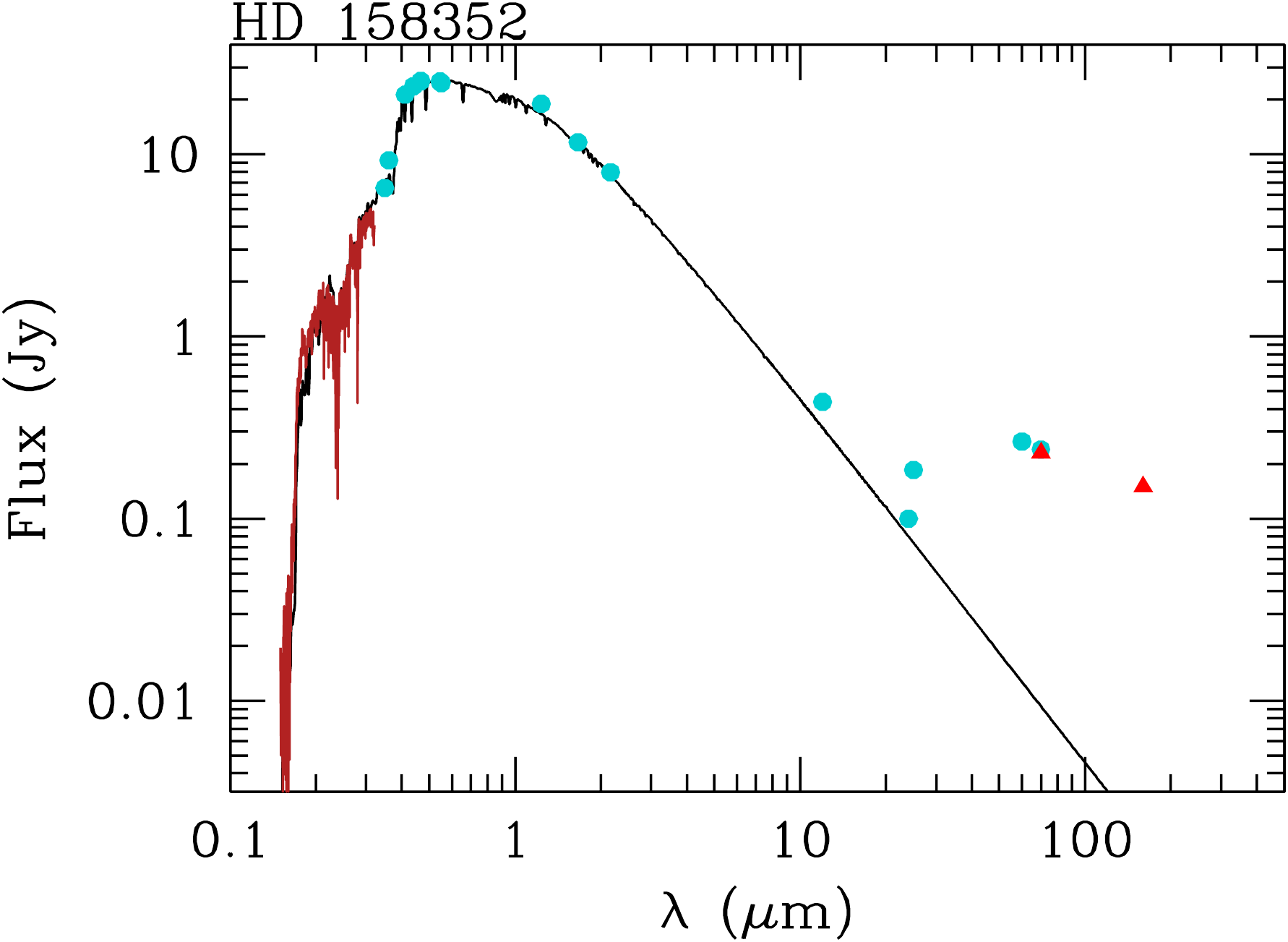}
\label{seds-debris}
\end{figure}

\begin{table*}[ht]
\begin{center}
\caption{Reddening values applied to the stellar atmosphere models in order to fit them to the optical photometry available,
$R_V$ = 3.1 is assumed.
}
\begin{tabular}{cccc}
\hline\hline
Star &   $E(B\!-\!V)$  & Star &$E(B\!-\!V)$\\
AB Aur        &   0.08 &HD 142666  &  0.30\\
HD 31648   &  0.05 &HD 144668  &  0.15\\
HD 35187   &  0.25 &HD 150193   & 0.50\\
HD 36112   &  0.05 &KK Oph A/B  &  0.52/0.90\\
CQ Tau       &  0.45 &51 Oph     &  0.12\\
HD 98922   &  0.20 &HD 163296  &  0.15\\
HD 97048   &  0.37 &HD 169142   & 0.00\\
HD 100453 &  0.00 &HD 179218  &  0.10\\
HD 100546 &  0.03 &------------& ------\\
HD 104237 &  0.05 &49 Cet   &    0.07\\
HD 135344B&   0.12 &HD 32297  &   0.20\\
HD 139614   & 0.00 &HR 1998   &   0.00\\
HD 141569   & 0.12 &HR 4796A   &  0.00\\
HD 142527   & 0.26 &HD 158352  &  0.00\\
\hline
\end{tabular}
\label{Av}
\end{center}
\end{table*}

\begin{table*}[ht]
\begin{center}\caption{References used for the construction of the SEDs:
2MA = 2MASS Point Source Catalog;
A73 = \citet{A73}; 
B92 = \citet{B92}; 
B11 = \citet{B11};
C76 = \citet{C76}; 
C95 =\citet{C95}; 
D90 = \citet{D90c}; 
D13 = \citet{Donaldson2013};
E01 = \citet{EI2001};
F92 = \citet{F92};
F98 = \citet{F98};
FSC = IRAS Faint source catalogue (cc = colour corrected);
G11 = \citet{G11};
H89 = \citet{H89};
H92 = \citet{H92};
H93 = \citet{H93};
H94 = \citet{H94};
H94b =\citet{H94b};
H98 = \citet{H98};
IRS = Spitzer IRS data \citep{Houck2004};
J96 = \citet{J96}; 
K85 = \citet{K85};
L05 = \citet{L05}; 
L90 = \citet{L90};
M94 = \citet{M94};
M97 = \citet{M97}; 
M98 = \citet{M98}, transformed from Geneva into Johnson system using transformation formula by \citet{Harmanec2001}; 
M99 = \citet{M99};
M00 = \citet{M00};
M01 = \citet{Meeus2001};
OU01 = \citet{OU2001};
PSC = IRAS Point Source Catalogue ;
R08 = \citet{R08};
SI = SIMBAD;
S96 = \citet{S96};
S01 = Sandell \& Weintraub (unpublished JCMT data); 
S04 =\citet{S04};
S06 = \citet{S06};
S11 = \citet{S11}; 
T81 = \citet{T81}; 
T85 = \citet{T85}; 
T86 = \citet{T86}; 
T01 =\citet{T01};
V89 = \citet{V89};
W01 = \citet{W01} 
W88 = \citet{W88}; 
W92 = \citet{W92}; 
W95 = \citet{W95};
W07 = \citet{W07}
}
\begin{tabular}{lllrlllllll}
\hline
        &Optical        & Near-infrared & Mid-infrared  & \textit{IRAS} &(sub-)mm \\
        &Johnson/Cousins& JHKLM   &NQ       &12-100 \mic & 350-2700 \mic\\
\hline
AB Aur   & C76 &A73,C76 &H92 &W92 &M94, M97, S11\\
HD 31648  & -- &2MA, A73 &M98 &PSC &M97, S11\\ 
HD 35187  & S96 &S96 &-- &PSC &S96, W95, S11\\
HD 36112  & M98 &M98 &M98 &PSC & S11\\
CQ Tau   & OU01 &E01 & -- &PSC & B11, M00, M97, G11, T01\\
HD 97048  & K85 &T86 &B92 &W92 &H93\\
HD 98922 & W01 & 2MA& IRS &-- &--\\
HD 100453 & M98 &F92 &M98 &PSC &--\\
HD 100546 & ESO &H89 &M98 &PSC &H94b\\
HD 104237 & ESO &H89 &M98 &FSC &H94b\\
HD 135344B& C95 &C95 &M98 &FSC &S96, C95, S11\\
HD 139614 & V89 &V89 &M98 &FSC &S96\\
HD 141569A& ESO &S96 &M98 &FSC &S96, W95, S11\\
HD 142527 & M98 &M98 &M98 &PSC &W95\\
HD 142666 & S96 &S96 &M98 &FSC &S96, S11\\
HD 144668 & T81 &T81 &H94 &W92 &S01, H92, S11\\
HD 150193 & K85 &K85, D90c &B92, D90c & M97, J96& S11\\
KK Oph   & H92& H92 & H92 &PSC & S11, H94, S11\\
51 Oph   & M98 & W88  &M98 &PSC &S96\\
HD 163296 & M98 &T85 &B92 &H92 &M94, S11\\
HD 169142 & V89 &S96 &M98 &PSC &S96, S11\\
HD 179218 & M99 & M99 & L90 & -- &M00 \\
\hline
49 Cet & SI, H98 & E01,2MA & W07 &FSC & W07, M01\\
HD 32297 & D13 & D13 & D13 & D13 & D13 \\
HR 1998 & SI & 2MA & F98 &FSC, S06, F98 &--\\
HR 4796A & SI, H98 & 2MA & F98 & PSC, L05, M01& S04\\
HD 158352 & SI, H98 & 2MA &-- &R08, FSC cc &--\\
\hline
\end{tabular}
\label{literature}
\end{center}
\end{table*}

\section{PACS observation identifications}

In Table~\ref{obsids} we present the \textit{Herschel}/PACS observation log, listing the observation IDs and integration times and observation dates for
every target in our sample. Concatenated scan pairs (or quartets) are denoted by a `/' between numbers at the end
of the observation ID, whilst distinct observations of the same target are separated by a `,'.

\begin{table*}[h]
\begin{center}
\caption{Observation log.}
\begin{tabular}{llcc}
\hline\hline 
Star & Observation ID & Integration times & Obs date \\
     &                &  (s)      & (YY-MM-DD)        \\
\hline
 AB Aur    & 1342228443/4 & 336/336 & 2011-09-07\\
 HD 31648  & 1342193131,1342217510/1 & 220,276/276 &2010-03-31, 2010-03-30\\
 HD 35187  & 1342217498/499/500/501 & 276/276/276/276  & 2011-03-30\\
 HD 36112  & 1342217502/3/4/5 & 276/276/276/276 & 2011-03-30\\
 CQ Tau    & 1342218557/58/59/60 & 276/276/276/276 & 2011-04-11 \\
 HD 97048  & 1342188847,1342223488/9 & 220,276/276 & 2010-01-02, 2011-06-23\\
 HD 98922  & 1342249132 & 52 & 2012-08-06\\
 HD 100453 & 1342188853,1342222616/7 & 220,276/276 & 2010-01-02, 2011-06-14\\
 HD 100546 & 1342188879,1342223466/7 & 220,276/276 & 2010-01-03, 2011-06-22\\ 
 HD 104237 & 1342188848 & 220 & 2010-01-02\\
 HD 135344B& 1342215603/4/5/6 & 276/276/276/276 & 2011-03-07\\
 HD 139614 & 1342215599/600/601/602 & 276/276/276/276 & 2011-03-07\\
 HD 141569 & 1342215382/3 & 276/276 & 2011-03-06\\
 HD 142527 & 1342216045/6/7/8 & 276/276/276/276 & 2011-03-14\\
 HD 142666 & 1342215470/1/2/3 & 276/276/276/276 & 2011-03-07\\
 HD 144668 & 1342262481/2/3/4 & 160/160/160/160 & 2013-01-29\\
 HD 150193 & 1342216497/8  & 276/276 & 2011-03-21\\
 KK Oph    & 1342205976/7/8/9 & 276/276/276/276 & 2010-10-07\\
 51 Oph    & 1342193054,1342205974/5 & 220,276/276 & 2010-03-30, 2010-10-06/07\\
 HD 163296 & 1342228401/2 & 276/276 & 2011-09-10\\ 
 HD 169142 & 1342183656 & 159   & 2009-09-11    \\ 
 HD 179218 & 1342220085/6/7/8 & 276/276/276/276 & 2011-05-07\\
\hline
 49 Cet    & 1342188485,1342224377/78/79/80 & 220,1122/1122/1122/1122 & 2009-12-23, 2011-07-18 \\ 
 HD 32297  & 1342193125,1342217452/3 & 220,276/276 & 2010-03-31, 2011-03-30 \\
 HR 1998   & 1342206320/1,1342205200/1 & 276/276,445/445 & 2010-10-11, 2010-09-27 \\ 
 HR 4796A  & 1342188519,1342213852/3 & 220,276/276 & 2009-12-25, 2011-02-08 \\
 HD 158352 & 1342183652 & 159     & 2009-09-11   \\ 
\hline
\hline
\end{tabular}
\tablefoot{HD 158352 and HD 169142 were observed in ``staring mode'', a PACS observing mode that became obsolete in the course of 2010.}
\label{obsids}
\end{center}
\end{table*}

\section{Extended sources profiles}

As previously shown for AB Aur, here we present plots of the radial profiles of 
our extended sources compared to the radial profile of the point source $\alpha$ Boo.

\clearpage
\begin{figure}
\caption{Radial profiles for extended sources compared with a discless point source. One pixel is equivalent to 1 \arcsec.}
\vspace*{1.0cm}
\hspace*{0.3cm}\includegraphics[scale=0.4]{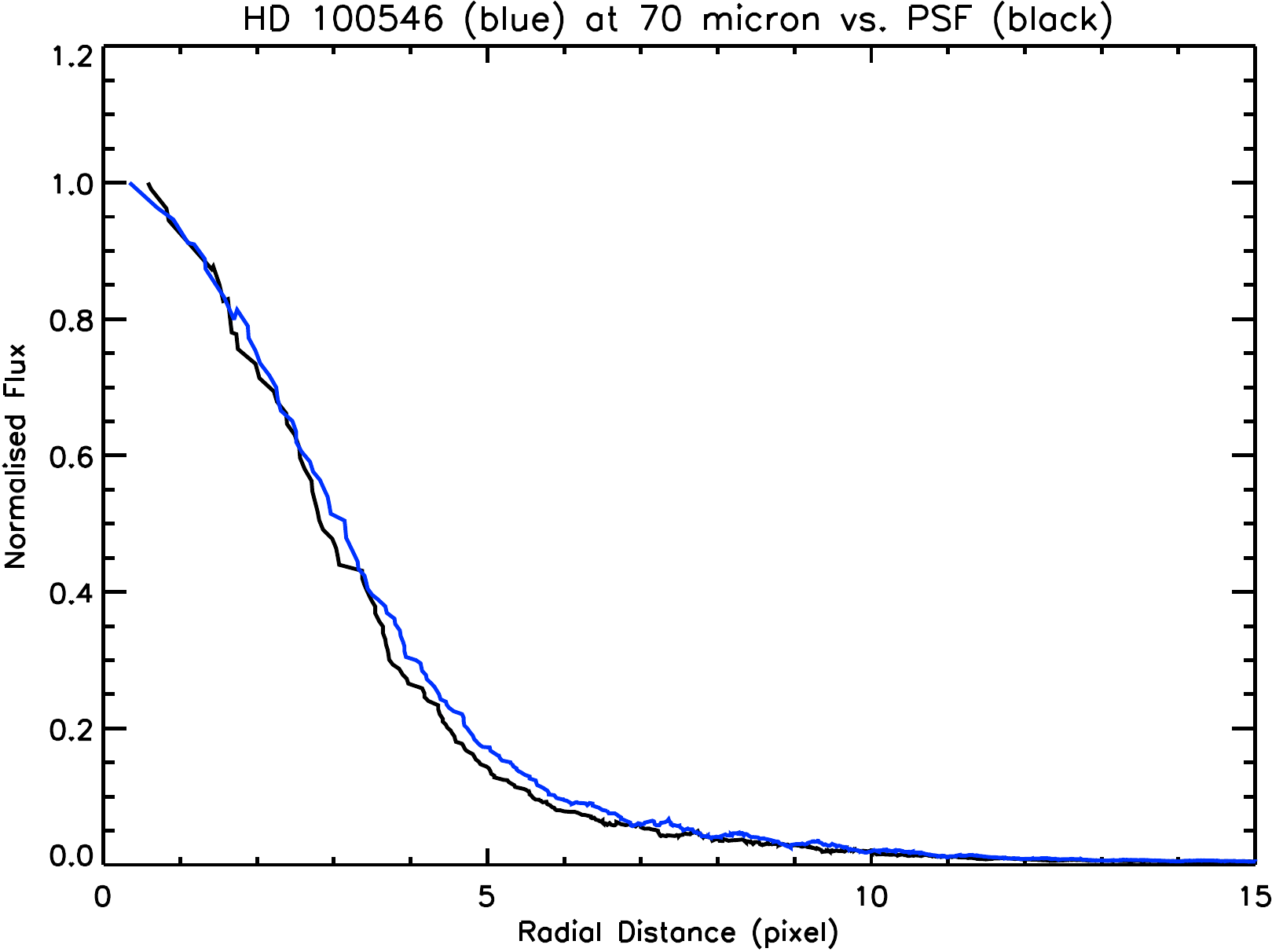}\hspace*{0.3cm}\includegraphics[scale=0.4]{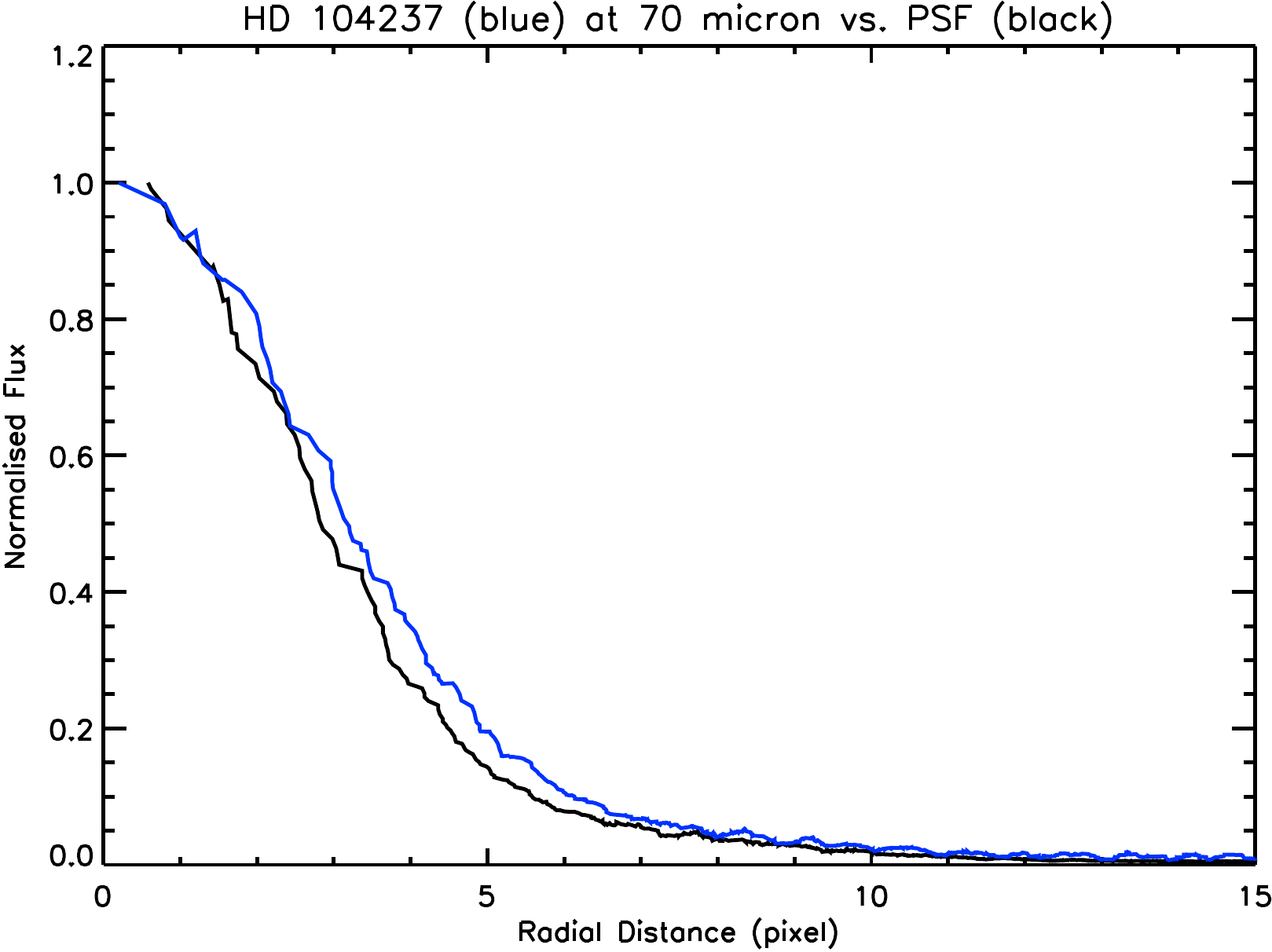}
\hspace*{0.3cm}\includegraphics[scale=0.4]{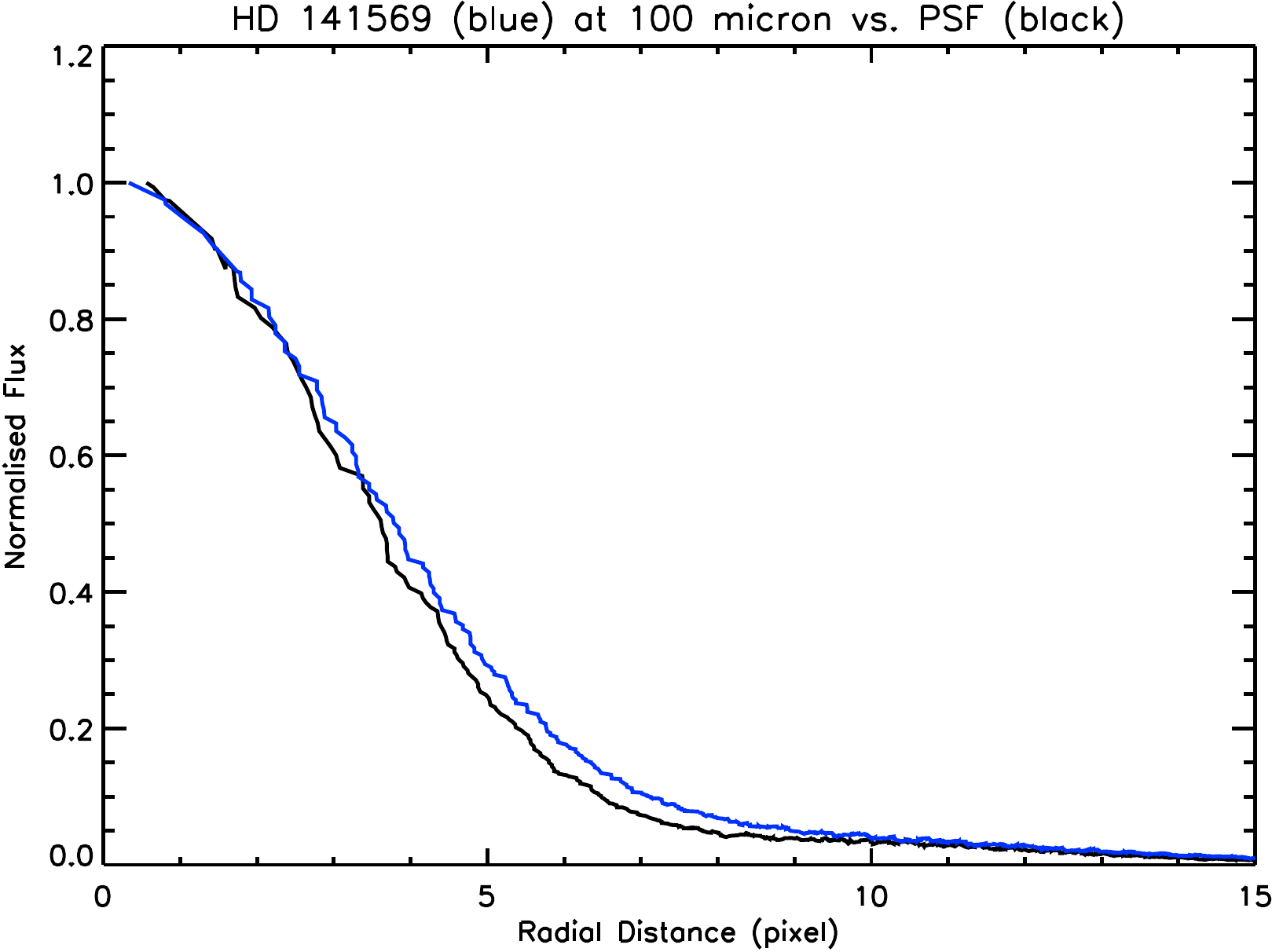}\hspace*{0.3cm}\includegraphics[scale=0.4]{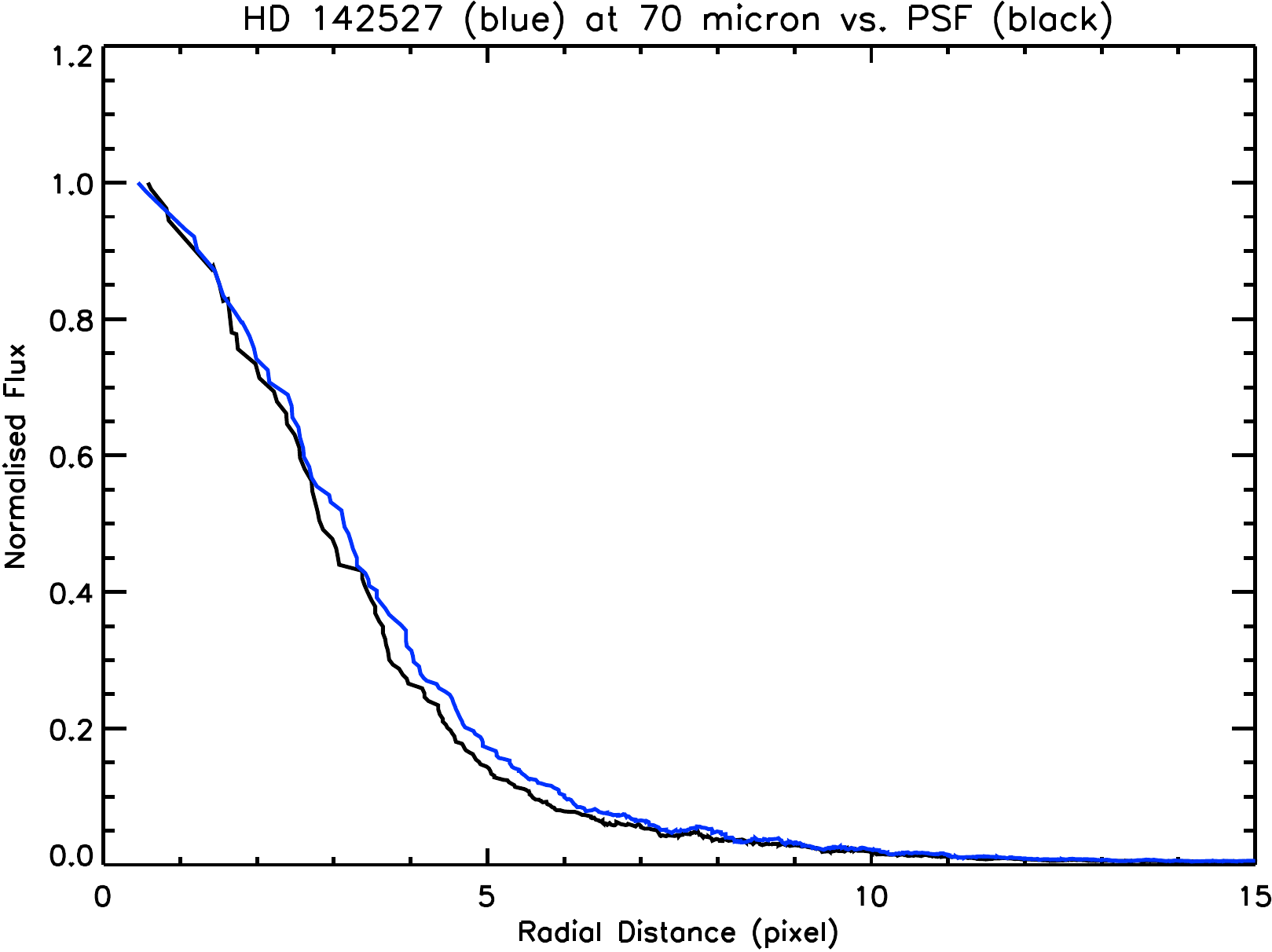}
\hspace*{0.3cm}\includegraphics[scale=0.4]{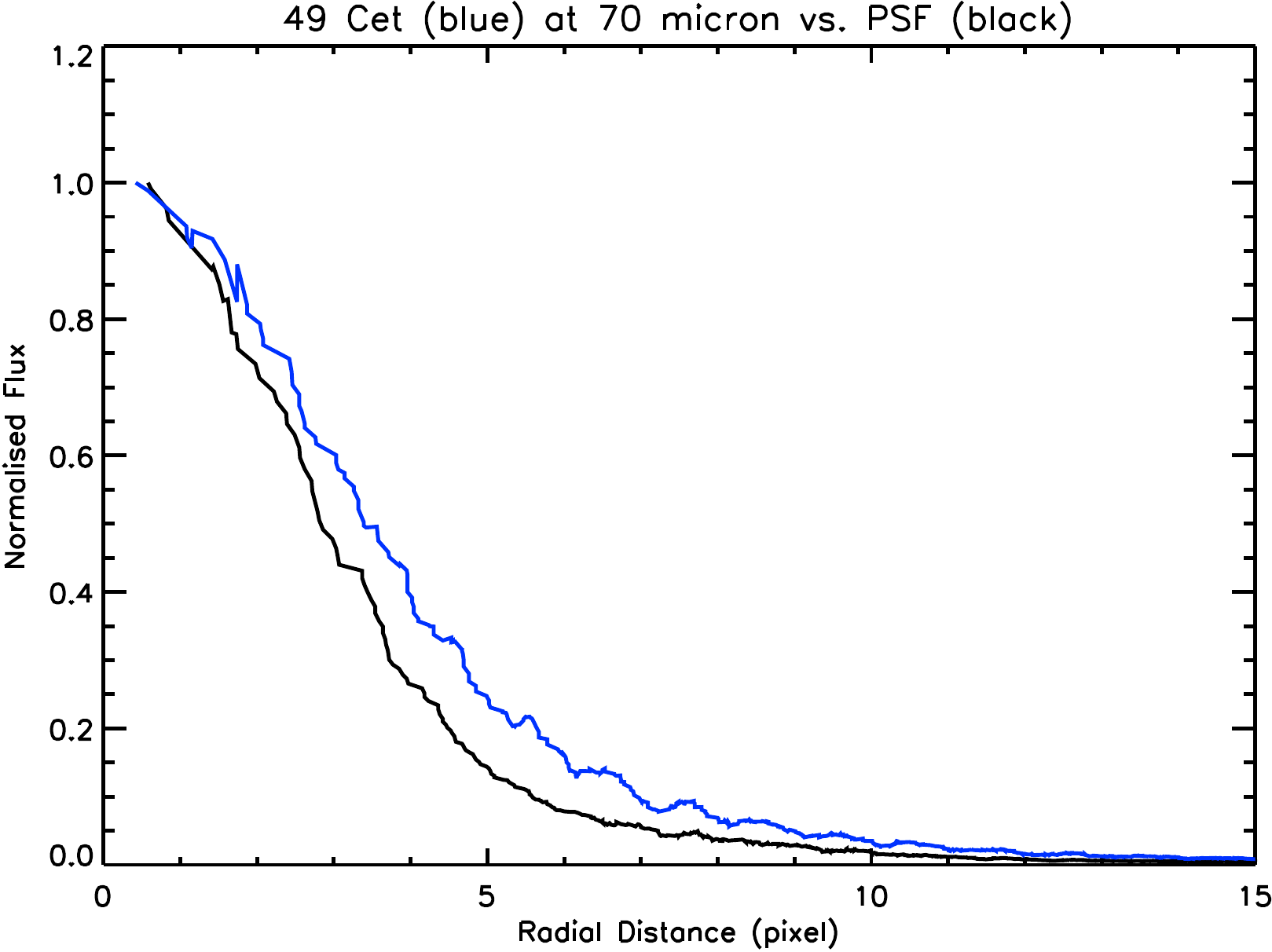}\hspace*{0.3cm}\includegraphics[scale=0.4]{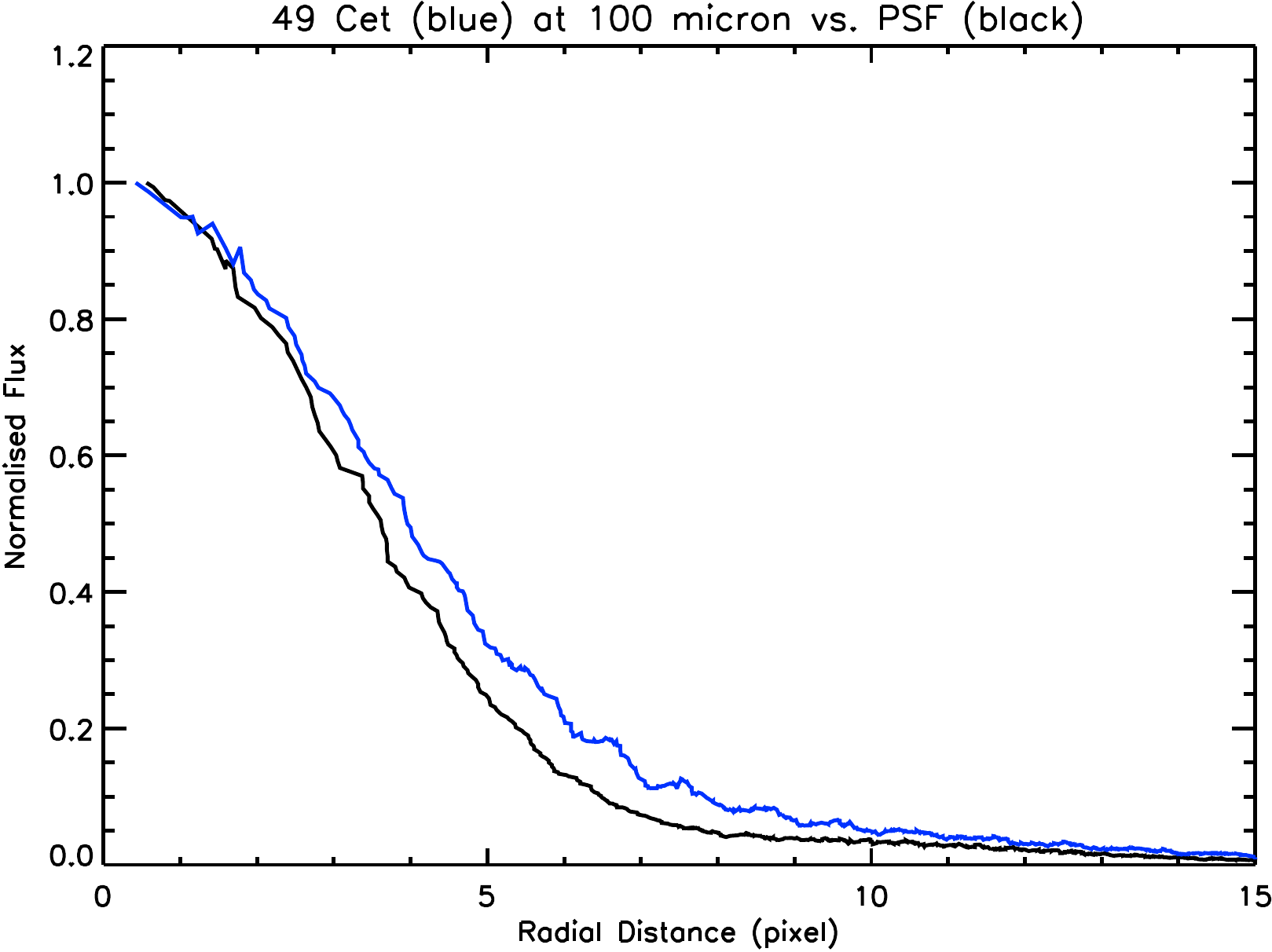}
\label{extended_profiles}
\end{figure}

\clearpage

\end{appendix}

\end{document}